\documentclass[a4paper,11pt]{article}

\usepackage{jheppub}
\usepackage{graphicx,color,amsmath}

\def\beq{\begin{eqnarray}}
\def\eeq{\end{eqnarray}}
\def\[{\left[}
\def\]{\right]}

\def\gsim{\lower.7ex\hbox{$\;\stackrel{\textstyle>}{\sim}\;$}}
\def\lsim{\lower.7ex\hbox{$\;\stackrel{\textstyle<}{\sim}\;$}}
\newcommand{\thdm}[0]{2HDM }
\newcommand{\thdmws}[0]{\rm 2HDM}

\newcommand{\bxsga}{B\to X_s \gamma}

\newcommand{\brbxsga}{{\rm BR}(B\to X_s \gamma)}

\newcommand{\mw}{M_W}
\newcommand{\mh}{m_{H^\pm}}

\newcommand{\muw}{\mu_W}
\newcommand{\mub}{\mu_b}

\title{\boldmath  Off-diagonal terms in  Yukawa textures of the Type-III 2-Higgs doublet model
and light charged Higgs boson phenomenology }

\author[a]{J. Hern\' andez--S\' anchez}
\emailAdd{jaimeh@ece.buap.mx}
 \affiliation[a]{Fac. de Cs. de la
Electr\'onica, Benem\'erita Universidad Aut\'onoma de Puebla, Apdo. Postal 542, 72570 Puebla, Puebla, M\'exico and Dual C-P Institute of High Energy Physics, M\'exico.}

\author[b]{S. Moretti}
\emailAdd{s.moretti@soton.ac.uk}
\affiliation[b]{School of Physics and Astronomy, University of Southampton, Highfield, Southampton SO17 1BJ, United Kingdom, and Particle Physics Department, Rutherford Appleton Laboratory, Chilton, Didcot, Oxon OX11 0QX, United Kingdom}

\author[c]{R. Noriega-Papaqui}
\emailAdd{rnoriega@uaeh.edu.mx}
\affiliation[c]{\'Area Acad\'emica de Matem\'aticas y F\'{\i}sica, Universidad Aut\'onoma del Estado
de Hidalgo, Carr. Pachuca-Tulancingo Km. 4.5, C.P. 42184, Pachuca, Hgo. and Dual C-P Institute of High Energy Physics, M\'exico.}

\author[d]{A. Rosado}
\emailAdd{rosado@ifuap.buap.mx}
\affiliation[d]{ Instituto de
F\'{\i}sica, BUAP, Apdo. Postal J-48, C.P. 72570 Puebla, Pue.,
M\'exico.}

\abstract{
We discuss flavor-violating constraints and consequently possible charged Higgs boson phenomenology emerging from a four-zero Yukawa texture embedded within the Type-III  2-Higgs Doublet Model (2HDM-III).
Firstly, we show in detail how we can obtain several kinds of 2HDMs when some parameters in the Yukawa texture are absent. Secondly, we present a comprehensive study of the main $B$-physics constraints
 on such parameters induced by flavor-changing processes, in particular on the off-diagonal terms
of such a texture: i.e., from
$\mu -e$ universality in $\tau$ decays, several leptonic $B$-decays ($B \to \tau \nu$, $D \to \mu \nu$ and $D_s \to {l} \nu$), the semi-leptonic transition $B\to D \tau \nu$, plus
$B \to X_s \gamma$, including $B^0-\bar B^0$ mixing, $B_s \to \mu^+ \mu^-$ and the radiative decay $Z\to b \bar{b}$.
Thirdly, having selected the surviving 2HDM-III parameter space,
we show that the $H^- c \bar{b}$ coupling
can be very large over sizable expanses of it, in fact,
a very different situation with respect to
2HDMs with a flavor discrete symmetry (i.e., ${\mathcal{Z}}_2$) and very similar to the case of the Aligned-2HDM (A2HDM) as well as of models with three or more Higgs doublets. Fourthly, we study in detail the ensuing $H^\pm$ phenomenology
at the Large Hadron Collider (LHC), chiefly the $c\bar b \to H^+$  production mode and the $H^+\to c\bar b$ decay
channel while assuming $\tau^+\nu_\tau$ decays in the former and $t\to bH^+$ production in the latter, showing that significant scope exists in both cases.}

\keywords{Higgs Physics, Beyond Standard Model}
\arxivnumber{1212.6818}

\begin{document}
\maketitle
\flushbottom

\section{Introduction}
\label{sec:intro}

The main problem in flavor physics Beyond the Standard Model
(BSM) \cite{stanmod} is to control the presence of Flavor
Changing Neutral Currents (FCNCs) that have been
observed to be highly suppressed by a variety of experiments. Almost all BSM scenarios that describe
physics in energy regions higher than the Electro-Weak (EW) scale have
contributions with FCNCs at tree level, unless some symmetry is
introduced in the scalar sector to suppress them. One of the most
important extensions of the SM is the 2-Higgs Doublet Model (2HDM)
\cite{Barger:1989fj,kanehunt,Aoki:2009ha}, due to
its wide variety of dynamical features and the fact that it can
represent a low-energy limit of general models like the Minimal
Supersymmetric Standard Model (MSSM). There are several realizations of the
2HDM, called Type I, II, X and Y (acronymed as 2HDM-I \cite{Haber:1978jt}, 2HDM-II \cite{Donoghue:1978cj}, 2HDM-X and 2HDM-Y \cite{Barnett:1983mm, Grossman:1994jb, hep-ph/9603445,Akeroyd:1994ga}) or \emph{inert} Types, wherein
(part of) the scalar particle content does not acquire a Vacuum Expectation Value (VEV)
\cite{Ma:2008uza,Ma:2006km, Barbieri:2006dq,LopezHonorez:2006gr}.
In the most general version of a 2HDM, the fermionic couplings of the neutral scalars are non-diagonal in flavor and, therefore, generate unwanted FCNC phenomena. Different ways
to suppress FCNCs have been developed, giving rise to a variety of specific implementations of the \thdmws.
The simplest and most common approach is to impose a $\mathcal{Z}_2$ symmetry forbidding all non-diagonal  terms in flavor space
in the Lagrangian
\cite{Glashow:1976nt}. Depending on the charge assignments under this symmetry, the model is called Type I, II, X and Y  or \emph{inert}.
There are other suggestions for the most general 2HDM: (i) the alignment in flavor space of the Yukawa couplings of the two scalar doublets, which guarantees the absence of tree-level FCNC interactions \cite{Pich:2009sp,Zhou:2003kd};  (ii) the Lepton Flavor Violating (LFV) terms introduced
as a deviation from the Model II Yukawa interactions in
\cite{Kanemura:2005hr,Kanemura:2004cn}; (iii) the 2HDM-III with  a particular  Yukawa `texture',  forcing the non-diagonal Yukawa couplings to be proportional to the geometric mean of the two fermion masses, $g_{ij}\propto \sqrt{m_i m_j} \chi_{ij}$ \cite{Cheng:1987rs,Atwood:1996vj,DiazCruz:2004pj,DiazCruz:2009ek, Mahmoudi:2009zx}\footnote{It is well known that, through
Yukawa textures \cite{fritzsch,Fritzsch:2002ga}, it is possible to build
a matrix that preserves the expected Yukawa couplings that depend on
the fermion masses. From a phenomenological point of view, the Cheng-Sher {\it ansatz} \cite{Cheng:1987rs} has been very
useful to describe the phenomenological content of the corresponding Yukawa matrix
and the salient features of the hierarchy of quark masses.}; (iv) recently, a Partially Aligned 2HDM (PA2HDM) was presented and a four-zero texture is employed
therein too, so that this newly suggested scenario includes (i) and (iii) as particular cases \cite{HernandezSanchez:2011ti}.

Therefore, the mechanism through which the FCNCs are
controlled defines the actual version of the model and the consequently different
phenomenology that can be contrasted with experiment. In particular, we focus here on the
version where the Yukawa couplings depend on the
hierarchy of masses. This version is the one where the mass matrix has a four-zero
texture form \cite{fritzsch,Fritzsch:2002ga}. This matrix is based on the phenomenological
observation that the off-diagonal elements must be small in order to
dim the interactions that violate flavor, as experimental results
show. Although the phenomenology of Yukawa couplings constrains the
hierarchy of the mass matrix entries, it is not enough to determine
the strength of the interaction with scalars. Another assumption on
the Yukawa matrix is related to the additional Higgs doublet. In
versions I and II  a discrete symmetry is introduced on the Higgs
doublets, fulfilled by the scalar potential, that leads to the
vanishing of most of the free parameters. However, version III,
having a richer phenomenology, requires a slightly more general
scheme. Interesting phenomenological implications of  2HDMs with a four-zero texture for the charged Higgs boson sector \cite{DiazCruz:2004pj, DiazCruz:2009ek, BarradasGuevara:2010xs} and neutral Higgs boson sector \cite{GomezBock:2005hc, HernandezSanchez:2011fq} have been studied. In these works one estimated  the order of the parameters $\chi_{ij} = {\cal O}(1)$, including the off-diagonal terms of the Yukawa texture. In contrast, a complete and detailed analysis that includes off-diagonal terms of the Yukawa texture in presence of the recent data of processes at low energy has been omitted in previous works. Therefore, in this paper, we present  a comprehensive study of the main flavor constraints on the parameters that come from a four-zero Yukawa texture considering the off-diagonal terms and present their relevance for charged Higgs boson phenomenology at the Large Hadron Collider (LHC). The presence of a charged scalar $H^{\pm}$ is in fact one of the most distinctive features of a two-Higgs doublet extended scalar sector. In the following, we analyze its phenomenological impact in low-energy flavor-changing processes within the 2HDM-III with a four-zero texture and constrain the  complex parameters $\chi_{ij}$ therein using present data on different leptonic, semi-leptonic/semi-hadronic and hadronic decays.

If $m_{H^\pm} <  m_t-m_b$, such particles would most copiously
(though not exclusively  \cite{Aoki:2011wd,JPG}) be produced in the decays of top quarks via
$t\to  H^\pm b$ \cite{tbH}.
Searches in this channel have been performed by the Tevatron experiments, assuming
the decay modes $H^\pm\to cs$ and  $H^\pm\to \tau\nu$ \cite{:2009zh,Aaltonen:2009ke}.
Since no signal has been observed, constraints are obtained on the parameter space $[m_{H^\pm}, \tan\beta]$,
where $\tan\beta=v_2/v_1$ (i.e., the ratio of the VEVs of the two Higgs doublets).
Searches in these channels have now been carried out also at the LHC:
 for $H^\pm\to cs$ with 0.035 fb$^{-1}$ by ATLAS \cite{ATLAS:search} and
for $H^\pm\to \tau\nu$ with 4.8 fb$^{-1}$ by ATLAS \cite{ATLAS_Htau} plus with 1 fb$^{-1}$ by CMS \cite{CMS_Htau}.
These are the first searches for $H^\pm$ states at this collider. The constraints on $[m_{H^\pm}, \tan\beta]$
from the LHC searches for $t\to  H^\pm b$
are now more restrictive than those obtained from the corresponding Tevatron searches.

The phenomenology of $H^\pm$ states in models with three or more Higgs doublets,
called Multi-Higgs Doublet Models (MHDMs), was first studied comprehensively in \cite{Grossman:1994jb}, with an
emphasis on the constraints from low-energy processes (e.g., the decays of mesons).
Although the phenomenology of $H^\pm$ bosons at high-energy colliders in MHDMs and 2HDMs has many similarities,
the possibility of $m_{H^\pm}< m_t-m_b$ together with an enhanced Branching Ratio (BR) for
$H^\pm\to cb$ would be a distinctive feature of MHDMs. This scenario, which was
first mentioned in \cite{Grossman:1994jb} and studied in more detail originally in
\cite{Akeroyd:1994ga,Akeroyd:1995cf,Akeroyd:1998dt} and most recently in \cite{Akeroyd:2012yg}, is of immediate interest for the
ongoing searches for $t\to  H^\pm b$ with $H^\pm\to cs$ by the LHC \cite{ATLAS:search}.
Although the current limits on $H^\pm\to cs$  can also be applied to
the decay $H^\pm \to cb$ (as discussed in \cite{Logan:2010ag} in the context of the Tevatron searches), a further
improvement in sensitivity to  $t\to H^\pm b$ with $H^\pm \to cb$ could be obtained by tagging the
$b$ quark which originates from $H^\pm$ decays \cite{Akeroyd:1995cf,DiazCruz:2009ek,Logan:2010ag,Akeroyd:2012yg}.
{Large values of BR$(H^\pm \to cb)$ are also possible in certain 2HDMs,
such as  the ``flipped 2HDM''  with Natural Flavor Conservation
(NFC) \cite{Akeroyd:1994ga,Aoki:2009ha,Logan:2010ag}.
However, in this model one would generally expect $m_{H^\pm} \gg m_t$, due to the
constraint from $b\to s \gamma$  ($m_{H^\pm} > 295$ GeV \cite{Borzumati:1998nx,Misiak:2006zs}) so that
$t\to H^\pm b$ with $H^\pm \to cb$ would not proceed. However, in our version of the 2HDM-III  there are additional
new physics contributions which enter $b\to s \gamma$, thus weakening the constraint on $m_{H^\pm}$ \cite{DiazCruz:2009ek}.  }
We will estimate the increase in sensitivity to BR$(H^\pm \to cb)$ and
to the fermionic couplings of $H^\pm$ in the 2HDM-III scenario. Further, always in the latter, we will re-visit the possibility
of direct $H^\pm$ production from $cb$-fusion, where the on-shell $H^\pm$ state eventually decays to $\tau\nu_\tau$ pairs,
 its only resolvable signature in the context of fully hadronic machines.

We now proceed as follows. The formulation of the general \thdm with a four-zero texture for the Yukawa matrix is recalled in section~\ref{model}.
The phenomenological consequences of having a charged Higgs field are analyzed in the next section
in processes at low energy, extracting the corresponding constraints on the aforementioned new physics parameters $\chi_{ij}$, by
discussing the constraints derived from tree-level leptonic and semi-leptonic/semi-hadronic decays, while in section IV we discuss the
ensuing light charged Higgs phenomenology at the LHC. Finally, we elaborate our conclusions in section V.


\section{The Yukawa sector of the 2HDM-III with a four-zero Yukawa texture}\label{model}

In this section, we will discuss the main characteristics
of the general Higgs potential and the using of a specific
four-zero texture in the Yukawa matrices within the 2HDM-III. In this connection, notice that, when
a flavor symmetry in the Yukawa sector is implemented, discrete
symmetries in the Higgs potential are not needed, so that the most
general Higgs potential must be introduced.

\subsection{The general Higgs potential in the 2HDM-III}

The 2HDM includes two Higgs scalar doublets of hypercharge $+1$:
$\Phi^\dag_1=(\phi^-_1,\phi_1^{0*})$ and
$\Phi^\dag_2=(\phi^-_2,\phi_2^{0*})$. The most general $SU(2)_L \times U(1)_Y $
invariant  scalar potential  can be written as~\cite{Gunion:2002zf}
\begin{eqnarray}
V(\Phi_1,\Phi_2)&=&\mu^2_1(\Phi_1^\dag
\Phi_1)+\mu^2_2(\Phi^\dag_2\Phi_2)-\left(\mu^2_{12}(\Phi^\dag_1\Phi_2)+{\rm
H.c.}\right) + \frac{1}{2}
\lambda_1(\Phi^\dag_1\Phi_1)^2 \\ \nonumber
&& +\frac{1}{2} \lambda_2(\Phi^\dag_2\Phi_2)^2+\lambda_3(\Phi_1^\dag
\Phi_1)(\Phi^\dag_2\, \Phi_2)
+\lambda_4(\Phi^\dag_1\Phi_2)(\Phi^\dag_2\Phi_1) \\ \nonumber
&& +
\left(\frac{1}{2} \lambda_5(\Phi^\dag_1\Phi_2)^2+\left(\lambda_6(\Phi_1^\dag
\Phi_1)+\lambda_7(\Phi^\dag_2\Phi_2)\right)(\Phi_1^\dag \Phi_2)+
{\rm H.c.}\right), \label{potential}
\end{eqnarray}
where all parameters are assumed to be real\footnote{The  $\mu^2_{12}$, $\lambda_5$, $\lambda_6$ and $\lambda_7$ parameters are complex in general, but we will assume that they are real for simplicity.}. Regularly,  in  the 2HDM Type I and II the terms proportional to $\lambda_6$ and $\lambda_7$ are absent, because the  discrete symmetry $\Phi_1\to \Phi_1$ and
$\Phi_2\to -\Phi_2$ is imposed in order to avoid dangerous FCNC effects.
However, in our model, where mass matrices with a four-zero texture are considered, as intimated,
it is not necessary to implement the above discrete symmetry. Thence, one must keep the terms proportional to $\lambda_6$ and $\lambda_7$.  These parameters play an important
role in one-loop processes, where self-interactions
of Higgs bosons could be relevant \cite{HernandezSanchez:2011fq}. Besides,  the parameters $\lambda_6$ and $\lambda_7$ are essential to obtain the decoupling limit of the model in which only one CP-even scalar is light, as hinted by current Tevatron and LHC data \cite{latest_Higgs}. While these terms exist, there are two independent energy scales, $v$ and $\Lambda_{\rm 2HDM}$ (the scale at which
additional BSM physics is required to control persisting divergences in the Higgs masses and self-couplings), and the spectrum of Higgs boson masses is such that
$m_{h^0}$
 is of order $v$ whilst $m_{H^0}$, $m_{A^0}$ and $m_{H^\pm}$ are all of the order of $\Lambda_{\rm 2HDM}$
\cite{Gunion:2002zf}. Then,  the heavy Higgs bosons decouple in the limit  $\Lambda_{\rm 2HDM}\gg v$, according to the decoupling theorem \cite{kanehunt}. Conversely, when the scalar potential does respect the discrete symmetry, it is impossible to have two independent energy scales \cite{Gunion:2002zf}. This implies that all of the physical scalar masses lie at the EW scale $v$.
Being that $v$ is already fixed by experiment though, a very heavy Higgs boson can only arise by means of a large dimensionless
coupling constant $\lambda_i$. In this case, the decoupling theorem is not valid, thus opening the possibility for the
appearance of non-decoupling effects. Moreover, since the scalar potential contains some terms that violate the $SU(2)$
custodial symmetry, non-decoupling effects can arise in one-loop induced Higgs boson couplings \cite{Kanemura}.

The scalar potential (\ref{potential}) has been diagonalized to
generate the mass-eigenstates fields. The charged components of the
doublets lead to a physical charged Higgs boson and the
pseudo-Goldstone boson associated with the $W$ gauge field:
\begin{eqnarray}
&&G^\pm_W=\phi^\pm_1c_\beta+\phi^\pm_2s_\beta,\\
&&H^\pm=-\phi^\pm_1s_\beta+\phi^\pm_2c_\beta,
\end{eqnarray}
with
\begin{equation}
m^2_{H^\pm}=\frac{\mu^2_{12}}{s_\beta c_\beta}-\frac{1}{2} v^2 (\lambda_4+\lambda_5+t^{-1}_\beta \lambda_6+t_\beta \lambda_7),
\end{equation}
where we have introduced the short-hand notations, $t_\beta=\tan\beta$, $s_\beta=\sin\beta$ and $c_\beta=\cos \beta$.
Besides, the imaginary part of
the neutral components $\phi^0_{iI}$ defines the neutral CP-odd
state and the pseudo-Goldstone boson associated with the $Z$
gauge boson. The corresponding rotation is given by:
\begin{eqnarray}
&&G_Z=\phi^0_{1I}c_\beta+\phi^0_{2I}s_\beta, \\
&&A^0=-\phi^0_{1I}s_\beta+\phi^0_{2I}c_\beta,
\end{eqnarray}
where
\begin{equation}
m^2_{A^0}=m^2_{H^\pm}+\frac{1}{2} v^2(\lambda_4- \lambda_5).
\end{equation}
Finally, the real part of the neutral components of the
$\phi^0_{iR}$ doublets defines the CP-even Higgs bosons $h^0$ and $H^0$.
The mass matrix is given by:
\begin{equation}
M_{Re}=\left( \begin{array}{ccc} m_{11} & m_{12} \\
m_{12} & m_{22}\\
\end{array}\right),
\end{equation}
where
\begin{eqnarray}
&&m_{11}=m^2_{A} s^2_\beta + v^2 (\lambda_1 c_\beta^2+s^2_\beta \lambda_5+2 s_\beta c_\beta  \lambda_6),\\
&&m_{22}=m^2_{A} c^2_\beta + v^2 (\lambda_2 s_\beta^2+c^2_\beta \lambda_5+2 s_\beta c_\beta  \lambda_7), \\
&&m_{12}=-m^2_{A} s_\beta c_\beta + v^2 \Big( (\lambda_3+\lambda_4) s_\beta c_\beta+ \lambda_6 c_\beta^2+ \lambda_7 s^2_\beta \Big).
\end{eqnarray}
The physical CP-even states, $h^0$ and $H^0$, are written as
\begin{eqnarray}
&&H^0=\phi^0_{1R}c_\alpha+\phi^0_{2R}s_\alpha, \\
&&h^0=-\phi^0_{1R}s_\alpha+\phi^0_{2R}c_\alpha,
\end{eqnarray}
where
\begin{equation}
\tan 2\alpha=\frac{2m_{12}}{m_{11}-m_{22}},
\end{equation}
and
\begin{equation}
m^2_{H^0,h^0}=\frac{1}{2}\left(m_{11}+m_{22}\pm
\sqrt{(m_{11}-m_{22})^2+4m^2_{12}}\right).
\end{equation}

\subsection{The Yukawa sector in the 2HDM-III with a four-zero texture}

We shall follow Refs.~\cite{DiazCruz:2004pj, ourthdm3b}, where a
specific four-zero texture has been implemented for the Yukawa
matrices within the 2HDM-III. This allows one to express the couplings
of the neutral and charged Higgs bosons in terms of the fermion
masses, Cabibbo-Kobayashi-Maskawa
(CKM) mixing angles and certain dimensionless parameters,
which are to be bounded by current experimental constraints. Thus,
in order to derive the interactions of the charged Higgs boson, the
Yukawa Lagrangian is written as follows:
{\small
\beq
 {\cal{L}}_{Y} && = -\Bigg(
Y^{u}_1\bar{Q}_L {\tilde \Phi_{1}} u_{R} +
                   Y^{u}_2 \bar{Q}_L {\tilde \Phi_{2}} u_{R} +
Y^{d}_1\bar{Q}_L \Phi_{1} d_{R}  \nonumber \\
&&                    + Y^{d}_2 \bar{Q}_L\Phi_{2}d_{R} +Y^{{l}}_{1}\bar{L_{L}}\Phi_{1}l_{R} +
Y^{{l}}_{2}\bar{L_{L}}\Phi_{2}l_{R} \Bigg),
\label{lag-f}
\eeq }
\noindent where $\Phi_{1,2}=(\phi^+_{1,2},
\phi^0_{1,2})^T$ refer to the two Higgs doublets, ${\tilde
\Phi_{1,2}}=i \sigma_{2}\Phi_{1,2}^* $, $Q_{L}$ denotes the
left-handed fermion doublet, $u_{R} $ and $d_{R}$ are the
right-handed fermion singlets and, finally, $Y_{1,2}^{u,d}$ denote the
$(3 \times 3)$ Yukawa matrices. Similarly, one can see
the corresponding left-handed fermion doublet  $L_{L}$, the right-handed fermion singlet  $l_{R}$ and the
Yukawa matrices $Y_{1,2}^{{l}}$ for leptons.

After spontaneous EW Symmetry Breaking (EWSB), one can derive the
fermion mass matrices from eq. (\ref{lag-f}), namely
\begin{equation}
M_f= \frac{1}{\sqrt{2}}(v_{1}Y_{1}^{f}+v_{2}Y_{2}^{f}), \qquad
f = u, d, l.
\label{masa-fermiones}
\end{equation}
\noindent
We will assume that both Yukawa matrices $Y^f_1$ and $Y^f_2$ have the
four-texture form and are Hermitian \cite{Fritzsch:2002ga,DiazCruz:2004pj}. Following this convention, the
fermions mass matrices have the same form, which can be written as:
\begin{eqnarray}
M_f=
\left( \begin{array}{ccc}
0 & C_{f} & 0 \\
C_{f}^{*} & \tilde{B}_{f} & B_{f} \\
0 & B_{f}^{*} & A_{f}
\end{array}\right).
\end{eqnarray}
When $\tilde{B}_{q}\to 0$ one recovers the six-texture form.
We also consider the hierarchy
$\mid A_{q}\mid \, \gg \, \mid \tilde{B}_{q}\mid,\mid B_{q}\mid ,\mid C_{q}\mid$,
which is supported by the observed fermion masses in the SM.

The mass matrix is diagonalized through
the bi-unitary matrices $V_{L,R}$, though each Yukawa matrices is
not diagonalized by this transformation. The diagonalization is
performed in the following way:
\begin{equation}
\bar{M}_f = V_{fL}^{\dagger}M_{f}V_{fR}. \label{masa-diagonal}
\end{equation}

The fact that $M_f$ is Hermitian, under the considerations given
above, directly implies that $V_{fL} =
V_{fR}$, and the mass eigenstates for the fermions are given by
\begin{equation}
u = V_{u}^{\dagger}u', \qquad d = V_{d}^{\dagger}d', \qquad l =
V_{l}^{\dagger}l'. \label{redfields}
\end{equation}
Then, eq. (\ref{masa-fermiones}) in this basis takes the form
\begin{equation}
\bar{M}_f=\frac{1}{\sqrt{2}}(v_{1}\tilde{Y}_{1}^{f}+v_{2}
\tilde{Y}_{2}^{f}), \label{diag-Mf}
\end{equation}
where $\tilde{Y}_{i}^{f}=V_{fL}^{\dagger}Y_{i}^{f}V_{fR}$.
In order to compare the kind of new physics coming from our Yukawa texture
with some more traditional 2HDMs (in particular with the 2HDM-II),  in  previous works \cite{DiazCruz:2004pj, DiazCruz:2009ek, BarradasGuevara:2010xs,GomezBock:2005hc, HernandezSanchez:2011fq},  some of us have adopted the following re-definitions:  \\
2HDM-II-like
\begin{eqnarray}
\tilde{Y}_{1}^{d}&=&\frac{\sqrt{2}}{v\cos\beta}\bar{M}_{d}-\tan\beta\tilde{Y}_{2}^{d}, \nonumber \\
\tilde{Y}_{2}^{u}& =&\frac{\sqrt{2}}{v\sin\beta}\bar{M}_{u}-\cot\beta\tilde{Y}_{1}^{u}, \nonumber \\
\tilde{Y}_{1}^{{l}}&=& \tilde{Y}_{1}^{d} (d \to {l}).
\label{rotyukawas-1}
\end{eqnarray}
These re-definitions are convenient because we can get the Higgs-fermion-fermion coupling in the 2HDM-III as $g_{\rm 2HDM-III}^{ff \phi }= g_{\rm 2HDM-II}^{ff\phi} + \Delta g^{ff\phi} $, where $g_{\rm 2HDM-II}^{ff\phi}$ is the coupling in the 2HDM-II and $\Delta g^{ff\phi} $ is the contribution of the four-zero texture. If $\Delta g^{ff\phi}  \to 0$ we can recover the 2HDM-II. However, these re-definitions are not unique. In fact, there are others possibilities since from eq. (\ref{diag-Mf}) one can reproduce the 2HDM-I, 2HDM-X or 2HDM-Y  as we can obtain for  any version of 2HDM the following relation: $g_{\rm 2HDM-III}^{f f \phi }= g_{\rm 2HDM-any}^{f f \phi} + \Delta' g^{f f \phi} $. The other possible  re-definitions are:\\
2HDM-I-like
\begin{eqnarray}
\tilde{Y}_{2}^{d}&=&\frac{\sqrt{2}}{v\sin\beta}\bar{M}_{d}-\cot\beta\tilde{Y}_{1}^{d}, \nonumber \\
\tilde{Y}_{2}^{u}& =&\frac{\sqrt{2}}{v\sin\beta}\bar{M}_{u}-\cot\beta\tilde{Y}_{1}^{u}, \nonumber \\
\tilde{Y}_{2}^{{l}}&=& \tilde{Y}_{2}^{d} (d \to {l}).
\label{rotyukawas-2}
\end{eqnarray}
2HDM-X-like
\vspace*{-0.5cm}
\begin{eqnarray}
\tilde{Y}_{2}^{d}&=&\frac{\sqrt{2}}{v\sin\beta}\bar{M}_{d}-\cot\beta\tilde{Y}_{1}^{d}, \nonumber \\
\tilde{Y}_{2}^{u}& =&\frac{\sqrt{2}}{v\sin\beta}\bar{M}_{u}-\cot\beta\tilde{Y}_{1}^{u}, \nonumber \\
\tilde{Y}_{1}^{{l}}&=& \tilde{Y}_{1}^{d} (d \to {l}).
\label{rotyukawas-3}
\end{eqnarray}
2HDM-Y-like
\vspace*{-0.5cm}
\begin{eqnarray}
\tilde{Y}_{1}^{d}&=& \frac{\sqrt{2}}{v\cos\beta}\bar{M}_{d}-\tan\beta\tilde{Y}_{2}^{d}, \nonumber \\
\tilde{Y}_{2}^{u}& =&\frac{\sqrt{2}}{v\sin\beta}\bar{M}_{u}-\cot\beta\tilde{Y}_{1}^{u}, \nonumber \\
\tilde{Y}_{2}^{{l}}&=& \tilde{Y}_{2}^{d} (d \to {l}).
\label{rotyukawas-4}
\end{eqnarray}

After spontaneous EWSB  and including the
diagonalizing matrices for quarks and Higgs bosons\footnote{The
details of both diagonalizations are presented in
Ref.~\cite{DiazCruz:2004pj}.}, the interactions of the charged Higgs bosons
$H^\pm$ and neutral Higgs bosons $\phi^0$ ($\phi^0 = h^0, \, H^0, \, A^0$ ) with quark pairs for any parametrization 2HDM-(I,II,X,Y)-like have the following form:
{\small
\begin{eqnarray}
\hspace{-2cm}\label{QQH} {\cal{L}}^{\bar{f}_i f_j \phi} & = & - \frac{g}{
2\sqrt{2} M_W}  \Bigg[  \sum^3_{l=1}\bar{u}_i \left\{  (V_{\rm CKM})_{il} \left[ X
\, m_{d_{l}} \, \delta_{lj} -f(X) \left(\frac{\sqrt{2}
M_W}{g}\right) \left( \tilde{Y}^d_{n(X)} \right)_{lj}  \right] \right. (1+  \gamma^{5})
\nonumber \\
& & + \left. \left[ Y \, m_{u_{i}} \, \delta_{il} -f(Y)
\left(\frac{\sqrt{2} M_W}{g}\right) \left(
\tilde{Y}^u_{n(Y)} \right)_{il}^{\dagger}  \right]
(V_{\rm CKM})_{lj}  (1-  \gamma^{5}) \right\} \, d_{j} \, H^{+} \\
&& + \bar{\nu}_i   \bigg[ Z \, m_{{l}_{i}} \, \delta_{ij} -f(Z) \left(\frac{\sqrt{2}
M_W}{g}\right) \left( \tilde{Y}^{l}_{n(Z)} \right)_{ij}  \bigg] ( 1+  \gamma^{5})  {l}_j H^+ + h.c. \Bigg]
\nonumber \\
& & -\frac{g}{2 M_W} \Bigg( m_{d_i}
\bar{d_{i}} \bigg\{ \left[ \xi_H^d  \delta_{ij}- \frac{(\xi_h^d+ X \xi_H^d  )}{f (X)}
\frac{\sqrt{2} }{g}
\left(\frac{m_W}{m_{d_{i}}}\right)(\tilde{Y}_{n(X)}^d)_{ij}\right] H^{0}
\nonumber \\
&  &+ \left[\xi_h^d  \delta_{ij}+ \frac{(\xi_H^d- X \xi_h^d  )} {f (X)}\frac{\sqrt{2} }{g }
\left(\frac{m_W}{m_{d_{i}}}\right)(\tilde{Y}_{n(X)}^d)_{ij}\right] h^{0}
\nonumber \\
& &+ i
\left[-X \delta_{ij}+  f(X) \frac{\sqrt{2} }{g }
\left(\frac{m_W}{m_{d_{i}}}\right)(\tilde{Y}_{n(X)}^d)_{ij}\right]
\gamma^{5} A^{0} \bigg\} d_{j} \nonumber \\
& &+ m_{u_i} \bar{u}_{i} \bigg\{ \left[ \xi_H^u \delta_{ij}+\frac{(\xi_h^u- Y \xi_H^u  )}{f(Y)}
\frac{\sqrt{2} }{g }
\left(\frac{m_W}{m_{u_{i}}}\right)(\tilde{Y}_{n(Y)}^u)_{ij}\right] H^{0}
\nonumber \\
&  &+
\left[\xi_h^u  \delta_{ij}- \frac{(\xi_H^u+ Y \xi_h^u  )}{f (Y)} \frac{\sqrt{2} }{g }
\left(\frac{m_W}{m_{u_{i}}}\right)(\tilde{Y}_{n(Y)}^u)_{ij}\right]h^{0}
\nonumber \\
& &+ i
\left[-Y \delta_{ij} + f(Y) \frac{\sqrt{2} }{g }
\left(\frac{m_W}{m_{u_{i}}}\right)(\tilde{Y}_{n(Y)}^u)_{ij}\right]
\gamma^{5} A^{0} \bigg\} u_{j}  \nonumber \\
& & + m_{{l}_i}
\bar{{l}_{i}} \bigg\{ \left[ \xi_H^{l} \delta_{ij}- \frac{(\xi_h^{l}+ Z \xi_H^{l}  )}{f(Z)}
\frac{\sqrt{2} }{g}
\left(\frac{m_W}{m_{d_{i}}}\right)(\tilde{Y}_{n(Z)}^{l})_{ij}\right] H^{0}
\nonumber \\
&  &+ \left[\xi_h^{l} f(Z) \delta_{ij}+ \frac{(\xi_H^{l}- Z \xi_h^{l}  )}{f(Z)} \frac{\sqrt{2} }{g }
\left(\frac{m_W}{m_{d_{i}}}\right)(\tilde{Y}_{n(Z)}^{l})_{ij}\right] h^{0}
\nonumber \\
& &+ i
\left[-Z \delta_{ij}+  f(Z) \frac{\sqrt{2} }{g }
\left(\frac{m_W}{m_{d_{i}}}\right)(\tilde{Y}_{n(Z)}^{l})_{ij}\right]
\gamma^{5} A^{0} \bigg\} {l}_{j}
 \Bigg), \nonumber
\end{eqnarray}
}
where $V_{\rm CKM}$ denotes the mixing matrices of the quark sector, the functions $f_(x)$ and $n(x)$
are given by:
\vspace*{-0.1cm}
\begin{eqnarray}
f (x) & = & \sqrt{1+x^2}, \nonumber \\
n (x) & = & \left \{ \begin{matrix} 2 & \mbox{if } x = \tan \beta,
\\ 1 & \mbox{if  } x = \cot \beta. \end{matrix}\right.
\end{eqnarray}
the parameters  $X$, $Y$, $Z $ are given in
Refs. \cite{Grossman:1994jb, hep-ph/9603445,Akeroyd:1994ga,Akeroyd:2012yg,Borzumati:1998nx,Aoki:2009ha,Pich:2009sp}
and the factors
$\xi_\phi^f$ are presented in Ref. \cite{Aoki:2009ha}. Following this notation we can list the parameters for the framework  2HDM-(I,II,X,Y)-like  through Tab.
\ref{couplings}.

Following the analysis in~\cite{DiazCruz:2004pj} one can derive a better approximation for the product
$V_q\, Y^{q}_n \, V_q^\dagger$, expressing the
rotated matrix $\tilde {Y}^q_n$, in the form
\vspace*{-0.1cm}
\begin{eqnarray}
\left[ \tilde{Y}_n^{q} \right]_{ij}
= \frac{\sqrt{m^q_i m^q_j}}{v} \, \left[\tilde{\chi}_{n}^q \right]_{ij}
=\frac{\sqrt{m^q_i m^q_j}}{v}\,\left[\chi_{n}^q \right]_{ij}  \, e^{i \vartheta^q_{ij}},
\label{cheng-sher}
\end{eqnarray}
\noindent
where the $\chi$'s are unknown dimensionless parameters of the model, they come
from the election of a specific texture of the Yukawa matrices.  It is important to mention that eq. (\ref{cheng-sher}) is a consequence of the diagonalization process of Yuwaka matrices, assuming the hierarchy among the fermion masses (see Ref.  \cite{DiazCruz:2004pj}), namely, the {Cheng-Sher ansatz} is a particular case of this parametrization. Besides, in order to have an acceptable model, the
parameters $\chi$'s could be  $O(1)$ but not more, generally. Recently we have calculated the $\chi^2$ fit of Yukawa matrices including the CKM matrix, and we find that the parameters off-diagonal are  $O(1)$ (e.g.,  $\chi_{23}^{f} \leq 10$), therefore we cannot ignore all of these \cite{felixetal}. Besides, in Ref. \cite{Bijnens:2011gd},  they study the general 2HDMs  considering  renormalization group evolution of the Yukawa couplings and the cases when the ${\cal Z}_2$-symmetry is broken, called non-diagonal models (e.g., the models with a structure incorporating the {Cheng-Sher ansatz}). It is interesting to note that it is
actually the off-diagonal elements in the down-sector that become large whereas the ones in
the up-sector $\chi^u(\mu) \leq 0.1$, assuming the conservative criterion $\chi^f \leq 0.1$, where $\mu$ is the renormalization scale.  On the other hand, the FCNC  processes at low energy are going to determine bounds for these parameters with high precision, aspect  which is studied in this work. In order to perform our phenomenological study, we find it convenient
to rewrite the Lagrangian given in eq.~(\ref{QQH})  in terms of the
coefficients $ \left[\tilde{\chi}_{n}^q \right]_{ij}$, as follows:
{\small
\begin{eqnarray}
\label{LCCH}\nonumber
{\cal{L}}^{\bar{f}_i f_j \phi}  & = & -
\frac{g}{ 2\sqrt{2} M_W} \Bigg[  \sum^3_{l=1} \bar{u}_i \Bigg[  (V_{\rm CKM})_{il} \left( X \, m_{d_{l}} \, \delta_{lj}
-\frac{f(X)}{\sqrt{2} }  \,\sqrt{m_{d_l} m_{d_j} } \, \tilde{\chi}^d_{lj}  \right)   (1+\gamma^5)
\nonumber \\
& & + \left( Y \, m_{u_{i}} \, \delta_{il}
  -\frac{f(Y)}{\sqrt{2} }  \,\sqrt{m_{u_i} m_{u_l} } \, \tilde{\chi}^u_{il} \right)
  (V_{\rm CKM})_{lj}  (1-\gamma^5)  \Bigg] \, d_{j} \, H^{+}\\ \nonumber
  && + \bar{\nu}_i   \bigg( Z \, m_{{l}_{i}} \, \delta_{ij} -\frac{f(Z)}{\sqrt{2} }  \,\sqrt{m_{{l}_i} m_{d_j} } \, \tilde{\chi}^{l}_{ij}   \bigg)
  ( 1+  \gamma^{5})  {l}_j H^+ + h.c. \Bigg]   \nonumber \\
  & & - \frac{g}{2 M_W} \Bigg[   \bar{d_i} \Bigg(
\left[  m_{d_i}  \xi_H^d
\delta_{ij} - \frac{(\xi_h^d+X \xi_H^d)}{ f(X)  }
\frac{\sqrt{m_{d_i} m_{d_j}} }{\sqrt{2}} \tilde{\chi}_{ij}^d\right]H^{0}
\nonumber \\
                &   & +
\left[m_{d_i}\xi_h^d \delta_{ij} + \frac{(\xi_H^d-X \xi_h^d)}{ f(X)  }
\frac{\sqrt{m_{d_i} m_{d_j}} }{\sqrt{2}} \tilde{\chi}_{ij}^d\right] h^{0}
\nonumber \\
                &   & +i \left[-m_{d_i} X \delta_{ij} + f(X) \frac{\sqrt{m_{d_i} m_{d_j}}}{\sqrt{2}} \tilde{\chi}_{ij}^d\right]
\gamma^{5}  A^0  \Bigg)  d_j \nonumber \\
                &   &u_i \Bigg( \left[m_{u_i}\xi_H^u \delta_{ij} + \frac{(\xi_h^u-Y \xi_H^u)}{ f(Y) }
\frac{\sqrt{m_{u_i} m_{u_j}}}{\sqrt{2}} \tilde{\chi}_{ij}^u\right]H^{0}
\nonumber \\
                &   & + \left[ m_{u_i} \xi_h^u \delta_{ij} - \frac{(\xi_H^u+Y \xi_h^u)}{ f(Y)}
\left(\frac{\sqrt{m_{u_i} m_{u_j}}}{\sqrt{2}}\right)\tilde{\chi}_{ij}^u\right]h^{0}
\nonumber \\
                &   & + i \left[- m_{u_i} Y \delta_{ij} +
f(Y) \frac{\sqrt{m_{u_i} m_{u_j}}}{\sqrt{2}} \tilde{\chi}_{ij}^u\right]
\gamma^{5} A^0\Bigg) u_j \nonumber \\
& & +    \bar{{l}_i} \Bigg(
\left[  m_{{l}_i}  \xi_H^{l} \delta_{ij} - \frac{(\xi_h^{l}+Z \xi_H^{l})}{ f(Z)  }
\frac{\sqrt{m_{{l}_i} m_{{l}_j}} }{\sqrt{2}} \tilde{\chi}_{ij}^{l} \right]H^{0}
\nonumber \\
                &   & +
\left[m_{{l}_i}\xi_h^{l} \delta_{ij} + \frac{(\xi_H^{l}-Z \xi_h^{l})}{ f(Z)  }
\frac{\sqrt{m_{{l}_i} m_{{l}_j}} }{\sqrt{2}} \tilde{\chi}_{ij}^{l} \right] h^{0}
\nonumber \\
                &   & +i \left[-m_{{l}_i} Z \delta_{ij} + f(Z) \frac{\sqrt{m_{{l}_i} m_{{l}_j}}}{\sqrt{2}} \tilde{\chi}_{ij}^{l}\right]
\gamma^{5}  A^0  \Bigg)  {l}_j  \Bigg],
\end{eqnarray}  }
where we have redefined $\left[ \tilde{\chi}_{1}^u \right]_{ij} =
\tilde{\chi}^u_{ij}$, $\left[ \tilde{\chi}_{2}^d \right]_{ij} =
\tilde{\chi}^d_{ij}$ and $\left[ \tilde{\chi}_{2}^{l} \right]_{ij} =
\tilde{\chi}^{l}_{ij}$.
Then, from eq.~(\ref{LCCH}), the couplings $\bar{f_i} f_j \phi^0$, $\bar{u}_i d_j H^+$ and $u_i \bar{d}_j
H^-$ are given by:
{\small
\begin{eqnarray}
\label{coups1}
g_{h^0 \bar{f_i}f_j} &=& -\frac{ig}{ 2 M_W}
(m_{f_i}h_{i j}^f ),  \quad g_{H^0 \bar{f_i}f_j} = -\frac{ig}{ 2 M_W} ( m_{f_i} H_{i j}^f ),
\quad  g_{A^0 \bar{f_i}f_j}= -\frac{ig}{ 2 M_W} (m_{f_i} A_{i j}^f  \gamma_5), \nonumber \\
g_{H^+\bar{u_i}d_j} &=& -\frac{ig}{2 \sqrt{2} M_W}
(S_{i j} +P_{i j} \gamma_5), \quad g_{H^- u_i \bar{d_j}}= -\frac{ig
}{2 \sqrt{2} M_W}  (S_{i j} -P_{i j} \gamma_5). \label{quark}
\end{eqnarray}
}
where $h_{ij}^f$, $H_{ij}^f$, $A_{ij}^f$, $S_{i j}$ and $P_{i j}$ are defined as:
\begin{eqnarray}\label{hHA}
h_{ij}^d & = & \xi_h^d \delta_{ij} + \frac{(\xi_H^d-X \xi_h^d)}{\sqrt{2} f(X)} \sqrt{\frac{m_{d_j}}{m_{d_i}}} \tilde{\chi}_{ij}^d, \quad
h_{ij}^{l} = h_{ij}^d (d\to {l}, \, \,  X \to Z), \nonumber \\
H_{ij}^d & = & \xi_H^d \delta_{ij} - \frac{(\xi_h^d+X \xi_H^d)}{\sqrt{2} f(X)} \sqrt{\frac{m_{d_j}}{m_{d_i}}} \tilde{\chi}_{ij}^d, \quad
H_{ij}^{l} = H_{ij}^d (d\to {l}, \, \,  X \to Z),  \\ \nonumber
A_{ij}^d & = & -X \delta_{ij} + \frac{f(X)}{\sqrt{2} } \sqrt{\frac{m_{d_j}}{m_{d_i}}} \tilde{\chi}_{ij}^d, \quad
A_{ij}^{l} = A_{ij}^d (d\to {l}, \, \,  X \to Z), \nonumber \\
h_{ij}^u & = & \xi_h^u \delta_{ij} - \frac{(\xi_H^u+Y \xi_h^u)}{\sqrt{2} f(Y)} \sqrt{\frac{m_{u_j}}{m_{u_i}}} \tilde{\chi}_{ij}^u, \nonumber \\
H_{ij}^u & = & \xi_H^u \delta_{ij} + \frac{(\xi_h^u-Y \xi_H^u)}{\sqrt{2} f(Y)} \sqrt{\frac{m_{u_j}}{m_{u_i}}} \tilde{\chi}_{ij}^u, \nonumber \\
A_{ij}^u & = & -Y \delta_{ij} + \frac{f(Y)}{\sqrt{2} } \sqrt{\frac{m_{u_j}}{m_{u_i}}} \tilde{\chi}_{ij}^u,  \nonumber \\
S_{i j} & = &   m_{d_{j}} \, X_{ij}
 +  m_{u_{i}} \, Y_{ij}, \quad P_{i j}  =  m_{d_{j}} \, X_{ij}
 -  m_{u_{i}} \, Y_{ij},
\end{eqnarray}
with
\begin{eqnarray}
X_{i j} & = &   \sum^3_{l=1}  (V_{\rm CKM})_{il} \bigg[ X \, \frac{m_{d_{l}}}{m_{d_j}} \, \delta_{lj}
-\frac{f(X)}{\sqrt{2} }  \,\sqrt{\frac{m_{d_l}}{ m_{d_j} }} \, \tilde{\chi}^d_{lj}  \bigg],
\nonumber \\
Y_{i j} & = &  \sum^3_{l=1}  \bigg[ Y  \, \delta_{il}
  -\frac{f(Y)}{\sqrt{2} }  \,\sqrt{\frac{ m_{u_l}}{m_{u_i}} } \, \tilde{\chi}^u_{il}  \bigg]  (V_{\rm CKM})_{lj}.
  \label{Xij}
\end{eqnarray}
For the case of leptons $S_{ij}^{l} = P_{ij}^{l}$  we have
\begin{eqnarray}
S_{i j}^{l} & = &     m_{{l}_{j}} \, Z_{ij}^{l}, \nonumber \\
Z_{i j}^{l}& = &   \bigg[Z \, \frac{m_{{l}_{i}}}{m_{{l}_j}} \,
\delta_{ij} -\frac{f(Z)}{\sqrt{2} }  \,\sqrt{\frac{m_{{l}_i}}{m_{{l}_j}}  }
\, \tilde{\chi}^{l}_{ij}  \bigg].
\label{Zij}
\end{eqnarray}
Then,  the couplings ${l}_i^- \nu_{{l}_j} H^+$ and ${l}_i^+ \nu_{{l}_j}
H^-$ are given by
{\small
\begin{eqnarray}
\label{coups2} g_{H^+ {l}_i^- \nu_{{l}_j}} &=&
-\frac{ig}{ \sqrt{2} M_W} S_{i j}^{l}
\left( \frac{1 + \gamma_5}{2}  \right), \quad g_{H^- {l}_i^+ \nu_{{l}_j}} =
-\frac{ig}{ \sqrt{2} M_W} S_{i j}^{l} \left( \frac{1 - \gamma_5}{2} \right). \label{lepton}
\end{eqnarray}
}
\begin{table}[tbp]
\centering
\begin{tabular}{|c|c|c|c|c|c|c|c|c|c|}
\hline
 2HDM-III& $X$ &  $Y$ &  $Z$ & $\xi^u_h $  & $\xi^d_h $ & $\xi^d_{l} $  & $\xi^u_H $  & $\xi^d_H $ & $\xi^{l}_H $\\ \hline
2HDM-I-like~
&  $-\cot\beta$ & $\cot\beta$ & $-\cot\beta$ & $c_\alpha/s_\beta$ & $c_\alpha/s_\beta$ & $c_\alpha/s_\beta$
& $s_\alpha/s_\beta$ & $s_\alpha/s_\beta$ & $s_\alpha/s_\beta$\\
2HDM-II-like
& $\tan\beta$ & $\cot\beta$ & $\tan\beta$ & $c_\alpha/s_\beta$ & $-s_\alpha/c_\beta$ & $-s_\alpha/c_\beta$
& $s_\alpha/s_\beta$ & $c_\alpha/c_\beta$ & $c_\alpha/c_\beta$\\
2HDM-X-like
& $-\cot\beta$ & $\cot\beta$ & $\tan\beta$ &  $c_\alpha/s_\beta$ & $c_\alpha/s_\beta$ & $-s_\alpha/c_\beta$
& $s_\alpha/s_\beta$ & $s_\alpha/s_\beta$ & $c_\alpha/c_\beta$\\
2HDM-Y-like
& $\tan\beta$ & $\cot\beta$ & $-\cot\beta$ & $c_\alpha/s_\beta$ & $-s_\alpha/c_\beta$ & $c_\alpha/s_\beta$
& $s_\alpha/s_\beta$ & $c_\alpha/c_\beta$ & $s_\alpha/s_\beta$\\
\hline
\end{tabular}
\caption{ Parameters $X$, $Y$ and $Z$  defined in the  Yukawa interactions of eq.   (\ref{QQH}) for four versions of the 2HDM-III with a four-zero texture, which come from eqs. (\ref{rotyukawas-1})--(\ref{rotyukawas-4}). Here $s_\alpha= \sin\alpha$, $c_\alpha= \cos\alpha$,
$s_\beta= \sin\beta$ and $c_\beta= \cos\beta$. }
\label{couplings}
\end{table}
In order to compare these couplings with previous works \cite{Grossman:1994jb, hep-ph/9603445,Akeroyd:1994ga,Akeroyd:2012yg,Borzumati:1998nx,Aoki:2009ha},  we find it convenient to  define the couplings
$\bar{u}_i d_j H^+$ and $u_i \bar{d}_j H^-$ in terms of the
matrix elements $X_{ij}$, $Y_{ij}$ and $Z_{ij}$.
Following the definitions (\ref{hHA})--(\ref{Zij}) we obtain the following compact expression
for the interactions of  Higgs bosons with the fermions:
\begin{eqnarray}
{\cal L}^{\bar{f}_i f_j \phi} & =&
-\left\{\frac{\sqrt2}{v}\overline{u}_i
\left(m_{d_j} X_{ij} {P}_R+m_{u_i} Y_{ij} {P}_L\right)d_j \,H^+
+\frac{\sqrt2m_{{l}_j} }{v} Z_{ij}\overline{\nu_L^{}}{l}_R^{}H^+
+{H.c.}\right\} \nonumber \\
 & & -\frac{1}{v} \bigg\{ \bar{f}_i m_{f_i} h_{ij}^f  f_j h^0 + \bar{f}_i m_{f_i} H_{ij}^f  f_j H^0 - i \bar{f}_i m_{f_i} A_{ij}^f  f_j \gamma_5 A^0\bigg\}.
\label{lagrangian-f}
\end{eqnarray}
 When the parameters $\chi_{ij}^f=0$,  we obtain  $X_{11}=X_{22}=X_{33}=X$ (similarly for $Y$ and $Z$)  and one recovers the Yukawa interactions given in Refs.~\cite{Grossman:1994jb, hep-ph/9603445,Akeroyd:1994ga,Akeroyd:2012yg,Aoki:2009ha}.  Besides, in order to hold consistencies with the MHDM/A2HDM \cite{Grossman:1994jb}, we suggest that this Lagrangian could represent a MHDM/A2HDM with additional flavor physics in the Yukawa matrices as well as the possibility of FCNCs at tree level.   Returning to the Lagrangian given in eq. (\ref{lagrangian-f}), when the parameters $\chi_{ij}^f$ are present, one can see that
$X_{11}\neq X_{22} \neq X_{33} \neq X $ and the criteria of flavor constraints on $X$ cannot be applied directly to $X_{ij}$ (the same for $ Y$ and $Z$), but the analyses for  low energy processes  are similar. Below we shall discuss more about various aspects of the model.  Finally, it should be pointed out that parameters $X$, $Y$, $Z$, $\xi_\phi^f$ and $\chi_{ij}$ are arbitrary complex numbers, opening the possibility of having new sources of CP violation with tree-level FCNCs.

 Previously, in Ref. \cite{Mahmoudi:2009zx},  the flavor constraints of the 2HDM-III with a six-zero texture were studied, finding interesting results that we can use.  However, we should compare their results and ours so as to distinguish the two parametrizations.   Firstly, the
six-zero texture assumed in  \cite{Mahmoudi:2009zx} has been disfavored by current data on the CKM mixing angles \cite{Fritzsch:2002ga,Roberts:2001zy}. Hence, we focus here onto the four-zero texture, which is still acceptable phenomenologically and of which we consider the non-diagonal terms of the Yukawa matrices. Secondly, in order to unify notations we relate the parameters $\lambda_{ij}^{F}$ of
 \cite{Mahmoudi:2009zx} with our parameters $X_{ij}$, $Y_{ij}$ and $Z_{ij}$ as given in eqs. (\ref{Xij}) and (\ref{Zij}), as
follows\footnote{We adopt the description of the Yukawa sector  presented in
Ref.~\cite{Mahmoudi:2009zx}.}:
\begin{eqnarray}
{\cal L}^{\bar{f}_i f_j H^+} & =&
-\bigg\{  \overline{u}_i
\left(\sum_l^3  (V_{\rm CKM})_{il} \rho^D_{lj} {P}_R-\sum_l^3 \rho_{il}^U (V_{\rm CKM})_{lj}  {P}_L\right)d_j \,H^+   \nonumber \\
&& +
 \rho_{ij}^{l} \overline{\nu_L^{}}{l}_R^{}H^+
+{H.c.}\bigg\}, \\
\rho^F_{ij}&=&\frac{\sqrt{2 m_{Fi} m_{Fj}}}{v} \lambda{ij}^F,
\label{lagrangian-stal}
\end{eqnarray}
where $\rho^F_{ij}$ was introduced following the Cheng-Sher ansatz, considering $\lambda_{ij}^F \sim O(1)$.
If we compare this with eq. (\ref{lagrangian-f}), after using eqs. (\ref{Xij}) and (\ref{Zij}), we obtain the following relations:
\begin{eqnarray}
\lambda_{ij}^D &=& \bigg[ X \, \sqrt{\frac{m_{d_{i}}}{m_{d_j}} }\, \delta_{ij}
-\frac{f(X)}{\sqrt{2} }  \, \tilde{\chi}^d_{ij}  \bigg], \nonumber \\
\lambda_{ij}^U &=& -\bigg[ Y \, \sqrt{\frac{m_{u_{i}}}{m_{u_j}} }\, \delta_{ij}
-\frac{f(X)}{\sqrt{2} }  \, \tilde{\chi}^u_{ij}  \bigg], \nonumber \\
\lambda_{ij}^{l} &=& \bigg[ Z \, \sqrt{\frac{m_{{l}_{i}}}{m_{{l}_j}} }\, \delta_{ij}
-\frac{f(X)}{\sqrt{2} }  \, \tilde{\chi}^{l}_{ij}  \bigg],
\label{stal}
\end{eqnarray}
\noindent
and
\noindent
\begin{eqnarray}
X_ {ij} & = & \sum_l (V_{CKM})_{il} \sqrt{\frac{m_{d_l}}{m_{d_j}}} \lambda_{lj}^D \nonumber, \\
Y_ {ij} & = & \sum_l  \sqrt{\frac{m_{u_l}}{m_{u_i}}} \lambda_{il}^U  (V_{CKM})_{lj}.
\end{eqnarray}
In essence, in the remainder of our work, we assume that our model could represent an effective flavor theory,
 wherein the Higgs fields necessarily participates in the flavor structure and has the same features as
those of renormalizable flavor models  \cite{Frigerio:2004jg,Frampton:2008ci,Fukuyama:2010mz,Aranda:2011dx}. In this type of
scenarios, a horizontal flavor
symmetry, continuous or discrete, is added to the SM gauge group symmetry in such a way
as to reproduce the observed mass and mixing angle patterns by only using renormalizable terms in the
Lagrangians. This requirement has two immediate and interesting consequences: firstly, there must
be more than one $SU(2)$ doublet scalar; secondly, at least some of them must transform non-trivially
under the flavor symmetry \cite{Aranda:2012bv,Branco:2010tx}.

\section{Flavor constraints on  the  2HDM-III with a four-zero Yukawa texture}

In this section we will analyze the most important FCNC processes that are sensitive to, in particular, charged
Higgs boson exchange, the primary interest of this paper, as well as effect of a (neutral) SM-like Higgs boson $h^0$ (we assume that its mass is $m_{h^0} = 125$ GeV).
Starting from measurements obtained from from these processes we constrain the new physics  parameters $\chi_{ij}^f$ that come from four-zero Yukawa texture.  Finally, we study the possibility of obtaining a light  charged Higgs boson
compatible with all such measurements.
We will address the various experimental limits in different subsections.

\subsection{$\mu - e$ universality in $\tau$ decays}
\begin{figure}[t]
\begin{center}
\includegraphics[origin=c, angle=0, scale=0.5]{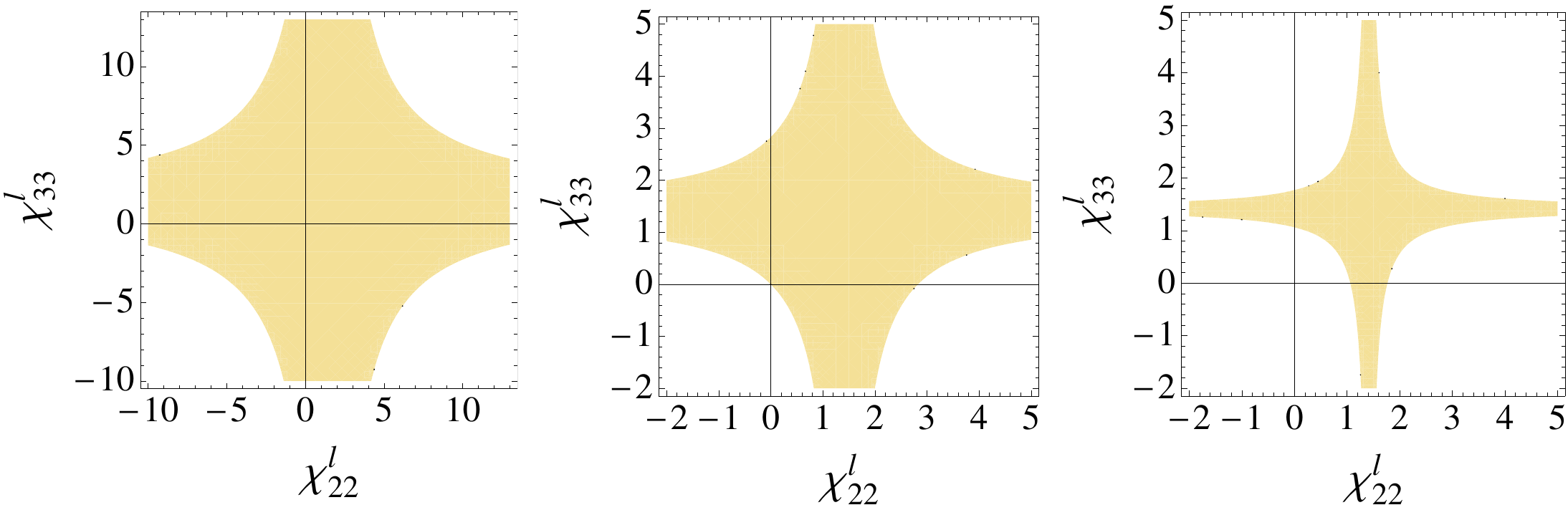}
\vspace*{-5mm}
\caption{Considering the constraint from $\mu - e$ universality in $\tau$ decays, we show the allowed region (orange color) for  $\chi_{22}^{l}$ and $\chi_{33}^{l}$ when $Z$ takes values of 10 (left-panel), 40 (center-panel) and  80 (right-panel). Here, 90 GeV $\leq m_{H^\pm} \leq 130$ GeV.}
\label{fig:emu}
\end{center}
\end{figure}

\begin{figure}[t]
\begin{center}
\includegraphics[origin=c, angle=0, scale=0.5]{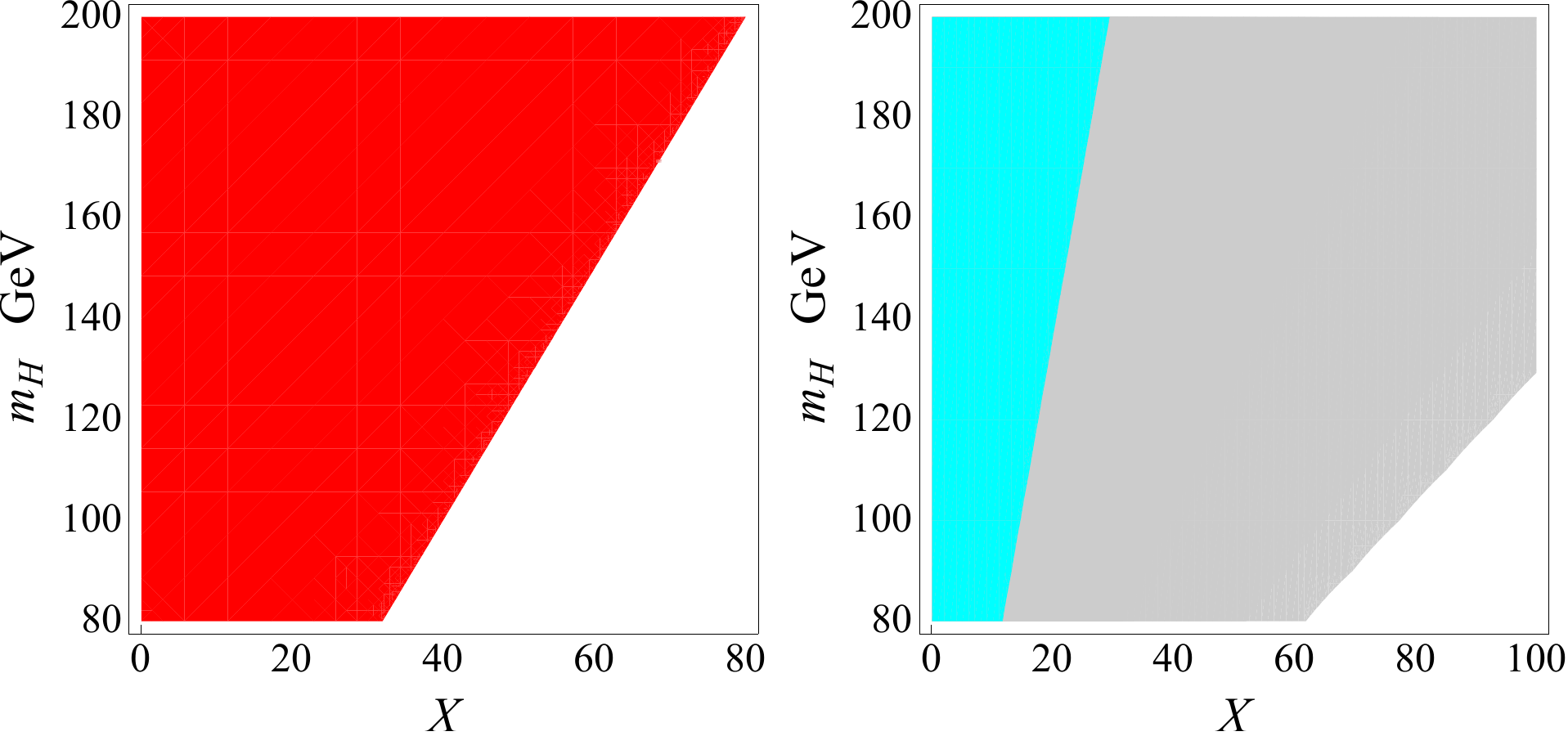}
\vspace*{-5mm}
\caption{Considering the constraint from $\mu - e$ universality in $\tau$ decays, we show the allowed region for  the plane $m_{H^\pm}-X$: left panel represent the case $\chi_{33}^{l} =0$ and $\chi_{22}^{l} =0$, right panel represents the case $\chi_{22}^{l} =0.1$ and $-20 \leq\chi_{33}^{l} \leq 20$ (blue region is for $5\leq |\chi_{ij}^{l}| $,  blue\&gray region is for $|\chi_{ij}^{l}| \leq 5$). The same occur when $\chi_{33}^{l} =0.1$ and $-20 \leq\chi_{22}^{l} \leq 20$.}
\label{fig:mhx}
\end{center}
\end{figure}

The $\tau$ decays  into $ \mu \bar{\nu}_{\mu} \nu_{\tau}$ and $ e \bar{\nu}_{e} \nu_{\tau}$
produce important constraints onto charged Higgs boson states coupling to
  leptons \cite{Logan:2009uf}, through the requirement of
 $\mu - e$ universality.  The consequent limits can be quantified through the following relation
  \cite{Tsai:1971vv, Aubert:2009qj}:
\beq
\left( \frac{g_{\mu}}{g_e} \right)_{\tau}^{2} =
\frac{BR(\tau \to \mu \bar{\nu}_{\mu} \nu_{\tau})}{BR(\tau \to e \bar{\nu}_{e} \nu_{\tau})}
\frac{g(m_e^2/m_{\tau}^2)}{g(m_{\mu}^2/m_{\tau}^2)} = 1.0036 \pm 0.0020
\eeq
where $g(x)= 1 -8x^2 +8x^3 - x^4 -12x^2 \, \log x$.
Following \cite{Grossman:1994jb}, in our case, the request of $\mu - e$ universality imposes the following relation
\begin{eqnarray}
\frac{BR(\tau \to \mu \bar{\nu}_{\mu} \nu_{\tau})}{BR(\tau \to e \bar{\nu}_{e} \nu_{\tau})}
\frac{f(m_e^2/m_{\tau}^2)}{f(m_{\mu}^2/m_{\tau}^2}
\simeq 1 +\frac{R^2}{4} -0.25R,
\end{eqnarray}
where $R$ is the scalar contribution  parametrized through the effective coupling, see eqs. (\ref{Zij}) and (\ref{coups2}),
\begin{eqnarray}
R = \frac{m_{\tau} m_{\mu}}{m_{H^\pm}^2} \, Z_{33} \, Z_{22} \, = \,
\frac{m_{\tau} m_{\mu}}{m_{H^\pm}^2}
\left[ Z -\frac{f(Z)}{\sqrt{2}} \chi_{33}^l \right]
\left[ Z -\frac{f(Z)}{\sqrt{2}} \chi_{22}^l \right].
\label{R}
\end{eqnarray}
One can see that $R$ is symmetric in the two parameters $\chi_{22}$ and $\chi_{33}$. Following the analysis of Ref. \cite{Jung:2010ik}, we can obtain the following explicit constraint:
\begin{eqnarray}
\frac{|Z_{22} Z_{33}|}{m_{H^\pm}^2} \leq 0.16 \, \, {\rm GeV}^{-1}  \qquad  (95\% ~{\rm CL}).
\end{eqnarray}
We show in Fig. \ref{fig:emu} the constraints on $\chi_{22}^{l}$ and $\chi_{33}^{l}$ with $ Z = 10$, 40, 80. One can see that, for small $Z$ values, the allowed region for $\chi_{22}^{l}$ and $\chi_{33}^{l}$  is large whereas, when $Z$ instead grows, the allowed region for theses parameters is smaller in comparison. The constraints
becomes most restrictive when $Z$ is large and we have a very light charged Higgs boson,
between 90 and 130 GeV.  The plot also shows that $\chi_{22}^{l}$ and $\chi_{33}^{l}$
could be  simultaneously $1$ and $-1$, respectively, and the more favorable region is the one where $\chi_{22}^{l} = \chi_{33}^{l} =1.5$ for  $0.5 \leq Z \leq 100$.
When $Z$  is large, if $\chi_{22}^{l} = 1$, one can see that
$0.5 \leq \chi_{33}^{l} \leq 2.5$ (the same happens when $\chi_{22}^{l}$ and $ \chi_{33}^{l} $
are interchanged).
Further, in Fig. \ref{fig:mhx} we present the plane $[m_{H^\pm},X]$  and  the allowed region is shown for the cases
$\chi_{33}^{l} =0$ and $\chi_{22}^{l} =0$ (left panel) and  $\chi_{22}^{l} =0.1$ and $-20 \leq\chi_{33}^{l} \leq 20$ (right panel).
 In the left panel we present the red region (without contributions from the parameters of flavor physics $|\chi_{ij}|$), which is allowed by $\mu - e$ universality in $\tau$ decays: e.g., for $m_{H^\pm}\leq 120$ GeV we must have the constraint  $X \leq 50$. In the right panel we show two regions: here, the blue(blue\&gray) one is allowed for the cases $5 \leq |\chi_{ij}|
(|\chi_{ij}|\leq5)$. The blue region is clearly the more restrictive one of the two and could become even smaller while the  $|\chi_{ij}|$'s grow. In the case shown, we can get
 that $m_{H^\pm}\leq 150$ GeV for $X\leq 20$. Conversely, with both regions combined, blue\&gray, which represent the portion of parameter space allowed for
 $0.8 \leq |\chi_{ij}|\leq 2$,  we see that the model is more favored, because it opens up
larger regions in the plane $[m_{H^\pm},X]$. One can see this, e.g., for $X\leq 80$, as the bound for the charged Higgs boson mass is now given by $m_{H^\pm} \geq 100$ GeV, that is, not dissimilar from
the previous case (when $X\leq20$).
\subsection{Leptonic meson decays }
The leptonic decay of  a charged meson, $M \to {l} \nu_{l}$, is sensitive to $H^+$ exchange due to the helicity suppression of the SM amplitude. The total decay width is given by \cite{Jung:2010ik,Deschamps:2009rh}:
\begin{eqnarray}
\Gamma(M_{ij} \to {l} \nu ) = G_F^2 m_{l} f_M^2 |V_{ij}|^2 \frac{m_{M_{ij}}}{8 \pi} (1+ \delta_{ em}) |1- \Delta_{ij}|,
\end{eqnarray}
where $i$,  $j$ represent the valence quarks of the meson, $V_{ij}$ is the relevant CKM matrix element, $f_M$ is the decay constant of the meson $M$ (the normalization of the meson decay constant correspond to $f_\pi=131 $ MeV),
$\delta_{em}$ denotes the electromagnetic radiative contributions and $\Delta_{ij}$ is the correction that comes from new physics information. In particular, for the 2HDM-III employing a four-zero Yukawa texture, the leptonic decays receive a contribution from charged Higgs
bosons in the following form:
\begin{eqnarray}
\Delta_{ij} = \bigg( \frac{m_M}{m_{H^\pm}} \bigg)^2 Z_{kk} \bigg( \frac{Y_{ij} m_{u_i}+ X_{ij} m_{d_j}}{V_{ij} (m_{u_i}+m_{d_j})} \bigg),\quad \, \, \, k=2,3.
\end{eqnarray}
In the more general 2HDM-III the  $\Delta{ij}$ correction can be a complex number.
As is pointed out in \cite{Jung:2010ik}, in some Two Higgs Doublet Models (2HDM's)  with natural flavor conservation  the correction $\Delta_{ij}$ (with $ \chi$'s $ =0$)  is predicted to be positive (in 2HDM-I) or
negative (in 2HDM-X), while
can have either sign in 2HDM-II and 2HDM-Y, depending on the decaying meson, whereas it
is absent in the {\it inert} Higgs scenario.

\begin{figure}[t]
\begin{center}
\includegraphics[origin=c, angle=0, scale=0.5]{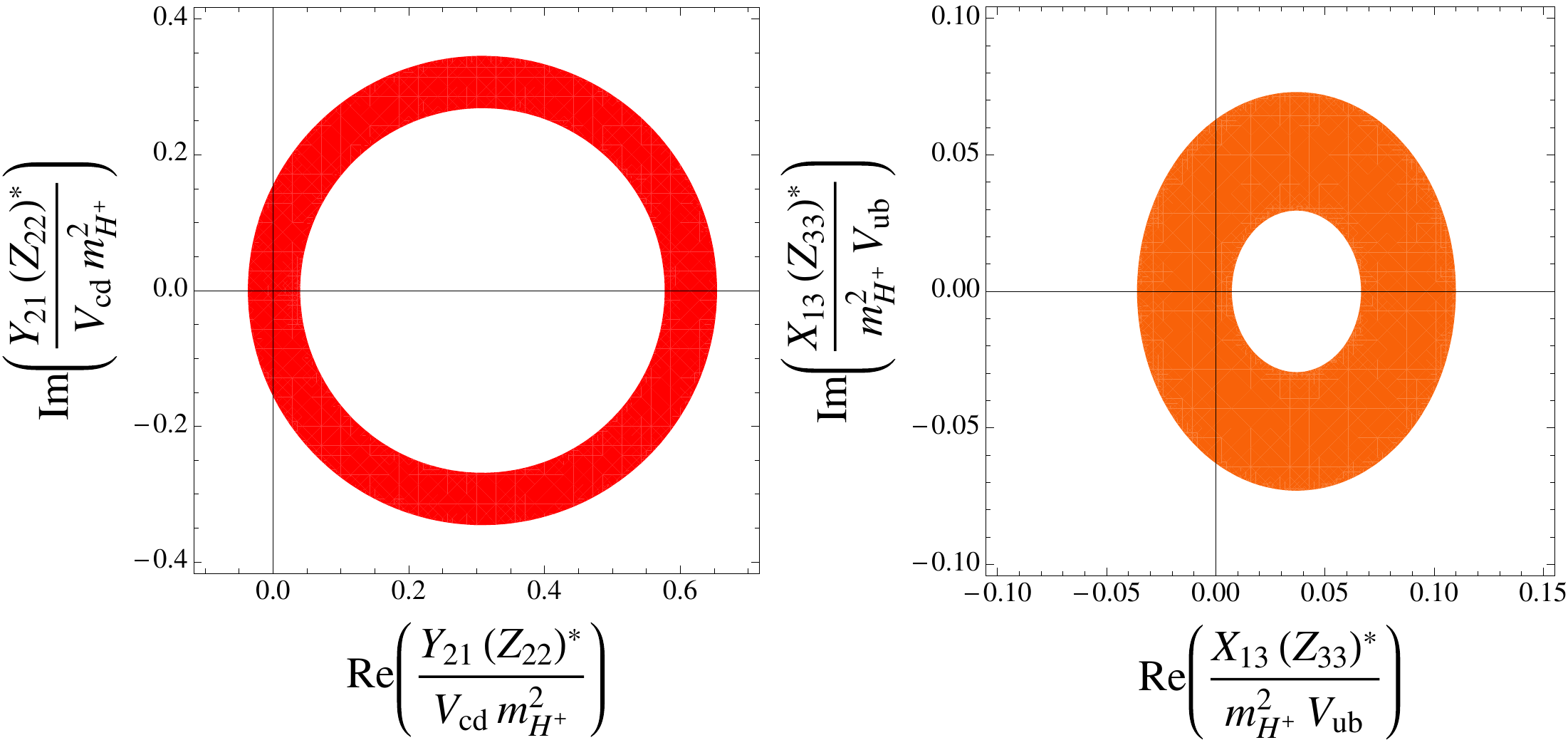}
\vspace*{-5mm}
\caption{Allowed region for the $Z_{22}^* Y_{21}/(m_{H^\pm}^2 V_{cd} )$ and
$Z_{33}^* X_{13}/(m_{H^\pm}^2 V_{ub} )$ complex planes
 from  $D\to  \mu \nu$ (left) and $B \to \tau \nu$ in units of GeV$^{-2}$.}
\label{fig:DB-plane}
\end{center}
\end{figure}

\begin{figure}[t]
\begin{center}
\includegraphics[origin=c, angle=0, scale=0.5]{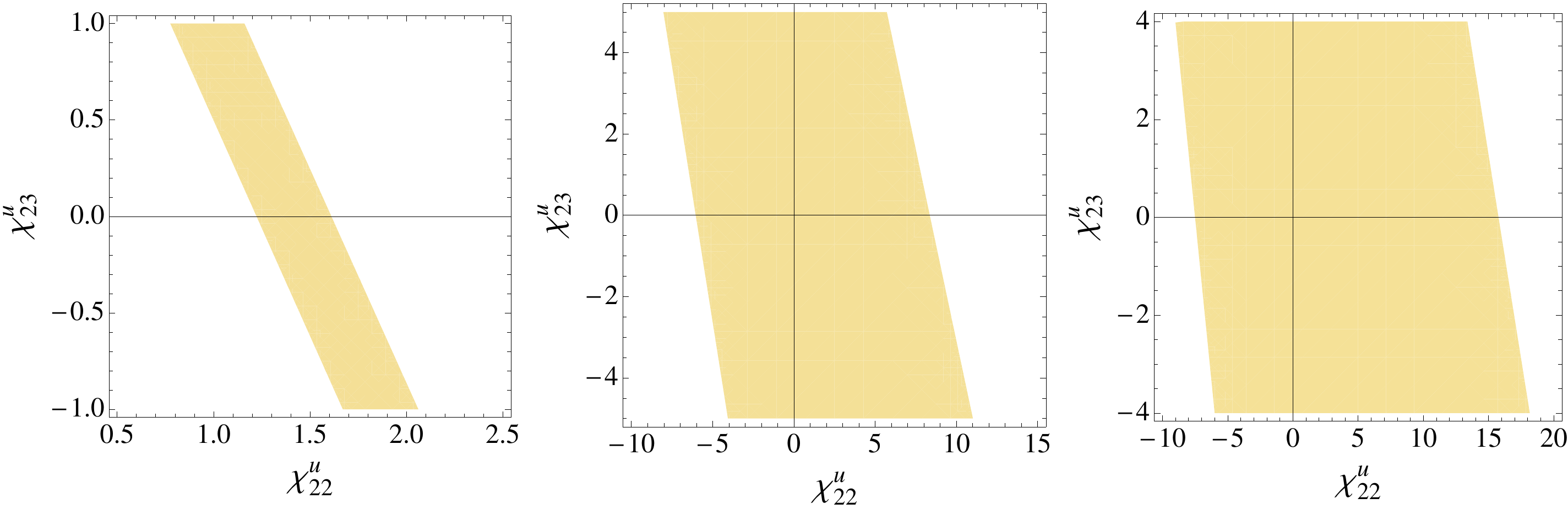}
\vspace*{-5mm}
\caption{The most constrained region for  $\chi_{22}^u$ and $\chi_{23}^u$ from  $D\to  \mu \nu$ for the following cases: $Z, \,Y>>1$ (left),
$Z >> Y$   (center) and $Y >>Z $ (right), with 80 GeV $\leq m_{H^\pm} \leq 160$ GeV.  We assume that $0.1 \leq \chi^{l}_{22} \leq 1.5$. }
\label{fig:Xu-D}
\end{center}
\end{figure}

\begin{figure}[t]
\begin{center}
\includegraphics[origin=c, angle=0, scale=0.5]{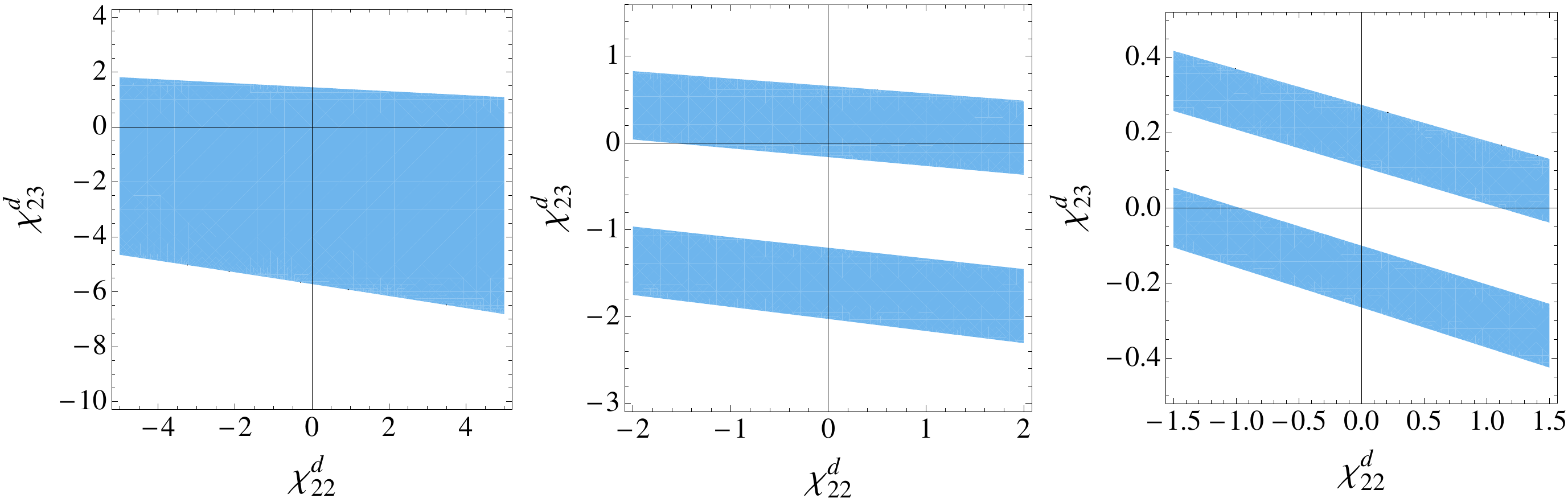}
\vspace*{-5mm}
\caption{The most constrained region for  $\chi_{22}^d$ and $\chi_{23}^d$ from  $B\to  \tau \nu$ for the following cases:
$X >> Z$ (left), $Z >> X $ (center) and $Z, \,X>>1$ (right), with 80 GeV $\leq m_{H^\pm} \leq 160$ GeV.
We assume that $0.1 \leq \chi^{l}_{22} \leq 1.5$. }
\label{fig:Xd-B}
\end{center}
\end{figure}

We focus on decays of heavy pseudoscalar mesons
$D \to \mu \nu$, $B \to \tau \nu$ and $D_s \to \mu \nu, \tau \nu$, which have been measured.
In $B$ and $D$ decays the function $\Delta_{ij}$, one can neglect the contribution proportional to the light quark mass
 because $m_u/m_b \leq m_d/m_c \sim O (10^{-3})$. Hence the functions $\Delta_{ij}$ for $D \to \mu \nu$ and $B \to \tau \nu$, respectively, are given by:
 \begin{eqnarray}
 \Delta_{cd}  & \approx& \frac{m_D^2}{m_{H^\pm}^2} Z_{22} \frac{Y_{21}}{V_{cd}}  \nonumber \\
 & = & \frac{m_D^2}{m_{H^\pm}^2} \bigg( Z - \frac{f (Z)}{\sqrt{2}} \chi_{22}^{l} \bigg)
 \bigg(  \big( Y - \frac{f (Y)}{\sqrt{2}} \chi_{22}^u \big) - \sqrt{\frac{m_t}{m_c}} \frac{V_{td}}{V_{cd}} \frac{f (Y)}{\sqrt{2}} \chi_{23}^u \bigg),  \\
  \Delta_{ub} & \approx & \frac{m_B^2}{m_{H^\pm}^2} Z_{33} \frac{X_{13}}{V_{ub}} \nonumber \\
  &= & \frac{m_B^2}{m_{H^\pm}^2} \bigg( Z - \frac{f (Z)}{\sqrt{2}} \chi_{33}^{l} \bigg)
  \bigg(  \big( X - \frac{f (X)}{\sqrt{2}} \chi_{33}^d \big) - \sqrt{\frac{m_s}{m_b}} \frac{V_{us}}{V_{ub}} \frac{f (X)}{\sqrt{2}} \chi_{23}^d \bigg).  \end{eqnarray}
Apparently, the factor $\sqrt{m_t/m_c}$ in $\Delta_{cd}$ could be considered as a dangerous term, which could make the theoretical predictions deviate from the experimental results, however, the term $V_{td}/V_{cd}$ reduces this possible effect. Similarly, this happens when
one wants to fit the four-zero texture of the Yukawa matrices with the CKM matrix.
Since experimental results of $B(D^+ \to \mu \nu )$, which were measured by CLEO collaboration \cite{Eisenstein:2008aa}, the authors of Ref. \cite{Jung:2010ik} found the following constraints at $95 \%$ C.L. for any model:
$ 0.8 \leq |1 - \Delta_{ub}|  \leq   2$ and  $0.87 \leq |1 - \Delta_{cd}| \leq   1.12$. Considering those constraints,
we can get the allowed circular bands in the $Z_{22}^* Y_{21}/(m_{H^\pm}^2 V_{cd} )$ and
$Z_{33}^* X_{13}/(m_{H^\pm}^2 V_{ub} )$ complex
planes. Our numerical analysis obtained from the decays $B \to \tau \nu$ and $D \to \mu \nu$ is shown
in Fig. \ref{fig:DB-plane}, which is consistent with the results of \cite{Jung:2010ik} when the parameters $\chi's$ are absent. For instance, we  also find the real solutions are $ Z_{33} X_{13}/(m_{H^\pm}^2 V_{ub} ) \in [-0.036, 0.008]$ GeV$^{-2}$ or $[0.064,0.108]$ GeV$^{-2}$
from the $B\to\tau\nu$,
and $Z_{22} Y_{21}/(m_{H^\pm}^2 V_{cd} ) \in [-0.037, 0.035]$ GeV$^{-2}$ or $[0.535,0.609]$ GeV$^{-2}$ from the $D\to\mu\nu$. In  Fig. \ref{fig:Xu-D}  we show the allowed region  for the plane $[\chi_{22}^u, \, \chi_{23}^u] $, assuming that
$\chi_{22}^{l} \in [0.1,1.5]$, and considering the bounds for
 $D \to \mu \nu$. One can see that $\chi_{23}^u \in [0.75, 1.25]$ when $\chi_{22}^u=1$ for   $30 \leq |Z| = |Y|$. For the cases $Z>>Y$ or $Y>>Z$ the permitted region is larger than  for $ 1<< |Z| = |Y|$ and $1\leq \chi_{23}^u$. Therefore,
 $1\leq \chi_{23}^u $ are allowed parameters for the leptonic decay of $D$ mesons and the consequences on the
 phenomenology of charged Higgs bosons could be an important probe of the flavor
structure of the Yukawa sector. Otherwise,
 for the low energy process  $B \to \tau \nu$ we can get  bounds for the parameters of
the Yukawa texture pertaining to the $d$-quark family. In Fig. \ref{fig:Xd-B}, we show the allowed regions in the plane $[\chi_{22}^d, \chi_{23}^d]$ for the following cases:  $X >> Z$  (left panel),
 $Z >> X $ (center panel) and $Z,X >>1$ (right panel), with 80 GeV $\leq m_{H^\pm} \leq 160$ GeV and  considering
 $0.1 \leq \chi^{l}_{22} \leq 1.5$. We can see that the non-diagonal parameter $\chi_{23}^d$ is more constrained than  $\chi_{23}^u$. For the case  $Z, \,X>>1$ and $\chi_{22}^d=1$,  for $Z=X=20$ we found $\chi_{23}^d \in [-0.35,-0.2]$ or $\chi_{23}^d \in [0,0.2]$, so that this case could correspond to
a 2HDM-II-like scenario (see Tab. \ref{couplings})
 when $\tan \beta $ is large. This scenario is  more constrained when $X=Z \geq 40$ and the bound for $ |\chi_{23}^d| \leq 0.2 $ is obtained. Another interesting case is when  $Z >> X $ and  $\chi_{22}^d=1$, for $X=0.1$ and $Z =80$ we obtain the following allowed regions:
 $\chi_{23}^d \in [-1.8,-1.2]$ or $\chi_{23}^d \in [-0.2,0.6]$, in this scenario it is
therefore possible to obtain the constraint $|\chi_{23}^d| = 1 $.  When
 $X>>Z$ we get a wider permitted region for  $\chi_{23}^d $, defined in the interval $(-7,2)$.

\begin{figure}[t]
\begin{center}
\includegraphics[origin=c, angle=0, scale=0.4]{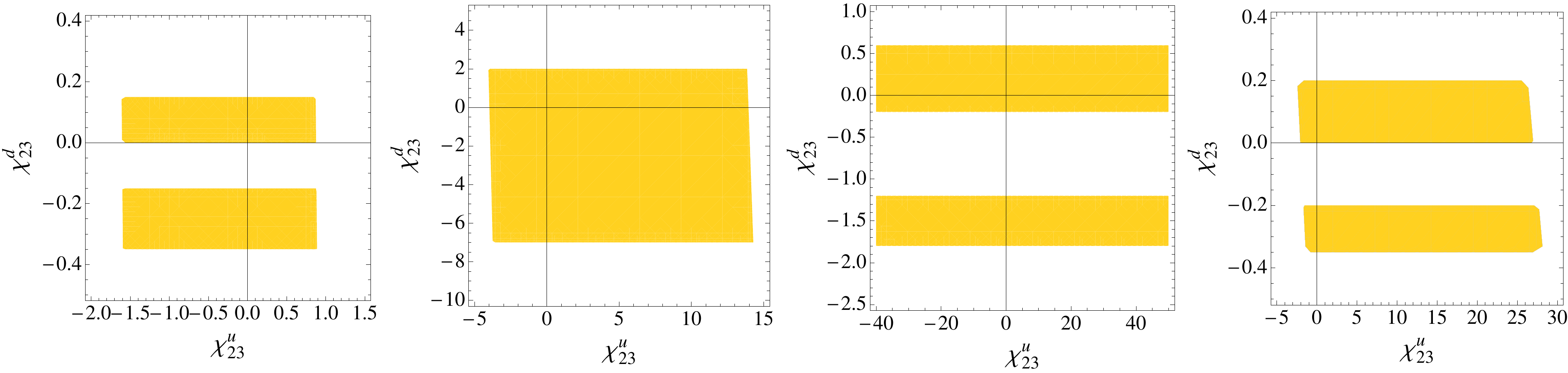}
\vspace*{-4mm}
\caption{The most constrained region for  $\chi_{23}^d$ vs. $\chi_{23}^u$ from  $B\to  \tau \nu$, $D \to \mu \nu$ and $D_s \to {l} \nu$ for
the following cases:  $|Z|=|X|= |Y| >> 1$ (left), $Z >> X,Y $ (center-left), $X >>Y, Z$  (center-right), and  $|Z|=|X| >> Y $ (right), with 80 GeV
$\leq m_{H^\pm} \leq 160$ GeV.
We assume that $0.1 \leq \chi^{l}_{22} \leq 1.5$ and $\chi^{u,d}_{22} = \chi^{u,d}_{33} =1$. }
\label{fig:XDs-B}
\end{center}
\end{figure}

 From $D_s \to \mu \nu, \tau \nu$ decays the constraint $0.97 \leq |1- \Delta_{cs} |\leq 1.16$ is obtained in Ref. \cite{Jung:2010ik} and one can get the real solutions  $\Delta_{cs}/m_{D_s}^2 \in [-0.044, 0.008]$ GeV $^{-2}$ or  [0.545, 0.598] GeV $^{-2}$.
 Here the ratio $m_s/m_c \approx 10 \%$, thus we cannot neglect
$s$-quark effects and the expression for  $\Delta_{cs} $ is:
 \begin{eqnarray}
 \Delta_{cs}  & = &  \bigg( \frac{m_{D_s}}{m_{H^\pm}} \bigg)^2 Z_{kk} \bigg( \frac{Y_{22} m_{c}+ X_{22} m_{s}}{V_{cs} (m_{c}+m_{s})} \bigg) \, \, \, ~~~~~~~~(k=2,3), \\ \nonumber
\end{eqnarray}
\begin{eqnarray}
 X_{22} & = & V_{cs} \bigg( X- \frac{f(X)}{\sqrt{2}} \chi_{22}^d \bigg)- \sqrt{\frac{mb}{ms}} V_{cb} \frac{f(X)}{\sqrt{2}}  \chi_{23}^d, \nonumber \\
 Y_{22} & = & V_{cs} \bigg( Y- \frac{f(Y)}{\sqrt{2}} \chi_{22}^u \bigg)- \sqrt{\frac{mt}{mc}} V_{ts} \frac{f(Y)}{\sqrt{2}}  \chi_{23}^u.
 \end{eqnarray}
 With this information, we can establish a correlation among the parameters that come from $D\to \mu \nu$ and $B\to \tau \nu$.
 Considering  the information from $B \to \tau \nu $, $D\to \mu \nu$ and $D_s \to \tau \nu, \, \mu \nu$, we show in Fig. \ref{fig:XDs-B} the constraints for  the non-diagonal terms of the Yukawa textures $  \chi_{23}^d $ and
   $\chi_{23}^u$, assuming  $0.1\leq \chi_{22}^{l} =  \chi_{33}^{l} \leq 1.5  $, as well as $  \chi_{22}^d =   \chi_{22}^u =1$. We present in Tab. \ref{bounds-chi23}  a set bounds for these parameters in  several scenarios, which are shown in Tab. \ref{couplings}. Combining results from the table, one can derive general constraints for   $  |\chi_{23}^d| \leq 0.15$ and $|\chi_{23}^u | \leq 1.5$ for almost all scenarios. Only in the  2HDM-Y-like version one can obtain a less stringent bound for $ \chi_{23}^u$.
\subsection{Semileptonic decays  $B\to D \tau \nu$}

Purely leptonic decays of mesons interwine EW
and QCD interactions. However, the role of strong
interaction materializes only in the presence of a decay constant,
to be assessed through theoretical methods. Semi-leptonic
decays are complicated to describe since they involve
form factors with a non-trivial dependence on the
momentum transfer.
If the form factors are known with sufficient accuracy,
semi-leptonic BRs start becoming stringent
constraints on new physics models.
The BaBar and Belle experiments  published
the first measurements of $B(B\to D \tau \nu)$ \cite{Matyja:2007kt, Aubert:2009at}.
Recently, using the full data set collected by BaBar, the update of
BR$(B\to D \tau \nu)$ and BR$(B\to D^* \tau \nu)$ was presented in \cite{Lees:2012xj},
from where it is clear that the 2HDM-II is disfavored. Since this model cannot explain $R(D)$ and $R(D^*)$ simultaneously (and for $B \to \tau \nu $ a high fine tuning is needed), where $R(D^*)$ are the ratios
\begin{eqnarray}
R(D^*) = {\rm BR}(B\to D^* \tau \nu)/ {\rm BR}(B\to D^* {l} \nu)
\end{eqnarray}
with
\begin{eqnarray}
R(D) & = &  0.44 \pm 0.058 \pm 0.042, \\ \nonumber
R(D^*) & = &  0.332 \pm 0.024 \pm 0.018.
\end{eqnarray}
However, lately, in Ref. \cite{Crivellin:2012ye}, it was shown that one can
simultaneously explain $R(D)$ and $R(D^*)$ in the 2HDM-III with a general flavor structure, where the non-diagonal terms from the $u$-quark sector are relevant.

\begin{figure}[t]
\begin{center}
\includegraphics[origin=c, angle=0, scale=0.45]{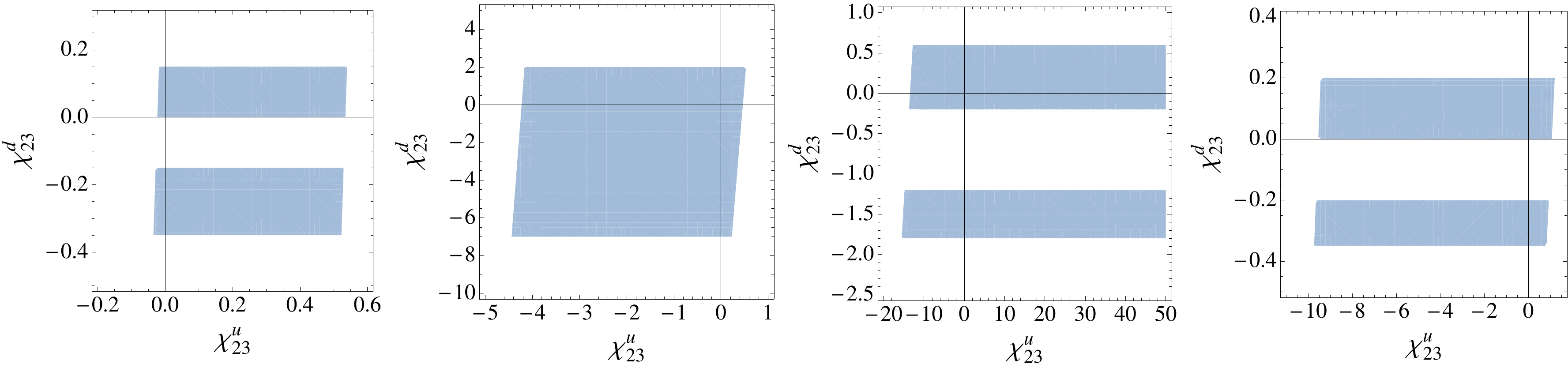}
\vspace*{-4mm}
\caption{The most constrained region for  $\chi_{23}^d$ vs. $\chi_{23}^u$ from  $B\to  \tau \nu$ and $B \to D \tau \nu$ for
the following cases:  $|Z|=|X|= |Y| >> 1$ (left), $Z >> X,Y $  (center-left), $X >>Y, Z$ (center-right), and $|Z|=|X| >> Y $ (right), with 80 GeV
$\leq m_{H^\pm} \leq 160$ GeV.
We assume that $0.1 \leq \chi^{l}_{22} \leq 1.5$ and $\chi^{u,d}_{22} = \chi^{u,d}_{33} =1$. }
\label{fig:XDB}
\end{center}
\end{figure}

Following the analysis of Ref. \cite{Deschamps:2009rh}, an interesting observable is the normalized BR,
$R_{B\to D \tau \nu} = {\rm BR} (B\to D \tau \nu)/ {\rm BR}  (B\to D e \nu)$,
which corresponds
to a $b \to c$ transition, with a CKM factor much
larger than the purely leptonic $B$ decay.   One can write
this term  as a second order polynomial in the charged Higgs boson coupling to fermions, as
\begin{eqnarray}
R_{B\to D \tau \nu} = a_0+ a_1  (m_B^2 - m_D^2) \delta_{23} + a_2 (m_B^2 - m_D^2)^2 \delta_{23}^2,
\label{RD}
\end{eqnarray}
where the factor $\delta_{23}$ is determined by the coupling $H^+ u_i\bar d_i$, where the general expression for $\delta_{ij}$ is given by
\begin{eqnarray}
\delta_{ij} = - \frac{Z_{33}}{m_{H^\pm}^2} \bigg( \frac{Y_{ij}m_{u_i}- X_{ij} m_{d_j}}{m_{u_i}- m_{d_j}}\bigg).
\end{eqnarray}
The polynomial coefficients $a_i$ in eq. (\ref{RD}) are given in Ref. \cite{Deschamps:2009rh} as:
\begin{eqnarray}
a_0 &= & 0.2970+0.1286 d\rho^2 + 0.7379 d\Delta, \nonumber \\
a_1& =& 0.1065+ 0.0546 d\rho^2 + 0.4631 d\Delta, \\ \nonumber
a_2 & = & 0.0178 +0.0010 d\rho^2 + 0.0077 d\Delta,
\end{eqnarray}
where $d \rho^2 = \rho^2 - 1.18$ and $d\Delta = \Delta -0.046$ are the variations of the semi-leptonic form factors
$\rho^2$ and $\Delta$ \cite{Barberio:2008fa, de Divitiis:2007uk}.
Similarly to the  leptonic process $D_s \to {l} \nu$, we can establish a correlation among the parameters that come from
$B \to D {l} \nu$ and $B\to \tau \nu$. In all cases, we consider simultaneously $R(D)$ and $R(D^*)$.
 One can then constraint the non-diagonal terms of the Yukawa texture, $  \chi_{23}^d $ and
   $\chi_{23}^u$, by assuming  $0.1\leq \chi_{22}^{l} =  \chi_{33}^{l} \leq 1.5  $ as well as $  \chi_{22}^d =   \chi_{22}^u =1$. We can then show in Tab. \ref{bounds-chi23}  a set of bounds for these parameters in  several scenarios. By combining results from this table, one can derive general constraints for   $  |\chi_{23}^d| \leq 0.15$ and $|\chi_{23}^u | \leq 1.5$ for several scenarios (see also Fig. \ref{fig:XDB}). Again, the 2HDM-Y-like version cannot offer  a bound for $ \chi_{23}^u  $ easily.
One can see that  the 2HDM-III with a Yukawa texture can avoid the constraints of the factor $R_{B\to D \tau \nu} $ and can thus appear as rather exotic physics, i.e., very different from the traditional 2HDMs with NFC. In particular, decay channels involving $H^\pm \to c b$
 could be relevant and be searched for in the transition $t \to H^\pm b$ \cite{DiazCruz:2009ek,Akeroyd:2012yg},
if the $H^\pm$ state is sufficiently light.
\begin{table}[tbp]
\centering
\begin{tabular}{|c||c|c|c|c|c|}
\hline
{\small 2HDM-III's} & {\small $\chi_{23}^d (B\to \tau \nu)$ } &   {\small $\chi_{23}^u(D_s\to {l} \nu)$ } & {\small  $\chi_{23}^u(B\to D \tau \nu)$ }& {\small $\chi_{23}^u$ (combination) } & {\small $(X, Y, Z) $ }\\ \hline
{\small 2HDM-I-like} & {\small (-0.35,-0.15) or} & {\small  (-1.5,0.9)  } &  {\small (-0.05,0.45) }  &  {\small (-0.05,0.45)  }
& {\small  (20, 20, 20) } \\  &{\small    (0,0.15) } & & & &  \\ \hline
{\small 2HDM-II-like } &  {\small  (-0.35,-0.2) or  } & {\small  (-2,27)   } &   {\small  (-9.6,1.2) }
&  {\small (-2,1.2)   } & {\small   (20, 0.1, 20) } \\ &{\small    (0,0.2) } & & & &  \\ \hline
{\small 2HDM-X-like } & {\small  (-7,2) } &  {\small  (-4,14) } &  {\small (-3.8,0.47)  } &  {\small  (-4,0.47) } &  {\small (0.1, 0 .1, 50)}
\\ &&&&& \\ \hline
{\small 2HDM-Y-like } &  {\small (-1.8,-1.2) or  } & {\small  (-40,50) } &  {\small  (-16,50) } &  {\small (-16,50) }
&  {\small (50, 0.5, 0.5) }\\ &{\small    (-0.2,0.6) } & & & &  \\ \hline
\end{tabular}
\caption{ Constraints from   $B \to D \tau \nu$, $D_s \to \tau \nu, \mu \nu$ and $B \to \tau \nu$
decays.
We show the allowed intervals  for $\chi_{23}^{u,d}$ constrained by each low energy process, according to  the
different scenarios presented in Tab. \ref{couplings} as well as the combination of constraints for the $\chi_{23}^{u}$ parameter. We assume $0.1\leq \chi_{22}^{l} =  \chi_{33}^{l} \leq 1.5  $ as well as $  \chi_{22}^d =   \chi_{22}^u =1$. Taking  80 GeV $\leq m_{H^\pm}\leq 160$ GeV and specific values for the $X,Y$ and $Z$ parameters given in Tab. \ref{couplings}. }
\label{bounds-chi23}
\end{table}
\subsection{$B \to X_s \gamma$ decays}

The radiative decay $B\to X_s \gamma $ has been calculated at Next-to-Next-to-Leading Order
(NNLO) in the SM, leading to the prediction
BR$(B\to X_s \gamma)_{\rm SM}= (3.15\pm 0.23) \times 10^{-4}$ \cite{Misiak:2006zs}. In the 2HDM the decay amplitude is known at NLO \cite{Borzumati:1998nx,Ciuchini:1997xe, Ciafaloni:1997un} and in the 2HDM with Minimal Flavor Violation (MFV) \cite{Degrassi:2010ne}  while, only very recently, NNLO
results have been presented for both a Type-I and Type-II 2HDM \cite{Hermann:2012fc}. The current average of the measurements by CLEO
\cite{Chen:2001fja}, Belle \cite{Abe:2001hk,Limosani:2009qg}, and BaBar \cite{Lees:2012ym,Lees:2012wg, Aubert:2007my} reads
 BR$(\bar{B} \to X_ \gamma)|_{E_\gamma > 1.6 \,\, {\rm GeV}} = (3.37\pm 0.23) \times 10^{-4}$.

 In this subsection we show the constraints on the
  off-diagonal terms of the four-zero Yukawa texture of the
2HDM-III through a general study of the processes $B\to X_s \gamma $.
We first start with a digression on Wilson coefficients entering the higher order calculations.

 \subsubsection{NLO Wilson coefficients at the scale $\mu_W$ }

To the first order in $\alpha_s$, the effective Wilson coefficients at the
scale $\mu_W = {\cal O}(\mw) $ can be written as \cite{Borzumati:1998nx,Xiao:2003ya}
\beq
 C^{\,{\rm eff}}_i(\mu_W) =  C^{0,\,{\rm eff}}_i(\mu_W)
             + \frac{\alpha_s(\mu_W)}{4\pi} C^{1,\,{\rm eff}}_i(\mu_W) \, .
\label{eq:wc-mw}
\eeq
The LO   contribution of our 2HDM-III version to the relevant Wilson coefficients at the matching energy scale $\mu_W$
take the form \cite{Borzumati:1998nx,Xiao:2003ya},
\beq
  \delta C^{0,eff}_{(7, 8) }(\muw)   =    \bigg|\frac{Y_{33}^{u} Y_{32}^{u*}}{V_{tb} V_{ts}}\bigg|  \, C_{(7,8),YY}^0(y_t)
  +       \bigg|\frac{X_{33}^{u} Y_{32}^{u*}}{V_{tb} V_{ts}}\bigg| \, C_{(7,8),XY}^0(y_t)  ,
\label{eq:c80mw}
\eeq
where  $y_t=m_t^2/\mh^2$, $ \delta C^{0,eff}_{(7, 8) }(\muw) =  C^{0,eff}_{(7, 8) }(\muw)- C^{0}_{(7, 8),SM }(\muw) $, and the coefficients $C^{0,1}_{(7, 8),SM }(\muw)$, $C_{(7,8),YY}^0 (y_t)$, $C_{(7,8),XY}^0 (y_t)$  are well known, which are given in \cite{Borzumati:1998nx,Xiao:2003ya}   and
\beq
 \bigg|\frac{Y_{33} Y_{32}^*}{V_{tb} V_{ts}}\bigg| & = & \bigg[ \bigg(  Y- \frac{f(y)}{\sqrt{2}} \chi_{33}^u\bigg) - \sqrt{\frac{m_c}{m_t}}
\bigg(\frac{V_{cb}}{V_{tb}} \bigg) \frac{f(Y)}{\sqrt{2}} \chi_{23}^u \bigg] \nonumber \\
& \times & \bigg[ \bigg(  Y- \frac{f(y)}{\sqrt{2}} \chi_{33}^u\bigg) - \sqrt{\frac{m_c}{m_t}} \bigg(\frac{V_{cs}}{V_{ts}} \bigg) \frac{f(Y)}{\sqrt{2}} \chi_{23}^u \bigg]^* , \label{coup-bsg1} \\
 \bigg|\frac{X_{33} Y_{32}^*}{V_{tb} V_{ts}}\bigg| & = & \bigg[ \bigg(  X- \frac{f(X)}{\sqrt{2}} \chi_{33}^d\bigg) - \sqrt{\frac{m_s}{m_b}}
\bigg(\frac{V_{ts}}{V_{tb}} \bigg) \frac{f(X)}{\sqrt{2}} \chi_{23}^d \bigg] \nonumber \\
& \times &
\bigg[ \bigg(  Y- \frac{f(y)}{\sqrt{2}} \chi_{33}^u\bigg) - \sqrt{\frac{m_c}{m_t}} \bigg(\frac{V_{cs}}{V_{ts}} \bigg) \frac{f(Y)}{\sqrt{2}} \chi_{23}^u \bigg]^*,
\label{coup-bsg2}
\eeq
 The NLO Wilson coefficients at the matching scale $\mu_W$ in the 2HDM-III can be written as \cite{Borzumati:1998nx}
\beq
 C_1^{1,\,{ eff}}(\muw) & = & 15 + 6\ln\frac{\mu_W^2}{\mw^2}\, , \\
 C_4^{1,\,{ eff}}(\muw) & = & E_0 + \frac{2}{3} \ln\frac{\mu_W^2}{\mw^2}
    + \bigg|\frac{Y_{33}^{u} Y_{32}^{u*}}{V_{tb} V_{ts}}\bigg|  \, E_H \, , \label{c41effmw} \\
 C_i^{1,\,{ eff}}(\muw) & = &   0 \hspace*{1.5truecm} (i=2,3,5,6)\, , \\
 \delta C_{(7,8)}^{1,\,{ eff}}(\muw) & =  & \bigg|\frac{Y_{33}^{u} Y_{32}^{u*}}{V_{tb} V_{ts}}\bigg|  \, C_{(7,8), YY}^{1}(\muw)
  +       \bigg|\frac{X_{33}^{u} Y_{32}^{u*}}{V_{tb} V_{ts}}\bigg| \, C_{(7,8),XY}^{1}(\muw), \label{c81effmw}
\eeq
where  the functions on the right-hand side of
eqs. (\ref{c41effmw}) and (\ref{c81effmw}) are given in Ref. \cite{Borzumati:1998nx,Xiao:2003ya}.
The  contributions  of our version 2HDM-III to the $B \to X_s \gamma$ decay are described by
the functions $C^{0,1}_{i,j}(\muw)$  ($i=7,8$ and $j=(YY,XY)$), as well as
the magnitude and
sign of the  couplings $Y_{33}^{u} $,  $Y_{32}^{u*}$ and $X_{33}^{u}$. Otherwise, in order to compare with previous results,
is convenient to write the
 eqs. (\ref{coup-bsg1}-\ref{coup-bsg2}) as:
\beq
 \bigg|\frac{Y_{33} Y_{32}^*}{V_{tb} V_{ts}}\bigg| & = & \bigg[ \lambda_{tt}^U - \sqrt{\frac{m_c}{m_t}}
\bigg(\frac{V_{cb}}{V_{tb}} \bigg) \frac{f(Y)}{\sqrt{2}} \chi_{23}^u \bigg] \bigg[ \lambda_{tt}^U - \sqrt{\frac{m_c}{m_t}} \bigg(\frac{V_{cs}}{V_{ts}} \bigg) \frac{f(Y)}{\sqrt{2}} \chi_{23}^u \bigg]^* ,\\
 \bigg|\frac{X_{33} Y_{32}^*}{V_{tb} V_{ts}}\bigg| & = & \bigg[ \lambda_{tt}^D - \sqrt{\frac{m_s}{m_b}}
\bigg(\frac{V_{ts}}{V_{tb}} \bigg) \frac{f(X)}{\sqrt{2}} \chi_{23}^d \bigg] \bigg[ \lambda_{tt}^U - \sqrt{\frac{m_c}{m_t}} \bigg(\frac{V_{cs}}{V_{ts}} \bigg) \frac{f(Y)}{\sqrt{2}} \chi_{23}^u \bigg]^*, \label{eq:xy-m30}
\eeq
where $\lambda_{tt}^D$ and $\lambda_{tt}^D$, expressed in eq. (\ref{stal}), are parameters defined in a version of the 2HDM-III without off-diagonal terms in the Yukawa texture \cite{Mahmoudi:2009zx, Xiao:2003ya}\footnote{In the version 2HDM-III of
\cite{Mahmoudi:2009zx, BowserChao:1998yp, Xiao:2003ya}, one has  $\lambda_{tt}^D= \lambda_{bb}$, $\lambda_{tt}^U= \lambda_{tt}$.}.  Again, when the off-diagonal terms of the
four-zero texture of Yukawa matrices are absent, we recover the results  mentioned.
When the heavy charged Higgs bosons is integrated out at the scale
$\muw$, the QCD running of the the Wilson coefficients $C_i(\muw)$ down to the
lower energy scale $\mu_b = {\cal O}(m_b)$.
Thence, for a complete NLO analysis of the radiative decay $\bxsga$ only the Wilson coefficient
$C^{\,{ eff}}_7(\mu_b)$ has to be known, which is:
\beq
C^{\,{ eff}}_7(\mu_b) =  C^{0,\,{ eff}}_7(\mu_b)
             + \frac{\alpha_s(\mu_b)}{4\pi} C^{1,\,{ eff}}_7(\mu_b) \, ,
\label{c7mub}
\eeq
where the functions $C^{0,\,{ eff}}_7(\mu_b) $  and $C^{1,\,{ eff}}_7(\mu_b)$ as functions of
$C^{0}_{i,j}(\mu_W)$ and their complete expressions are given in \cite{Borzumati:1998nx,Xiao:2003ya}.

\subsection{$\brbxsga$ }
\begin{figure}[t]
\begin{center}
\includegraphics[origin=c, angle=0, scale=0.5]{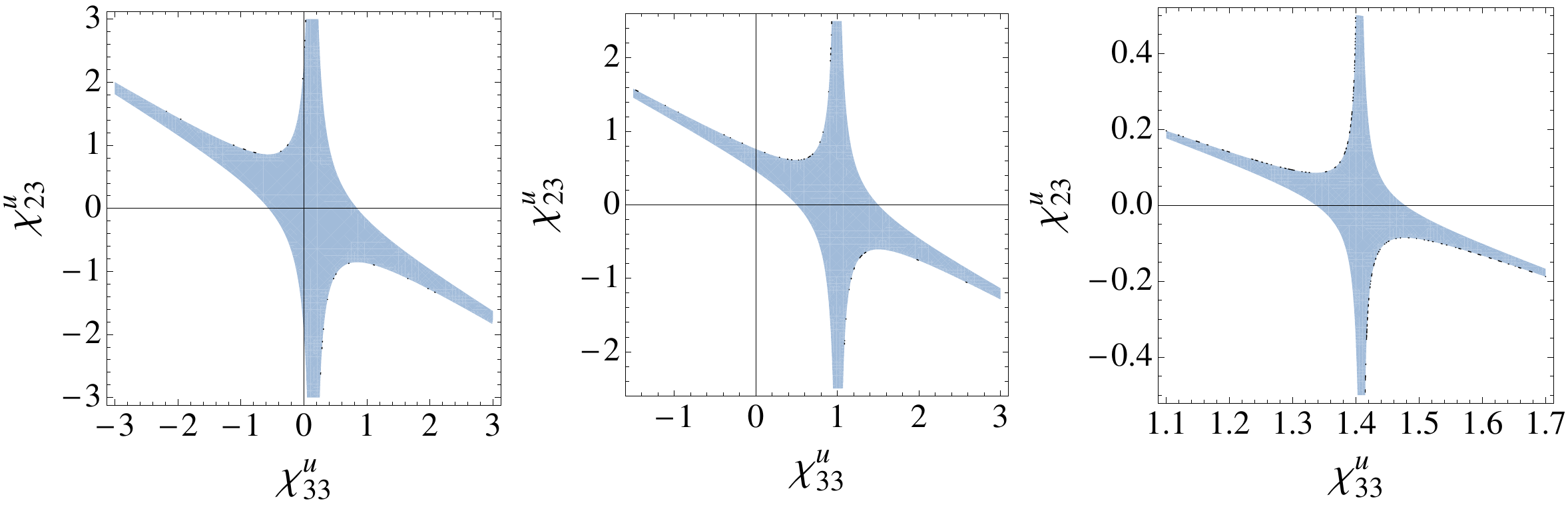}
\vspace*{-4mm}
\caption{The allowed  region for  $\chi_{33}^u$ vs. $\chi_{23}^u$ from  $B\to X_s \gamma$ (using the constraint
$\bigg|\frac{Y_{33} Y_{32}^*}{V_{tb} V_{ts}}\bigg|  < 0.25$) and  $B^0-\bar{B}^0$ mixing
(considering the constraint
$\bigg|\frac{Y_{33} Y_{31}^*}{V_{tb} V_{td}}\bigg|  < 0.25$) for
the following cases:  $ |Y| << 1$ (left),  $ Y= 1$ (center) and $Y=10$ (right), with 80 GeV $\leq m_{H^\pm} \leq 200$ GeV.   }
\label{fig:Bsg1}
\end{center}
\end{figure}

\begin{figure}[t]
\begin{center}
\includegraphics[origin=c, angle=0, scale=0.4]{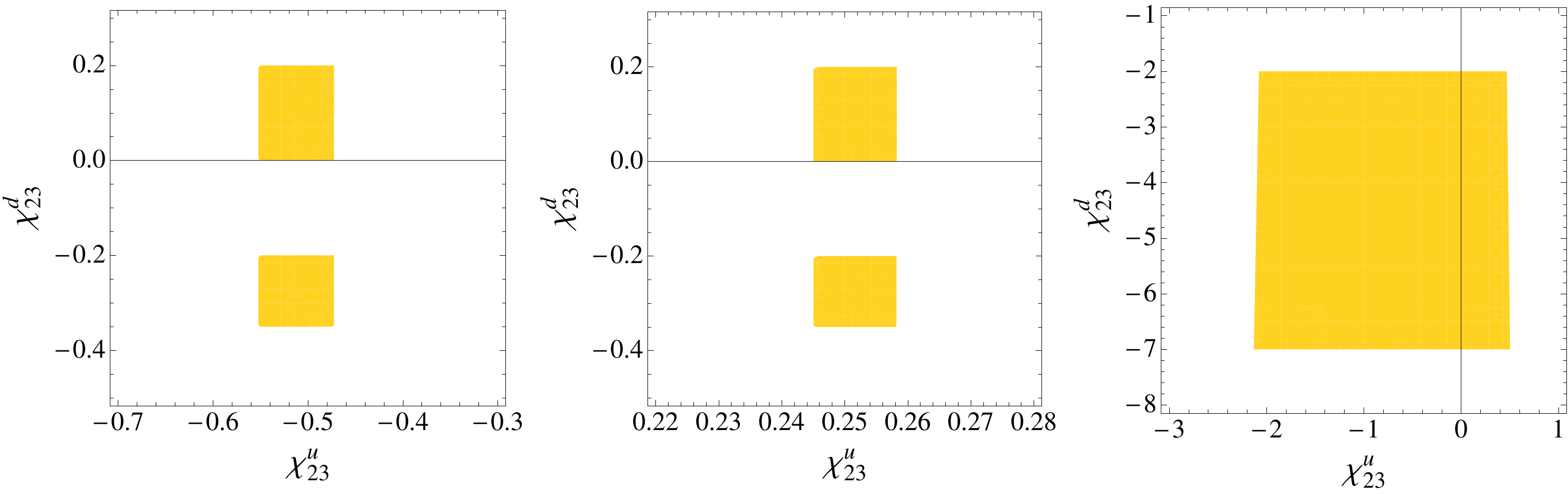}
\vspace*{-4mm}
\caption{The allowed  region for  $\chi_{23}^u$ vs. $\chi_{23}^d$ from  $B\to X_s \gamma$ (using the constraint
$ -1.7<Re\bigg[\frac{X_{33} Y_{32}^*}{V_{tb} V_{ts}}\bigg] < 0.7$) for
the following cases: $X=20$ and $ Y=0.1 $  (left),  $ X=Y= 20$ (center) and $X=Y=0.1$ (right), with 80 GeV $\leq m_{H^\pm} \leq 200$ GeV.
We assume $\chi_{33}^u=1$, $\chi_{33}^d=1$.  }
\label{fig:Bsg2}
\end{center}
\end{figure}

The BR of the inclusive radiative decay
$\bxsga$  at the LO level is given by \cite{Borzumati:1998nx,Xiao:2003ya}:
\beq
\brbxsga_{ LO} &=& { B}_{SL}
\left | \frac{ V_{ts}^*V_{tb} }{V_{cb}}\right |^2  \frac{6\alpha_{ em}}{\pi
\theta(z)}  \,  \left |C_7^{0,eff}(\mu_b)\right |^2   \label{br-lo}
\eeq
 and at the NLO level is
\beq
\brbxsga_{ NLO} &=& {B}_{SL}
\left | \frac{ V_{ts}^*V_{tb} }{V_{cb}}\right |^2  \frac{6\alpha_{ em}}{\pi \theta(z)
\kappa(z)} \,  \left [ |D|^2 + A + \Delta \right ]\, ,  \label{br-nlo}
\eeq
where ${ B}_{SL}=(10.74\pm 0.16)\%$ is the measured semi-leptonic
BR
of the $B$ meson \cite{Barberio:2008fa}, $\alpha_{ em}=1/137.036$ is the fine-structure constant,
$z=m_c^{pole}/m_b^{pole}$ is the ratio of the quark pole masses,
$\theta(z)$ and $\kappa(z)$ denote the phase space factor and the QCD
correction \cite{cm78} for the semi-leptonic $B$ decay  and  are given in \cite{Borzumati:1998nx,Xiao:2003ya}.
The term $D $ in eq. (\ref{br-nlo}) corresponds to the sub-processes
$b \to s \gamma$ \cite{Borzumati:1998nx}
\beq
D &=&  C^{\rm eff}_7(\mub)  +  V(\mub)\,, \label{eq:dbar}
\eeq
where the NLO Wilson coefficient $C^{\rm eff}_7(\mub)$ has been given in eq. (\ref{c7mub}),
and the function $V(\mub)$ is given by \cite{Borzumati:1998nx,Xiao:2003ya}.
In eq. (\ref{br-nlo}), term $A$ is the the correction coming from the
bremsstrahlung process $b \to s\gamma g$ \cite{ag91}.
Now we are ready to present numerical results of the BRs in the 2HDM-III.
Following the recent analysis of Refs.  \cite{Trott:2010iz,Jung:2010ik} and using  standard values  \cite{Borzumati:1998nx,Xiao:2003ya} for the charged Higgs boson mass
(80 GeV $\leq \mh \leq 300$ GeV), we can establish the following constraints:
\beq
 \bigg|\frac{Y_{33} Y_{32}^*}{V_{tb} V_{ts}}\bigg|  < 0.25, \,\,  \,\, \,\,  \,\, \,\, \,\,  \,\, \,\,  \,\, \,\,
  -1.7<Re\bigg[\frac{X_{33} Y_{32}^*}{V_{tb} V_{ts}}\bigg]< 0.7.
\eeq
Since $\bigg|\frac{Y_{33} Y_{32}^*}{V_{tb} V_{ts}}\bigg|  < 0.25$,  we show  in Fig. \ref{fig:Bsg1}  the allowed area in the plane
 $\chi_{33}^u-\chi_{23}^u$, for the cases $Y<<1$ (left panel), $Y=1$
(center panel) and $Y=10$ (right panel). One can then extract the bounds $-0.75\leq \chi_{23}^u \leq -0.15$ for $\chi_{33}^u=1$ and
$0.4\leq \chi_{23}^u \leq 0.9$ for $\chi_{33}^u=-1$, both when $Y<<1$.
Otherwise, using the second constraint $ -1.7<Re\bigg[\frac{X_{33} Y_{32}^*}{V_{tb} V_{ts}}\bigg] < 0.7$, we can obtain the interval permitted for $ \chi_{23}^u $, assuming the allowed interval for   $ \chi_{23}^d $ from $B\to \tau\nu$ and
  $\chi_{33}^u=1=$ $\chi_{33}^d=1$.  In Fig. \ref{fig:Bsg2} one can get the  allowed area for some scenarios of Tab. \ref{bounds-chi23}. We can, e.g., obtain $ \chi_{23}^u \in (-0.55,-0.48)$ for the case $X=20$ and $ Y=0.1 $  (left panel). An interesting scenario for the 2HDM-III is the 2HDM-X-like one, where the allowed region is larger than in other scenarios, with $ \chi_{23}^u \in (-2.2, 0.45)$  and
  $ \chi_{23}^d \in (-7, -2)$ (using the constraint coming from $B\to \tau \nu$), so that one can avoid the most restrictive constraints  hitherto considered.


\subsection{ $B^0-\bar{B}^0$ mixing}

Remembering that in Ref. \cite{Crivellin:2012ye}  the general flavor structure  for the 2HDM-III
has been found more consistent with various other $B$ physics constraints, we are motivated to
test this model  assumption also against limits coming from $B^0-\bar{B}^0$ mixing, where
 the charged Higgs boson contributes to the mass splitting
$\Delta M_{B_d}$.
Being that $\Delta M_{B_d}$ has been measured with very high precision \cite{Barberio:2008fa},
we utilize this quantity directly as in Ref.  \cite{Urban:1997gw,Xiao:2003ya}
\footnote{At LO  the quantity $x_d=\Delta M_{B_d}/\Gamma_B$  is used \cite{BowserChao:1998yp}.}.
 In the  2HDMs,
the NLO mass splitting $\Delta M_{B_d}$ is given by  \cite{Urban:1997gw,Xiao:2003ya}
\begin{eqnarray}
\Delta M_{B_d} = \frac{G_F^2}{6 \pi^2}  \mw^2 |V_{td}|^2 |V_{tb}|^2 S_{ 2HDM}(x_t,y_t) \eta_B(x_t,y_t)
(B_{B_d} f_{B_d}^2  m_B) , \label{dmd-m3}
\end{eqnarray}
where $x_t=\overline{m}^2_t(\mw)/\mw^2$,  $y_t=\overline{m}^2_t(\mw)/\mh^2$ and
\begin{eqnarray}
\eta_B(x_t,y_t)&=& \alpha_S(\mw)^{6/23} \left[ 1 + \frac{\alpha_S(\mw)}{4 \pi}
\left ( \frac{D_{ 2HDM}(x_t,y_t)}{S_{ 2HDM}(x_t,y_t)} - J_5 \right ) \right ]
\label{etabm3}
\end{eqnarray}
with
\begin{eqnarray}
S_{ 2HDM}(x_t,y_t)&=& \left [  S_0(x_t) + S_{WH}(x_t,y_t) + S_{HH}(x_t,y_t)\right ] ,
\label{eq:s2hdm} \\
D_{ 2HDM}&=& D_{ SM}(x_t) + D_H(x_t,y_t),\label{D2hdm}
\end{eqnarray}
here the high energy matching scale  $\mu=\mw$ is chosen.
The functions $D_{ SM}(x_t)$ and $D_H(x_t,y_t)$ of the eq. (\ref{D2hdm}) contain the SM and
new physics parts of the NLO QCD corrections to the mass difference
$\Delta M_{B_d}$ \cite{Urban:1997gw},
\begin{eqnarray}
D_{ SM}(x_t) &=& C_F \left [ L^{(1,SM)}(x_t) + 3 S_0(x_t) \right ]
+ C_A  \left [ L^{(8,SM)}(x_t) +  5 S_0(x_t) \right ]\;,
 \\
D_{H}(x_t,y_t) &=& C_F \left [ L^{(1,H)}(x_t,y_t) + 3 \left (
S_{WH}(x_t,y_t) + S_{HH}(x_t,y_t) \right ) \right ] \nonumber \\
& & + \;C_A \left [  L^{(8,H)}(x_t,y_t) + 5 \left ( S_{WH}(x_t,y_t)
+ S_{HH}(x_t,y_t) \right )  \right ]\,,
\end{eqnarray}
where $C_F=4/3$ and $C_A=1/3$ for $SU(3)_C$. The function $S_0(x_t)$ includes the
dominant top-box contribution in the SM and has been given in Refs. \cite{Urban:1997gw,Xiao:2003ya}. The functions
$S_{WH}(x_t,y_t)$ and $S_{HH}(x_t,y_t)$ incorporate the new physics contributions
from the box diagrams with one or two charged Higgs  bosons involved \cite{Urban:1997gw},
\begin{eqnarray}
S_{WH}(x_t,y_t)&=& \bigg|\frac{Y_{33}^* Y_{31}}{V_{tb} V_{td}}\bigg| \frac{y_t  x_t}{4}\bigg[ \frac{(2 x_t -8 y_t)
                  ln(y_t)}{(1-y_t)^2 (y_t-x_t)}+\frac{6x_t ln(x_t)}{(1-x_t)^2(y_t-x_t)}\nonumber \\
  && \ \ \ -\frac{8-2x_t}{(1-y_t)(1-x_t)}\bigg]\; ,  \\
S_{HH}(y_t) &=& \bigg|\frac{Y_{33}^* Y_{31}}{V_{tb} V_{td}}\bigg|^2\,\frac{y_t\;x_t}4\left [ \frac{1+y_t}{(1-y_t)^2}+
                \frac{2y_t ln[y_t]}{(1-y_t)^3}\right ],
\end{eqnarray}
with
\begin{eqnarray}
\bigg|\frac{Y_{33}^* Y_{31}}{V_{tb} V_{td}}\bigg| & = & \bigg[ \bigg(  Y- \frac{f(y)}{\sqrt{2}} \chi_{33}^u\bigg) - \sqrt{\frac{m_c}{m_t}}
\bigg(\frac{V_{cb}}{V_{tb}} \bigg) \frac{f(Y)}{\sqrt{2}} \chi_{23}^u \bigg]^*
\nonumber \\ &\times &
\bigg[ \bigg(  Y- \frac{f(y)}{\sqrt{2}} \chi_{33}^u\bigg) - \sqrt{\frac{m_c}{m_t}}
\bigg(\frac{V_{cd}}{V_{td}} \bigg) \frac{f(Y)}{\sqrt{2}} \chi_{23}^u \bigg] \nonumber \\
& = & \bigg[ \lambda_{tt}^U- \sqrt{\frac{m_c}{m_t}}
\bigg(\frac{V_{cb}}{V_{tb}} \bigg) \frac{f(Y)}{\sqrt{2}} \chi_{23}^u \bigg]^* \bigg[ \lambda_{tt}^U- \sqrt{\frac{m_c}{m_t}}
\bigg(\frac{V_{cd}}{V_{td}} \bigg) \frac{f(Y)}{\sqrt{2}} \chi_{23}^u \bigg]
\end{eqnarray}
where we have used eq. (\ref{stal}),
in order to compare with  previous results in the literature, where the non-diagonal terms are not considered
\cite{Mahmoudi:2009zx,Xiao:2003ya}, so that one can recover those results when $\chi_{23}^u = 0$.
Finally, the function $L^{(i,H)}$ ($i=1,8$)
describes the charged Higgs contribution \cite{Urban:1997gw}
\begin{eqnarray}
L^{(i,H)}(x_t,y_t) &= &2 \bigg| \frac{Y_{33}^* Y_{31}}{V_{tb} V_{td}}\bigg|\,WH^{(i)}(x_t,y_t)\,+
2 \bigg|\frac{Y_{33}^* Y_{31}}{V_{tb} V_{td}}\bigg|\,
\Phi H^{(i)}(x_t,y_t) \nonumber \\
& + & \bigg|\frac{Y_{33}^* Y_{31}}{V_{tb} V_{td}}\bigg|^2\,HH^{(i)}(y_t)\;.
\end{eqnarray}
The explicit expressions of the rather complicated functions $WH^{(i)}(x_t,y_t)$,
$\Phi H^{(i)}(x_t,y_t)$ and $HH^{(i)}(y_t)$ can be found in Ref. \cite{Urban:1997gw}.

Following the analysis of Ref. \cite{Xiao:2003ya} and considering the  areas allowed by the measured $\Delta M_{B_d}$ value
within a $2 \sigma$ error,  when we have a light charged Higgs (80 GeV $\leq m_{H^\pm} \leq 200$ GeV) one can  extract the limit
\begin{eqnarray}
\bigg|\frac{Y_{33}^* Y_{31}}{V_{tb} V_{td}}\bigg|  \leq 0.25,
\end{eqnarray}
which is consistent with the bounds obtained from $B\to X_s \gamma$ in the previous subsection.
We present then Fig. \ref{fig:Bsg1}, which is the allowed region in the plane  $[\chi_{33}^u$, $\chi_{23}^u$] for the cases $Y<<1$ (left panel), $Y=1$
(center panel) and $Y=10$ (right panel). One can get the bound $-0.75\leq \chi_{23}^u \leq -0.15$ for $\chi_{33}^u=1$, and
$0.4\leq \chi_{23}^u \leq 0.9$ for $\chi_{33}^u=-1$, both when $Y<<1$. This result reduces a lot the intervals  of Tab. \ref{bounds-chi23} for the cases presented.

\subsection{$ Z \to b \bar{b} $ }
Other stringent bounds on $\vert \tilde{\chi}_{33}\vert$
come from  radiative corrections to the process $Z \to b \bar{b}$, especially to the $Z$ decay
fraction into $b\bar{b}$ ($R_b$).
Following the formulas presented in Refs. \cite{Degrassi:2010ne,Haber:1999zh,Deschamps:2009rh,Jung:2010ik}, $R_b$ is parametrized as:
\begin{eqnarray}
R_b = \frac{\Gamma (Z \to b \bar{b})}{\Gamma (Z \to {\rm hadrons})} = \bigg( 1+ \frac{K_b}{k_b} \bigg)^{-1}
\end{eqnarray}
where
\begin{eqnarray}
k_b & = & \bigg[ ( \bar{g}_b^L- \bar{g}_b^R)^2 + ( \bar{g}_b^L+ \bar{g}_b^R)^2 \bigg] \bigg( 1+ \frac{3 \alpha}{4 \pi} Q^2_q \bigg) \nonumber \\
K_b & = & C_b^{QCD} \sum_{q \neq b,t } k_b
\end{eqnarray}
with $ C_b^{QCD}  = 1.0086$, which is a factor  that includes QCD corrections, and
\begin{equation}
\bar{g}_b^{L,R} = \bar{g}_{Zb\bar{b}}^{L,R}+ \delta \bar{g}^{L,R},
\end{equation}
where $g_{Zb\bar{b}}^{L,R}$ are the tree level  couplings and $\delta g^{L,R}$ are the radiative corrections that include the contributions of
new physics.
In models with two doublets $\delta g^L$ and $\delta g^R$ have contributions
from loops involving all the Higgses ($H^\pm$, $H^0$, $h^0$ and $A^0$). Then, following the calculation of Ref. \cite{Degrassi:2010ne},  we obtain that the dominant contributions for $\delta g^{L,R}$
come from the charged Higgs boson and  are given by
\begin{eqnarray}
 \delta g^L &=& \frac{\sqrt{2} G_F M_w^2}{ 16 \pi^2}
\left[  \frac{ m_t}{M_w} \left( Y-\frac{f(Y)}{\sqrt{2}}\chi_{33}^u\right)\right]^2 \nonumber \\
&\times&\left\lbrace \frac{R}{R-1} -\frac{R logR}{(R-1)^2} + f_2 (R)\right\rbrace   \label{gl}  ,\\
 \delta g^R &=& - \frac{\sqrt{2} G_F M_w^2}{ 16 \pi^2}
\left[  \frac{ m_b}{M_w}\left(X -\frac{f(X)}{\sqrt{2}}\chi_{33}^d\right)\right]^2 \nonumber \\
&\times&\left\lbrace \frac{R}{R-1} -\frac{R logR}{(R-1)^2} + f_2 (R)\right\rbrace , \label{gr}
\end{eqnarray}
where $R = m_t^2/m_{H^+}^2$ and the function $f_2 (R)$ governing the NLO corrections is given in   Refs.
\cite{Degrassi:2010ne} . Again, when  $\chi_{33}^{u,d} =0$, one gets the case  for the 2HDM-II.
According to the measured value of
$R_b$ \cite{partdat},
 \begin{equation}
 R_b = 0.21629 \pm 0.00066
 \end{equation}
we can get the experimental constraints for  $\delta R_b = |0.00066|$. Then,  from  eqs. (\ref{gl})--(\ref{gr})
  we obtain bounds for
$\vert \tilde{\chi}_{33}^{f} \vert$  and $X$, $Y$. From Ref. \cite{Jung:2010ik} we can also use  the combined limit from leptonic $\tau$ decays  and the global fit to (semi-)leptonic decays, which is given by:
\beq
\frac{|Y_{33}Z_{33}|}{m_{H^\pm}^2} < 0.005, \quad  |X_{33}| < 50.
\eeq
However, this constraint is already contained in the $b \to s \gamma$ ones, which are more restrictive, so that this last result does not modify
those obtained in the previous section.
\subsection{$B_s \to \mu^+ \mu^-$}

\begin{figure}[t]
\begin{center}
\includegraphics[origin=c, angle=0, scale=0.3]{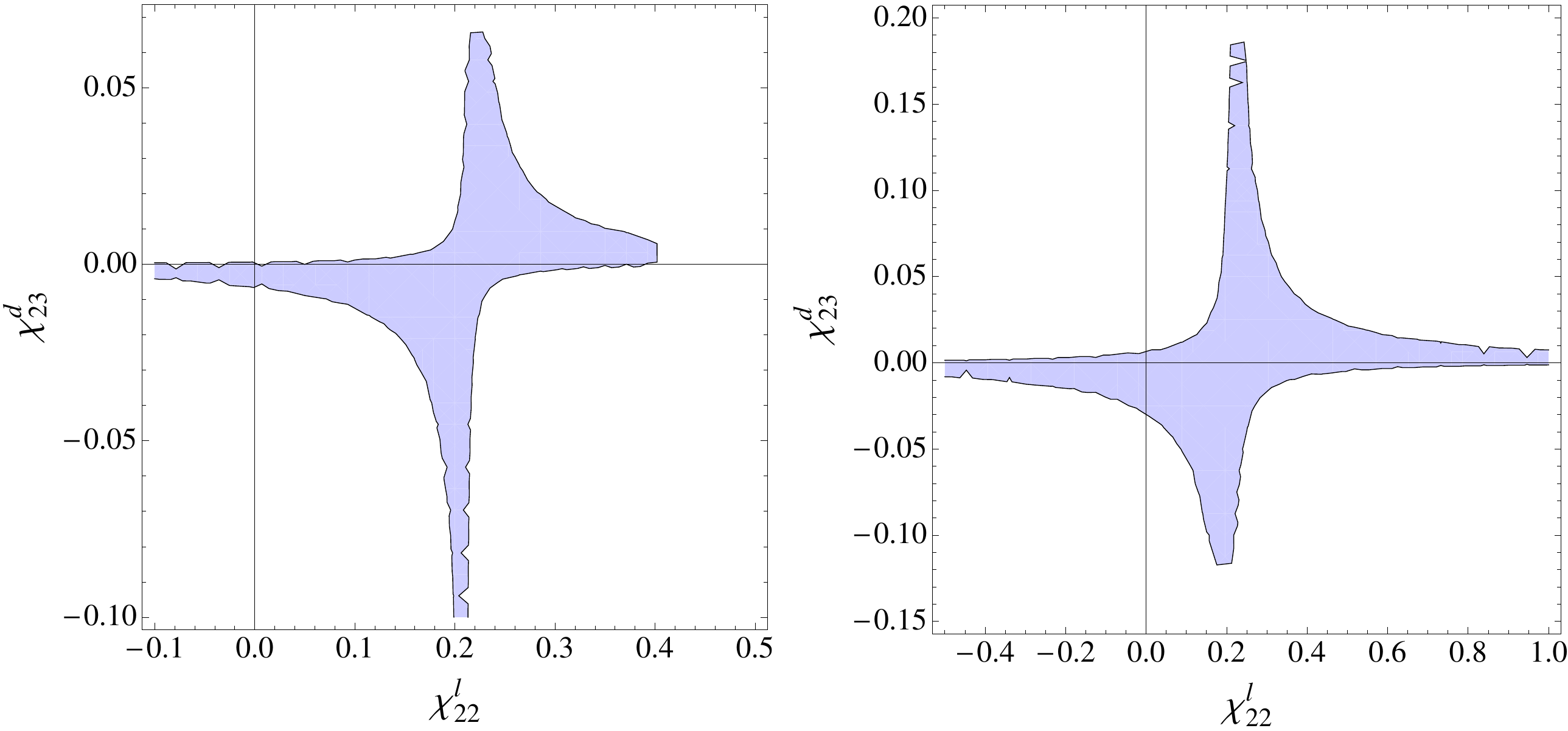}
\vspace*{-5mm}
\caption{The allowed region for $\chi_{22}^\ell$ vs. $\chi_{23}^d$ from $B_s \to \mu^+ \mu^-$, with $X= Z >> 1$, $m_{H^0} = 300$ GeV, 100 GeV $ \leq m_{H^\pm} \leq 350$ GeV, and for the following cases: $m_{A^0} = 100$ GeV (left) and $m_{A^0} = 300$ GeV (rigth).}
\label{fig:XZ20}
\end{center}
\end{figure}

\begin{figure}[t]
\begin{center}
\includegraphics[origin=c, angle=0, scale=0.3]{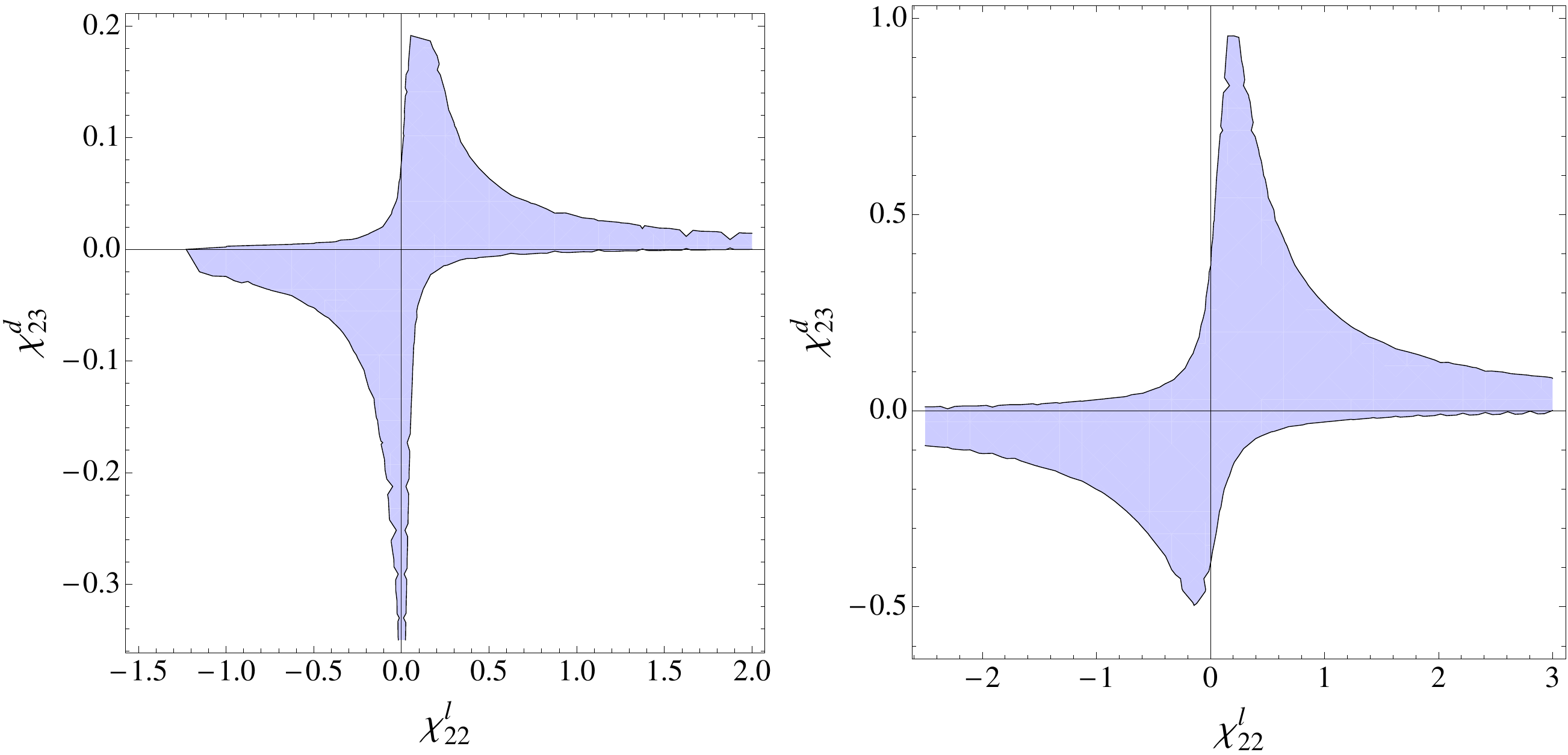}
\vspace*{-5mm}
\caption{The allowed region for $\chi_{22}^\ell$ vs. $\chi_{23}^d$ from $B_s \to \mu^+ \mu^-$, with $X >>Z $, $m_{H^0} = 300$ GeV, 100 GeV $ \leq m_{H^\pm} \leq 350$ GeV, and for the following cases: $m_{A^0} = 100$ GeV (left) and $m_{A^0} = 300$ GeV (rigth).}
\label{fig:XZm}
\end{center}
\end{figure}

A few months ago, the LHCb collaboration found the first evidence for
the $B_s \to \mu^+ \mu^-$ decay \cite{Aaij:2012nna}, with an experimental 
value for the BR given  by BR$(B_s \to \mu^+ \mu^-)= 3.2^{+1.5}_{-1.2} \times 10^{-9}$,
which imposes a lower bound on the parameters of our model. Recently,
the analysis of Higgs-mediated FCNCs in models with more than one Higgs 
doublet  has been performed \cite{Buras:2010mh} and it shows that the 
MFV case is more stable in suppressing FCNCs 
than the hypothesis of  NFC  when the quantum corrections are taken into 
account\footnote{A particular case of MVF in the 2HDM is the A2HDM \cite{Buras:2010mh}.}. 
In this work the scalar FCNC interactions  have been considered, as it happens in our model. 
On the other hand, in \cite{Huang:2000sm,Logan:2000iv}, the contribution for the 
2HDM-II is presented in the regime of large $\tan \beta $, which should be considered in our 
work in order to get these results when the $\chi_{ij}$ parameters are absent. As was presented in a similar case
 for another version of the 2HDM-III, where  both contributions, at tree and at one-loop level,   
were taken into account \cite{Crivellin:2013wna}\footnote{The version 2HDM-III presented in 
\cite{Crivellin:2013wna}  is the so-called 2HDM-II-like of our 2HDM-III, where the parameter 
$\epsilon_{ij}^f$ introduced in that reference is related to our parameter $\chi_{ij}^f$ in 
the following way:  $\epsilon_{ij} = \frac{\sqrt{m_i m_j} }{v}\chi_{ij}$. }. Then, following 
the calculation of the BR for the process $B_s \to \mu^+ \mu^-$  given in 
\cite{Buras:2010mh,Altmannshofer:2009ne,Dedes:2002er}, we have
\begin{eqnarray}
{\rm BR}(B_s \to \mu^+ \mu^-) = {\rm BR}(B_s \to \mu^+ \mu^-)_{\rm SM} \bigg( |1 + R_p|^2 + | R_s |^2  \bigg),
\end{eqnarray}
where
$R_s$  and $R_p$ contain, in the regime $X >>1$,  the corrections at one-loop level from 
charged Higgs bosons  and the contribution at tree level of the neutral Higgs bosons which come 
from the couplings $s \bar{b} \phi$ as well as $b \bar{s} \phi$ ($\phi= H$, $A$), as 
given in (\ref{coups1}), where the off-diagonal terms $\chi_{23}^d$  
contribute to $B_{s} \to \mu^+ \mu^-$\footnote{For the case $X>> 1$,  the contribution of the 
light neutral Higgs boson is neglected  because $h_{ij} << H_{ij} \sim A_{ij} \propto X$, as
justified by  taking the  values of Tab. \ref{couplings} for some specials cases and
$\alpha= \beta - \pi/2$.}. Therefore, $R_s$  and $R_p$ are
\begin{eqnarray}
R_{s}^{\phi sb} &=& \frac{4 \pi^2 }{ Y_0(x_t) g^2}  \bigg( \frac{1}{1 + m_s/m_b}\bigg)  \bigg(\frac{m_s}{m_b}\bigg) \frac{M_{B_s}^2}{V_{tb} V_{ts}^* }  \bigg[  \frac{H_{23}^d H_{22}^\ell}{m_{H^0}^2} \bigg], \label{rs1} \\
R_{p}^{\phi sb} &=& \frac{4 \pi^2 }{ Y_0(x_t) g^2}  \bigg( \frac{1}{1 + m_s/m_b}\bigg) \bigg(\frac{m_s}{m_b} \bigg)  \frac{M_{B_s}^2}{V_{tb} V_{ts}^* }   \bigg[ \ \frac{A_{23}^d A_{22}^\ell}{m_{A^0}^2}   \bigg], \label{rp1} \\
R_{s,p}^{loop} &=& \frac{X^2 }{ 8 Y_0(x_t) }  \bigg( \frac{1}{1 + m_s/m_b}\bigg)  \frac{M_{B_s}^2}{M_W^2 }   \bigg[  \frac{Log(r)}{r-1}   \bigg], \label{rsp2}
\end{eqnarray}
with $\phi_{2,3}^d$ and $\phi_{22}^\ell$ ($\phi = H$, $A$) as given in (\ref{hHA}), $r= m_{H^\pm}^2/ mt^2$, where $m_{b}$, $m_{s}$ and $m_t$ must be evaluated at their matching scales, and
\begin{eqnarray}
{\rm BR}(B_s \to \mu^+ \mu^-)_{SM} = \frac{G_F^2 \tau_{B_s}  }{ \pi }  \bigg( \frac{g^2}{16 \pi^2} \bigg)^2 M_{B_s} F_{B_s} ^2 m_{\mu}^2
\vert V_{tb} V_{ts}^{*} \vert^2  \sqrt{1 - \frac{4 m_{\mu}^2}{M_{B_s}^2}} Y_0(x_t)^2,
\end{eqnarray}
here, $Y_0(x_t)$ is the loop function given in \cite{Altmannshofer:2009ne}, $M_{B_s}$ and $\tau_{B_s}$ are the  mass and  lifetime of the $B_s$ meson, respectively, and $F_{B_s}$ = $242.0 (9.5)\, MeV$ is the $B_s$ decay constant\cite{Bazavov:2011aa}.
Otherwise, from (\ref{hHA}) and considering some of the cases given in Tab. \ref{couplings}, we can constrain
the non-diagonal terms  of the Yukawa texture $\chi_{23}^d$ and the diagonal terms  $\chi_{22}^\ell$ as well, for two cases: $X>> Z$ and $X,Z >>1$.  We present then Fig. \ref{fig:XZ20} the case $X,Z >>1$, taking $m_{H^0} = 300$ GeV, 100 GeV $ \leq m_{H^\pm} \leq 350$ GeV and for the following cases: $m_{A^0} = 100$ GeV (left panel) and $m_{A^0} = 300$ GeV (right panel). One can see that the process $B_s \to \mu^+ \mu^-$ imposes constraints  onto the parameter  $\chi^\ell_{22} $ and $\chi_{23}^d$ and that we obtain
the following bounds: $-0.1\leq \chi^\ell_{22} \leq 0.4 $
( $-0.4\leq \chi^\ell_{22} \leq 1 $) and $-0.1\leq \chi^d_{23} \leq 0.05 $
( $-0.4\leq \chi^d_{23} \leq 0.15 $) for $m_{A^0} = 100$ GeV ($m_{A^0} = 300$ GeV).  In Fig. \ref{fig:XZm}  we show  the allowed
region in the plane $[ \chi_{22}^\ell, \, \chi_{23}^d ]$ for  $X >>Z$, and  $m_{A^0} = 100$ GeV  (left panel) and $m_{A^0} = 300$ GeV
(right panel). One can get the  bounds $-1.5\leq \chi^\ell_{22} \leq 2 $
( $-3\leq \chi^\ell_{22} \leq 3 $) and $-0.4\leq \chi^d_{23} \leq 0.2 $
( $-0.5\leq \chi^d_{23} \leq 0.9 $) for $m_{A^0} = 100$ GeV ($m_{A^0} = 300$ GeV). One can see in Figs. 
\ref{fig:XZ20}--\ref{fig:XZm}
that, when $\chi^\ell_{22}$ is close to 1, we have that $\chi_{23}^d\leq 10^{-2}$. However, when we have
for such a parameter that $0.2 \leq \chi^\ell_{22} \leq 0.4$,
the off-diagonal term in the  Yukawa texture, $\chi_{23}^d$, can take values in the the interval $[-0.1,0.15]$ for $Z,X>>1$ or
$[-0.5, 1]$ for $X>>Z$.
These results constrain further the parameters
$\chi_{ij}^f$ and we present in Tab. \ref{comb} the combination of these results with the others of the previous sections.
We can then conclude from our results that the diagonal term $\chi^\ell_{22} $ and the off-diagonal term $\chi_{23}^d$ are very sensitive to the process $B_s \to \mu^+ \mu^-$  and we observe that  $\chi_{23}^u$ is also quite sensitive to the process $B \to X_s \gamma$. Besides, the possibility of light
charged Higgs bosons is still consistent with the experimental results for the process $B_s \to \mu^+ \mu^-$ in our version of the 2HDM-III.

\begin{table}[tbp]
\centering
\begin{tabular}{|c||c|c|c|c|}
\hline
{\small 2HDM-III's} & {\small $\chi_{23}^d $ } &   {\small $\chi_{23}^u$ } & {\small  $\chi_{22}^\ell$ }&  {\small $(X, Y, Z) $ }\\ \hline
{\small 2HDM-I-like} & {\small (0,0.2) } & {\small  (0.24,0.26)  } &  {\small (-0.4,1) } & {\small  (20, 20, 20) } \\   \hline
{\small 2HDM-II-like } &  {\small  (0,0.2)   } &   {\small  (-0.53,-0.49) }&  {\small (-0.4,1)   } & {\small   (20, 0.1, 20) } \\  \hline
{\small 2HDM-X-like } & {\small  (-7,2) } &  {\small  (-2.2,0.5) } &  {\small (-1.5,1.5) } &   {\small (0.1, 0 .1, 50)} \\ \hline
{\small 2HDM-Y-like } &  {\small (-0.2,0.6)  or  } & {\small (-0.53,-0.49) } &  {\small  (-1,1.5) } &  {\small (50, 0.5, 0.5) }\\  \hline
\end{tabular}
\caption{ Constraints from   $B \to D \tau \nu$, $D_s \to \tau \nu, \mu \nu$, $B \to \tau \nu$ and $B_s \to \mu^+ \mu^-$
decays.
We show the allowed intervals  for $\chi_{23}^{u,d}$ and $\chi_{22}^\ell$ constrained by each low energy process, according to  the
different scenarios presented in Tab. \ref{couplings} as well as the combination of constraints.
We assume $0.1\leq  \chi_{33}^{l} \leq 1.5  $ as well as $  \chi_{22}^d =   \chi_{22}^u =1$. Taking  80 GeV $\leq m_{H^\pm}\leq 350$ GeV and specific values for the $X,Y$ and $Z$ parameters given in Tab, \ref{couplings}. }
\label{comb}
\end{table}

\section{Light charged Higgs boson phenomenology}

Now we discuss some simple yet interesting phenomenology emerging in the 2HDM-III after all aforementioned
constraints are taken into account, in particular the decays of a light charged Higgs boson
(i.e., with mass below the top quark one). The expressions for the charged Higgs boson partial decay width $H^+ \to
u_i \bar{d}_j$ to massive quark-antiquark pairs are of the form
\begin{eqnarray}
\Gamma (H^+ \to u_i \bar{d}_j) &=& \frac{3 g^2}{32 \pi M_W^2 m_{H^+}^3 } \lambda^{1/2}(m_{H^+}^2, m_{u_i}^2, m_{d_j}^2) \nonumber\\
& & \times \, \bigg( \frac{1}{2}\bigg[ m_{H^+}^2-m_{u_i}^2-m_{d_j}^2\bigg] (S_{ij}^2+P_{ij}^2)- m_{u_i} m_{d_j}
(S_{ij}^2- P_{ij}^2 ) \bigg),
\label{decay-1}
\end{eqnarray}
where $\lambda$ is the usual kinematic factor $\lambda(a,b,c)=
(a-b-c)^2-4bc$. When we replace $\tilde{\chi}_{ud} \to 0$, the
formulas of the decay widths become those of the 2HDM with NFC: see, e.g.,
Refs.~\cite{kanehunt,Grossman:1994jb}. For  a scenario with light charged Higgs bosons,
for which we can then neglect decays into heavy fermions, the expression for the partial widths of the hadronic decay modes
of $H^\pm$'s  is reduced to:
\begin{equation}
\Gamma(H^\pm\to u_i\bar d_j)=\frac{3G_F m_{H^\pm}(m_{d_j}^2|X_{ij}|^2+m_{u_i}^2|Y_{ij}|^2)}{4\pi\sqrt 2}.
\label{width_ud}
\end{equation}
For the case of leptons, one has instead
\begin{equation}
\Gamma(H^\pm\to {l}^\pm_j\nu_i)=\frac{G_F m_{H^\pm} m^2_{{l}_j} |Z_{ij} |^2}{4\pi\sqrt 2}.
\label{width_tau}
\end{equation}
Again, when the parameters $\chi_{ij}^f$ are absent, we recover the results of the MHDM/A2HDM.
In $\Gamma(H^\pm\to u_i\bar d_j)$ the running quark masses should be evaluated at the scale of $m_{H^\pm}$ and
there are QCD vertex corrections which multiply the above partial widths by
$(1+17\alpha_s/(3\pi))$. In the 2HDM the parameter $\tan\beta$ determines the magnitude of the
partial widths. The BRs are well known and for the case of interest, $m_{H^\pm}< m_t$,
one finds that the dominant decay channel
is either $H^\pm \to cs$ or  $H^\pm \to \tau\nu$, depending on the value of $\tan\beta$.
In the 2HDM-I the BRs are independent of $\tan\beta$ instead and the BR($H^\pm \to \tau\nu$)
is about twice the BR($H^\pm \to cs$).

The magnitude of BR($H^\pm \to cb$) is always less than a few
percent in three (Type I, II and lepton-specific) of the four versions of the 2HDM with NFC, since the
decay rate is suppressed by the small CKM matrix element $V_{cb}\, (\ll V_{cs})$.
In contrast, a sizeable BR($H^\pm \to cb$) can be obtained in
the so called `flipped' 2HDM \cite{Barger:1989fj} for $\tan\beta > 3$. This possibility was not stated explicitly in \cite{Barger:1989fj} though, where
the flipped 2HDM was introduced. The first explicit mention of a large BR($H^\pm \to cb$) seems to have been in \cite{Grossman:1994jb}
and a quantitative study followed soon afterwards in \cite{Akeroyd:1994ga}.
As discussed in \cite{Akeroyd:2012yg} though, the condition $m_{H^\pm}< m_t$ in the flipped 2HDM
would require additional new physics in order to avoid the constraints on $m_{H^\pm}$
from $b\to s \gamma$, while this is not the case in the MHDM/A2HDM. Otherwise, according to the most restrictive flavor constraints obtained in section II, the interaction of charged Higgs bosons with the fermions in the 2HDM-III with a four-zero texture in the  Yukawa matrices  can be written in the same way as in the  MHDM/A2HDM plus an small deviation, namely:
\beq
g_{H^\pm u_i d_j}^{\rm 2HDM-III} =  g_{H^\pm u_i d_j}^{\rm MHDM/A2HDM} + \Delta g.
\eeq
Therefore the phenomenology of $H^\pm$ states of the 2HDM-III is very close to than
in the MHDM/A2HDM, although not necessarily the same. We will show in particular peculiar effects induced by the off-diagonal terms of the Yukawa texture studied here in processes considered recently in  \cite{Akeroyd:2012yg}.
\subsection{The dominance of the BR$(H^\pm \to cb)$}
A distinctive signal of a $H^\pm$ state from the 2HDM-III for $m_{H^\pm}< m_t-m_b$ would be a
sizable BR for $H^\pm \to cb$.
For $m_{H^\pm} < m_t-m_b$, the scenario of $|X|>> |Y|,|Z|$ in a 2HDM-III
gives rise to a ``leptophobic'' $H^\pm$ with
BR$(H^\pm\to cs)$+BR$(H^\pm\to cb) \sim 100\%$, as the BR($H^\pm \to \tau\nu$)
is negligible ($<<1\%$). The other decays of $H^\pm$ to quarks are subdominant,
with the BR($H^\pm\to us)\sim 1\%$ always and the BR($H^\pm\to t^*b$) becoming
sizeable only for $m_{H^\pm}\sim m_t$, as can be seen in the numerical analysis in
\cite{Logan:2010ag} (in the flipped 2HDM). Incidentally, note that the
case of $|X|>> |Y|,|Z|$ is obtained in the flipped 2HDM
for $\tan\beta > 3$, because $|X|= \tan\beta=1/|Y|=1/|Z|$ in this scenario.

Conversely, one can see that the configuration $Y >>$ $X$,$ Z$ (this imply that $Y_{ij} >>$ $X_{ij}$,$ Z_{ij}$, see eqs. (\ref{Xij})--(\ref{Zij}))
 is very interesting, because the decay $H^+ \to c \bar{b}$ is now dominant.
In order to show this situation, we calculate the dominant terms $m_c Y_{23}$, $m _c Y_{22}$ of the width $\Gamma(H^+ \to c \bar{b}, c\bar{s})$, respectively, which are given by:
\begin{eqnarray}
m_c Y_{cb} & = & m_c Y_{23}  = V_{cb}m_c \bigg(Y - \frac{f(Y)}{\sqrt{2}} \chi_{22}^u \bigg)- V_{tb} \frac{f(Y)}{\sqrt{2}} \sqrt{m_t m_c}
\chi_{23}^u  \nonumber  \\
&=&V_{cb}m_c \lambda_{22}^u+ V_{tb}  \sqrt{m_t m_c} \lambda_{23}^u, \\
m_c Y_{cs} & = & m_c Y_{22} = V_{cs}m_c \bigg(Y - \frac{f(Y)}{\sqrt{2}} \chi_{22}^u \bigg)- V_{ts} \frac{f(Y)}{\sqrt{2}} \sqrt{m_t m_c}
\chi_{23}^u \nonumber  \\
&=&V_{cs}m_c \lambda_{22}^u+  V_{ts}  \sqrt{m_t m_c} \lambda_{23}^u.
\end{eqnarray}
As $Y$ is large and $f(Y) = \sqrt{1+Y^2} \sim Y $, then the term $\bigg( Y - \frac{f(Y)}{\sqrt{2}} \chi_{22}^u \bigg)$ could be absent or small, when $\chi_{ij} = O(1)$. Besides,  the last term is very large because $ \propto\sqrt{m_t m_c}$, given that $m_t=173$ GeV, so that in the
end this is the dominant contribution. Therefore,
we can compute the ratio of two dominant decays, namely,  BR$(H^\pm\to cb)$ and BR$(H^\pm\to cs)$, which is given as follows:
\begin{equation}
R_{sb}=\frac{{\rm BR}(H^\pm\to cb)}{{\rm BR}(H^\pm\to cs)}\sim \frac{|V_{tb}|^2 }{|V_{ts}|^2 }.
\label{Rsb-1}
\end{equation}
In this case  BR$(H^\pm\to cb) \sim 100\%$ (for $\mh  < m_t-m_b$, of course) so that to verify
this prediction would really be the hallmark signal  of the 2HDM-III. Therefore, we can see
 that the non-diagonal term $\chi_{23}^u$ (or $\lambda_{23}^u$)  cannot be omitted and
this is an important result signalling new physics even beyond the standards 2HDMs. In a similar spirit, one can study other interesting channels, both in  decay and production and both at tree and one-loop
level \cite{DiazCruz:2004pj, DiazCruz:2009ek, BarradasGuevara:2010xs,GomezBock:2005hc, HernandezSanchez:2011fq}.

Another case is when $X >>$ $Y$,$ Z$, here, we get that the dominants terms are
$\propto m_b X_{23}$, $m _s X_{22}$, as
\begin{eqnarray}
m_b X_{cb} & = & m_b X_{23}  = V_{cb}m_b \bigg(X - \frac{f(X)}{\sqrt{2}} \chi_{33}^d \bigg)- V_{cs} \frac{f(X)}{\sqrt{2}} \sqrt{m_b m_s}
\chi_{23}^d \nonumber  \\
&=&V_{cb}m_b \lambda_{33}^d+ V_{cs}  \sqrt{m_s m_b} \lambda_{23}^d, \\
m_s X_{cs} & = & m_s X_{22} = V_{cs}m_s \bigg(X - \frac{f(X)}{\sqrt{2}} \chi_{22}^d \bigg)- V_{ts} \frac{f(X)}{\sqrt{2}} \sqrt{m_b m_s}
\chi_{23}^d \nonumber  \\
&=&V_{cs}m_s \lambda_{22}^d+  V_{cb}  \sqrt{m_s m_b} \lambda_{32}^d.
\end{eqnarray}
 In this case there are two possibilities. If $\chi= O(1)$ and positive then $\bigg(X - \frac{f(X)}{\sqrt{2}} \chi_{33}^d \bigg)$ is small and
 \beq
  R_{sb}\sim \frac{|V_{cs}|^2 }{|V_{cb}|^2 }.
  \label{Rsb-2}
  \eeq
  Here, the BR$(H^\pm \to cb)$ becomes large, again, this case too could be another exotic scenario of the 2HDM-III.
 The other possibility is when  $\chi= O(1)$ and negative, then
  \beq
  R_{sb}\sim \frac{m^2_b |V_{cb}|^2 }{m^2_s |V_{cs}|^2 },
  \label{Rsb-3}
  \eeq
  which is very similar to the cases studied recently in \cite{Akeroyd:2012yg}. In summary, one can see two possibilities to study the
  BR$(H^\pm \to cb)$: firstly, the scenarios given in eqs. (\ref{Rsb-1}) and (\ref{Rsb-2}),
which are peculiar to the 2HDM-III; secondly, the scenario very close to the MHDM/A2HDM expressed in eq. (\ref{Rsb-3}).

The CKM matrix elements are well measured, with $V_{cb}\sim 0.04$ (a direct measurement), $V_{cs}\sim 0.97$
(from the assumption that the CKM matrix is unitary),  $V_{ts}\sim 0.04$ and $V_{tb}\sim 0.999$ (from direct determination without assuming unitarity, as is possible from the single
top-quark-production cross section) \cite{partdat}. With this information, it is enough to determine in the cases represented by eqs.
 (\ref{Rsb-1}) and (\ref{Rsb-2}) the dominance of the channel
$H^\pm \to cb$ for light charged Higgs bosons. Otherwise, the situation given in eq. (\ref{Rsb-3}) is more delicate since the running quark masses $m_s$ and $m_b$ should be evaluated at the scale
$Q=m_{H^\pm}$ and this constitutes the main uncertainty in the ratio $R_{bs}$.
There is relatively little uncertainty for $m_b$,
with $m_b \,(Q=100 \,{\rm GeV})\sim 3 $ GeV. However,
there is more uncertainty in the value of $m_s$, although in recent years there has been much progress
in lattice calculations of $m_s$ and an average of six distinct unquenched calculations gives
$m_s=93.4\pm 1.1$ MeV \cite{Laiho:2009eu} at the scale of $Q=2$ GeV. A more conservative average of theses calculations,
$m_s = 94 \pm 3$ MeV,  is given in \cite{Colangelo:2010et}. In \cite{partdat} the
currently preferred range at $Q=2$ GeV is given as $80 \,{\rm MeV} < m_s < \, 130 $ MeV.
Using $m_s=93$ MeV at the scale $Q=2$ GeV (i.e. roughly the central value of the lattice average \cite{Laiho:2009eu,Colangelo:2010et})
one obtains $m_s \,(Q=100 \,{\rm GeV})\sim 55 $ MeV. Taking $m_s=80$ MeV and $m_s=130$ MeV
at $Q=2$ GeV one instead obtains  $m_s\sim 48$ MeV and $m_s\sim 78$ MeV, respectively, at $Q=100$ GeV.
    Smaller values of
$m_s$ will give a larger  BR$(H^\pm\to cb)$, as can be seen in the situation given in  eq.~(\ref{Rsb-3}).
Note that the value $m_s=55$ MeV is significantly smaller than the typical values $m_s\sim 150\to 200$ MeV which were often
used in Higgs phenomenology in the past two decades. In essence, we
 emphasize that the scenario  $|X|>> |Y|,|Z|$ with $m_{H^\pm}< m_t-m_b$ has a unique feature in that  the magnitude of $m_s$ is crucial for determining the relative probability of the two dominant decay channels of the $H^\pm$ state. Further, this is not the case for most other non-minimal Higgs sectors with $H^\pm$ states that are commonly studied in the literature.

In \cite{Akeroyd:1994ga} the magnitude of BR$(H^\pm\to cb)$ in the MHDM was studied in the plane of
$|X|$ and $|Y|$, for $|Z|=0$ and $0.5$, taking $m_s=0.18$ GeV and $m_b=5$ GeV.
With these quark masses the maximum value for the above ratio is $R_{bs}=1.23$, which corresponds to BR$(H^\pm\to cb)\sim 55\%$.
However, the values of $m_s=0.18$ GeV and $m_b=5$ GeV  are no longer realistic
(as it was subsequently noted in \cite{Akeroyd:1998dt}) and
two recent papers \cite{Logan:2010ag,Aoki:2009ha} have
in fact updated the magnitude of $R_{bs}$ (in the flipped 2HDM)
using realistic running quark masses at the scale $Q=m_{H^\pm}$.
In \cite{Logan:2010ag}, it appears that $m_s=0.080$ GeV at the scale $Q=m_{H^\pm}$
was used, which gives BR$(H^\pm\to cb)\sim 70\%$, in agreement with the results presented in  \cite{Akeroyd:2012yg}.
In \cite{Aoki:2009ha}, $m_s=0.077$ GeV at the scale $Q=m_{H^\pm}$
was used, with a maximum value for BR$(H^\pm\to cb)$ of $\sim 70\%$.
We note that none of these papers used the precise average $m_s=93.4\pm 1.1$ MeV \cite{Laiho:2009eu}
of the lattice calculations, which gives $m_s\sim 55$ MeV at the scale of
 $Q=m_{H^\pm}$. This smaller value of $m_s$ leads to a maximum value of
BR$(H^\pm\to cb)$ which is larger than
that given in \cite{Akeroyd:1994ga,Aoki:2009ha,Logan:2010ag}, as discussed below.

\begin{figure}[t]
\begin{center}
\includegraphics[origin=c, angle=0, scale=0.5]{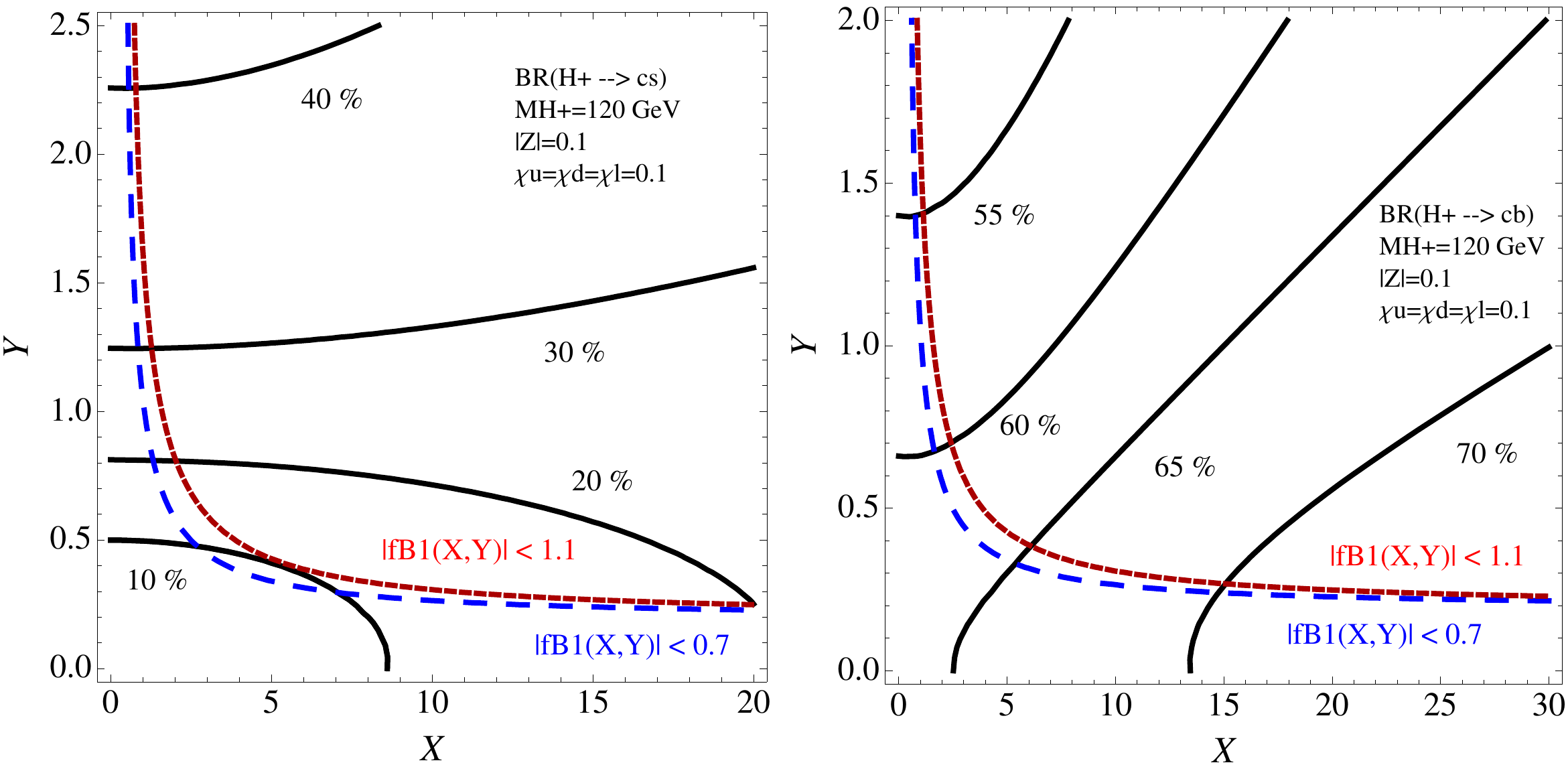}
\vspace*{-5mm}
\caption{ For the exotic scenario of eq. (\ref{Rsb-2}) we show  the contours of the BR$(H^\pm \to cs)$ (left) and  BR$(H^\pm \to cb)$ (right) in the plane $[X,Y]$ with $Z=$ 0.1,  $m_{A^0}=80$ GeV, $ m_{H^\pm} =120$ GeV and $\chi^f_{ij} = 0.1$.
The constraint $b\to s \gamma$
is shown as $|fB1(X,Y)| < 1.1 $ for $ Re[fB1(X,Y)]<0$, where $fB1 (X,Y)= \frac{X_{33} Y_{32}^*}{V_{tb} V_{ts}}$ is given in eq.
(\ref{eq:xy-m30}).  For $|fB1(X,Y)| < 0.7 $, it is when $ Re[fB1(X,Y)]<0$. We take $m_s( Q= \mh) = 0.055$ GeV. }
\label{fig:brcb}
\end{center}
\end{figure}

\begin{figure}[t]
\begin{center}
\includegraphics[origin=c, angle=0, scale=0.5]{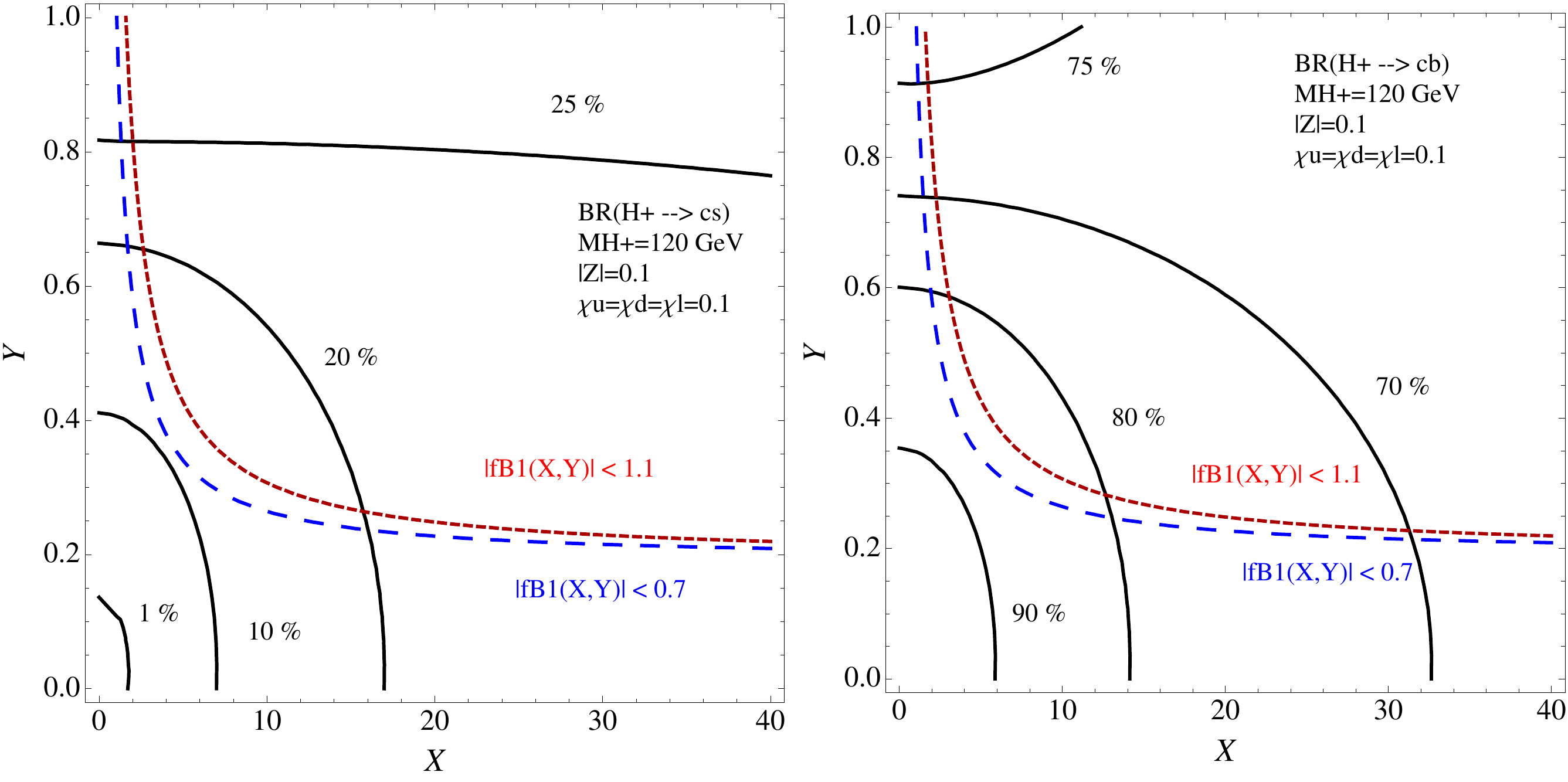}
\vspace*{-5mm}
\caption{ The same as Fig. \ref{fig:brcb} but for $m_{A^0}= 100 $ GeV.}
\label{fig:brcb2}
\end{center}
\end{figure}

\begin{figure}[t]
\begin{center}
\includegraphics[origin=c, angle=0, scale=0.6]{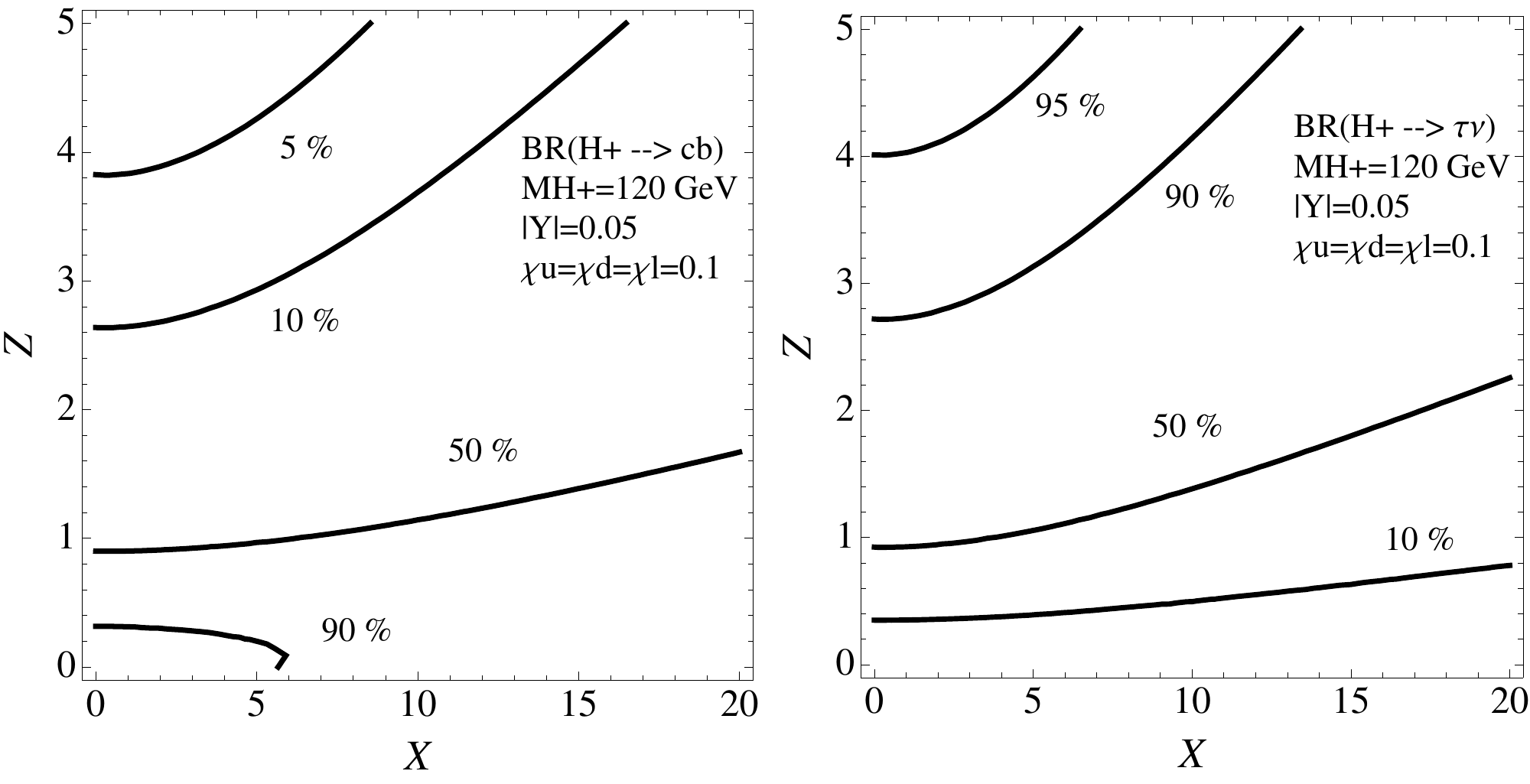}
\vspace*{-5mm}
\caption{ For the exotic scenario of the eq. (\ref{Rsb-1}), we show  the contours of
the BR$(H^\pm \to cb)$ (left) and
BR$(H^\pm \to \tau \nu)$ (right) in the plane $[X,Z]$ with $Z=$ 0.05, $Z>>X$,  $ m_{H^\pm} =120$ GeV and $\chi^f_{ij} = 0.1$.
 We take $m_s( Q= \mh) = 0.055$ GeV. }
\label{fig:brcb_xz}
\end{center}
\end{figure}

We now study the magnitude of $H^\pm \to cb$
as a function of the couplings $X,Y,Z$.
In Fig.~\ref{fig:brcb} we show the numerical study of
BR$(H^\pm\to cb)$ in the plane $[X,Y]$ in the 2HDM-III with $|Z|=0.1$, using $\chi_{ij}^f=0.1$, $m_s=0.055$ GeV and $m_b=2.95$ GeV
at the scale of $Q=m_{H^\pm}=120$ GeV.
With these values for the quark masses the maximum value is BR$(H^\pm\to cb)\sim 71\%  \, (91\%)$ considering $m_{A^0}= 80$ GeV
($m_{A^0}= 100$ GeV),
i.e., a significantly smaller (larger) value than BR$(H^\pm\to cb)\sim 81\%$ \cite{Akeroyd:2012yg}, and both larger than
BR$(H^\pm\to cb)\sim 55\%$  in \cite{Akeroyd:1994ga}.
In Figs.~\ref{fig:brcb} and \ref{fig:brcb2} we also display the bound from $b\to s \gamma$ (for $m_{H^\pm}=100$ GeV),
which is $\bigg|\frac{X_{33} Y_{32}^*}{V_{tb} V_{ts}}\bigg| < 1.1$ for $\frac{X_{33} Y_{32}^*}{V_{tb} V_{tsp}}$ being real and negative plus
$\bigg|\frac{X_{33} Y_{32}^*}{V_{tb} V_{ts}}\bigg| < 0.7$ for $\frac{X_{33} Y_{32}^*}{V_{tb} V_{ts}}$ being real and positive.
The parameter space for
BR$(H^\pm\to cb)> 70\% \, (90\%)$ roughly corresponds to $|X| > 10$  ($|X| > 5$) and $|Y| < 0.35$ for
$\bigg|\frac{X_{33} Y_{32}^*}{V_{tb} V_{ts}}\bigg|  < 0.7$.
In  Fig.~\ref{fig:brcb} and \ref{fig:brcb2}
we also show BR$(H^\pm\to cs)$. The latter could be 40\% (25\%), however, it cannot be maximized for
$|Y|>> |X|,|Z|$, because eq. (\ref{Rsb-1}) does not allow it, whereas  BR$(H^\pm\to \tau\nu)$ is maximized for  $|Z|>> |X|,|Y|$.
In Fig.~\ref{fig:brcb_xz} we show contours of BR$(H^\pm\to cb)$ and BR$(H^\pm\to \tau\nu)$
in the plane $[X,Z]$ for $m_{H^\pm}=120$ GeV, $m_{A^0}=100$ GeV
and $|Y|=0.05$. For this value of $|Y|$ the constraint from $b\to s\gamma$ is always satisfied
for the displayed range, $|X|<20$. One can see that the largest values of BR$(H^\pm\to cb)$
arise for $|Z|< 2$.

Prospects for $t\to H^\pm b$ with $H^\pm \to cb$ at the LHC have been
reviewed lately in Ref. \cite{Akeroyd:2012yg}
for the case of the MHDM/A2HDM, from which many results can however readily be adapted to our current studies.
Things go as follow.
The case of $m_{H^\pm} < m_t - m_b$ with a large BR($H^\pm \to cs$) can be tested in the decays of the
top quark via $t\to H^\pm b$ and was studied first in \cite{Barger:1989fj,Barnett:1992ut}.
Innumerable studies in this direction followed suit, far too many in fact for being listed here.
Also, we are concerned here primarily with $H^\pm \to cb$ decays.
The first discussion of $t\to H^\pm b$ followed by
$H^\pm \to cb$ was  given in \cite{Akeroyd:1995cf}. Recently,  $t\to H^\pm b$
with $H^\pm \to cb$ has been
studied in the context of the flipped 2HDM \cite{Logan:2010ag} and  the 2HDM without NFC
\cite{DiazCruz:2009ek} (other than in \cite{Akeroyd:2012yg}).

Light charged Higgs bosons ($H^\pm$) are being searched for (or simulated) in the decays of top quarks
($t\to H^\pm b$) at the Tevatron and at the LHC in a variety of modes \cite{:2009zh, Aaltonen:2009ke,Searches:2001ac,Ferrari:2010zz,ATLAS:search}. In particular,
separate searches are being
carried out for the decay channels $H^\pm \to cs$ and
$H^\pm \to \tau\nu$, with comparable sensitivity to the mass and fermionic couplings of $H^\pm$.
The searches for $H^\pm \to cs$ in \cite{Aaltonen:2009ke} and
\cite{ATLAS:search} look for a peak at $m_{H^\pm}$ in the dijet invariant mass distribution,
with the assumption that neither of the quarks is a $b$-quark.

In the MHDM/A2HDM realisations the
BR($H^\pm \to cb$) can be as large as $80\%$ for an $H^\pm$
that is light enough to be
generated in the $t\to H^\pm b$ decay. This should be contrasted to the case of the $H^\pm$
state belonging to more standard 2HDMs for which a large
BR($H^\pm \to cb$) is certainly possible
but one expects $m_{H^\pm} > m_t$ in order to comply with
the measured value of $b\to s\gamma$, which is not the case for the 2HDM-III. The latter scenario is therefore
on the same footing as the MHDM/A2HDM cases studied in  \cite{Akeroyd:2012yg}. Herein, in short, it was  suggested that a dedicated search for $t\to H^\pm b$ and $H^\pm \to cb$ would
probe values of the fermionic couplings of $H^\pm$ which are currently not excluded if one
required a $b$-tag in one of the jets originating from $H^\pm$, thus affording in turn
sensitivity to smaller values of  BR($t\to H^\pm b$) than those obtained to date (which use un-flavored jet samples).
Therefore, a dedicated search for $t\to H^\pm b$ and $H^\pm \to cb$ at the Tevatron and LHC would be
a well-motivated and at the same time simple extension of ongoing searches for $t\to H^\pm b$ with decay
$H^\pm \to cs$.

\subsection{The decay $H^\pm \to A W^*$ for $m_{A^0}< m_{H^\pm}$}
The above discussion has assumed that $H^\pm$ cannot decay into other (pseudo)scalars.
We now briefly discuss the impact of the decay channel $H^\pm\to A^0W^*$, which has been studied in
the 2HDM-II in \cite{Moretti:1994ds} and in other 2HDMs with small $|X|,|Y|$ and $|Z|$
in \cite{Akeroyd:1998dt}, in the light of direct searches at LEP (assuming $A^0\to b\overline b$) performed in \cite{Abdallah:2003wd}.
In a general non-SUSY 2HDM the masses of the scalars can be taken as free parameters. This is in contrast
to the MSSM in which one expects $m_{H^\pm}\sim m_{A^0}$ in most of the parameter space.
The scenarios of $m_{A^0}< m_{H^\pm}$ and $m_{A^0}> m_{H^\pm}$ are both possible in a 2HDM, but
large mass splittings among the scalars lead to sizeable contributions to EW precision observables
\cite{Toussaint:1978zm}, which
are parametrized by, e.g., the $S$, $T$ and $U$ parameters \cite{Peskin:1990zt}.
The case of exact degeneracy ($m_{A^0}=m_{H^0}=m_{H^\pm}$)
leads to values of $S$, $T$ and $U$ which are almost identical to those of the SM.
A recent analysis in a generic 2HDM \cite{Kanemura:2011sj}
sets $m_{H^0}=m_{A^0}$, $\sin(\beta-\alpha)=1$ and studies the maximum value of the
mass splitting $\Delta m= m_{A^0}-m_{H^\pm}$ (for earlier studies see \cite{Chankowski:1999ta}).
For $m_{A^0}=100$ GeV
the range $-70 \,{\rm GeV} < \Delta m < 20 \,{\rm GeV}$ is
allowed, which corresponds to $80 \,{\rm GeV} < m_{H^\pm} < 170 \,{\rm GeV}$.
For $m_{A^0}=150$ GeV the allowed range is instead
$-70 \,{\rm GeV} < \Delta m < 70 \,{\rm GeV}$, which corresponds to $80 \,{\rm GeV} < m_{H^\pm} < 220
\,{\rm GeV}$. Consequently, sizeable mass splittings (of either sign) of the scalars are possible.
Analogous studies in a MHDM have been performed in \cite{Grimus:2008nb}, with similar conclusions.

\begin{figure}[t]
\begin{center}
\includegraphics[origin=c, angle=0, scale=0.6]{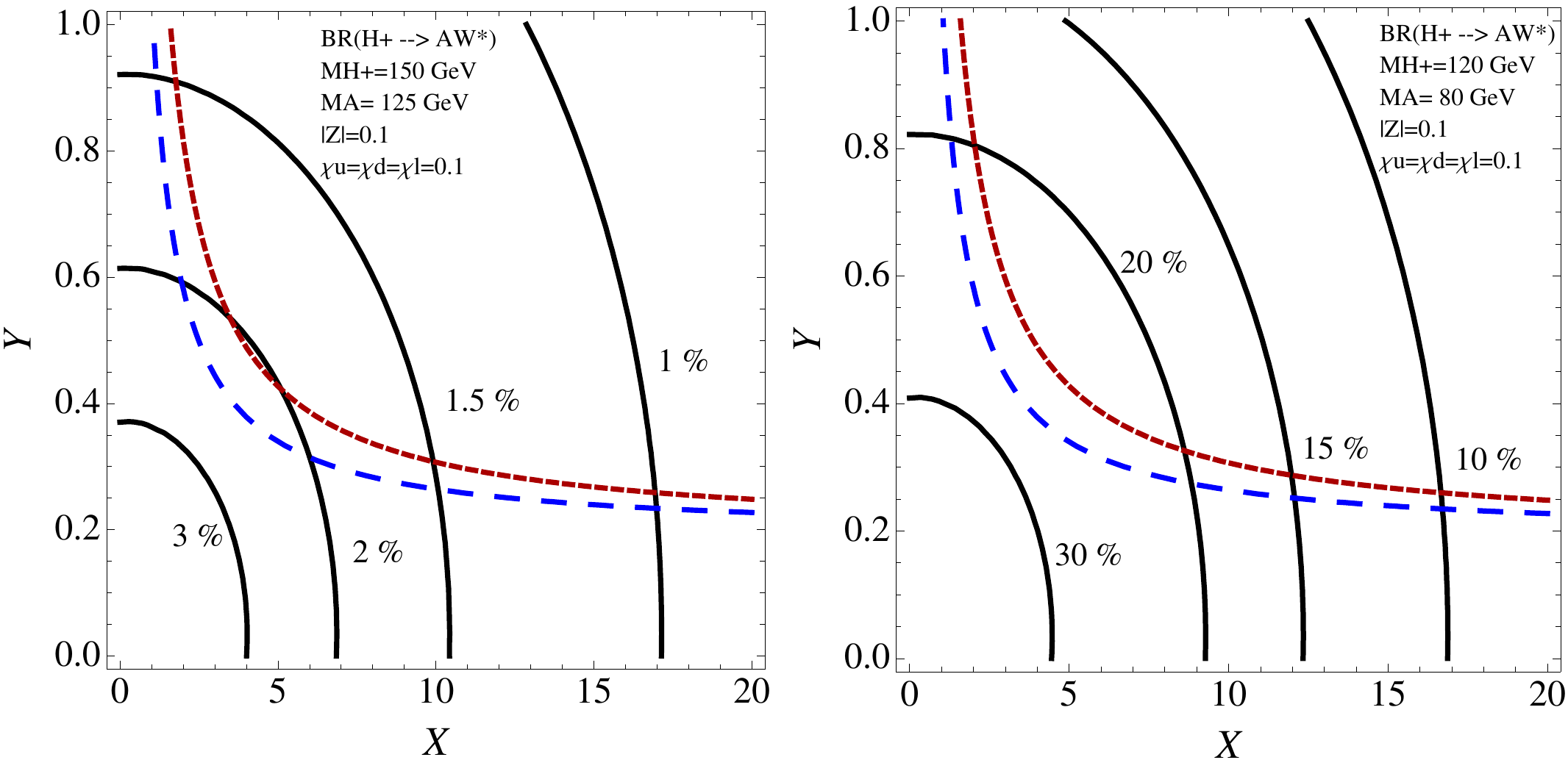}
\vspace*{-5mm}
\caption{  We show  the contours of the BR$(H^\pm \to AW^*)$ for   $\mh=150$ GeV, $m_{A^0}=125$ GeV (left)
and $\mh=120$ GeV, $m_{A^0}=80$ GeV (right) in the plane $[X,Y]$ with $Z=$ 0.1, $X>>Y$ and $\chi^f_{ij} = 0.1$ (all positive).  The constraint $b\to s \gamma$
is shown as $|fB1(X,Y)| < 1.1 $ for $ Re[fB1(X,Y)]<0$ (red-dashed), and $|fB1(X,Y)| < 0.7 $ is when $ Re[fB1(X,Y)]<0$ (blue-dashed).
 We take $m_s( Q= \mh) = 0.055$ GeV. }
\label{fig:braw}
\end{center}
\end{figure}

If $m_{A^0}< m_{H^\pm}$ then the decay channel $H^\pm\to A^0W^*$ can compete with the
above decays of $H^\pm$ to fermions, because the coupling $H^\pm A^0 W$ is
not suppressed by any small parameter. In Fig.~\ref{fig:braw} (left)  we show contours of BR($H^\pm\to A^0W^*$)
in the plane $[X,Y]$ with $|Z|=0.1$, $m_{A^0}=125$ GeV and $m_{H^\pm}=150$ GeV whereas in Fig.~\ref{fig:braw}
(right) we present the case $m_{A^0}=80$ GeV and $m_{H^\pm}=120$ GeV.
The contours are presented
in the parameter space of interest (i.e., $|Y|<1$ and $|X|>> 1$) because the contribution of the
term $m^2_c |Y_{22}|^2$ to the decay widths of $H^\pm$ to fermions could decrease. Comparing, e.g., Fig.~\ref{fig:braw} (right) to
Fig.~\ref{fig:brcb} (right)  one can see
that for $|X|\sim 4$  a BR$(H^\pm\to A^0W^*) \sim 30\%$ is competitive with a BR$(H^\pm\to cb) \sim 65\%$.
For smaller $m_{A^0}$ (e.g., less than $80 \,{\rm GeV}$) the contour of the BR$(H^\pm\to A^0W^*$)=$50\%$
would move to higher values of $|X|$.
Since the dominant decay of the $A^0$ is expected to be $A^0\to b\overline b$, the detection prospects in this channel
should also be promising because there would be more $b$ quarks from
$t\to H^\pm b$, $H^\pm \to A^0W^*$, $A^0\to b\overline b$ than from $t\to H^\pm b$ with $H^\pm \to cb$.
We note that there has been a search by the Tevatron for the channel $t\to H^\pm b$, $H^\pm \to A^0W^*$, $A^0\to \tau^+\tau^-$
\cite{Tev_search_HAW}, for the case
of $m_{A^0}<2m_b$ where $A^0\to b\overline b$ is not possible \cite{Dermisek:2008uu}.

At present there is much speculation about an excess of events around a mass of 125 GeV in the
search for the SM Higgs boson \cite{latest_Higgs}. An interpretation of these events as
originating from the process
$gg\to A^0 \to \gamma\gamma$ has been suggested in \cite{Burdman:2011ki}. In Fig.~\ref{fig:braw} (right)
we set $m_{A^0}=125$ GeV and $m_{H^\pm}=150$ GeV for this reason. Since the mass splitting between $H^\pm$ and $A^0$ is less than in
Fig.~\ref{fig:braw} (left), the contours move to lower values of $|X|$, but a BR$(H^\pm\to A^0W^*) =4\%$ is still possible
for $|X|<2$.
We note that if the excess of events at 125 GeV is attributed to a SM-like Higgs, then in the context of a 2HDM,
a candidate would be the lightest CP-even Higgs $h^0$ with a coupling to vector bosons of SM strength
(recent studies of this possibility can be found in
\cite{Ferreira:2011aa}). This scenario would correspond to $\sin(\beta-\alpha)\sim 1$ in a 2HDM
and therefore the coupling $H^\pm h^0 W$ (with a magnitude $\sim \cos(\beta-\alpha$) in a 2HDM) would be close
to zero. Hence the decay $H^\pm \to h^0 W^*$ would be suppressed by this small
coupling, as well as by the high virtuality of $W^*$. Several recent studies
\cite{Carmi:2012yp, Gabrielli:2012yz} fit the current data in all the Higgs search channels
to the case of a neutral Higgs boson with arbitrary couplings. A SM-like Higgs boson gives a good fit to the
data, although a slight preference for non-SM like couplings is emphasized in \cite{Gabrielli:2012yz}.
If the excess of events at 125 GeV turns out to be genuine {and} is well described by a non-SM like Higgs boson
of a 2HDM-III with a value of $\sin(\beta-\alpha)$ which is significantly less than unity, then the
BR($H^\pm \to h^0 W^*$) could be sizeable, with a magnitude given by Fig.~\ref{fig:braw} (right)
after scaling by $\cos^2(\beta-\alpha)$.

\section{Numerical Results}

\subsection{Indirect $H^\pm$ production}
We now quantify the magnitude of
$H^\pm \to cb$ events produced in the decays of $t$ quarks, and compare this
with the expected sensitivity at the LHC.
For the partial decay widths of $t\to W^\pm b$ and
$t\to H^\pm b$ we use the LO expressions (with $|V_{tb}|=1$) as follows:
\begin{eqnarray}
\Gamma(t\to W^\pm b)=\frac{G_F m_t}{8\sqrt 2 \pi}[m_t^2+2M_W^2][1-M_W^2/m_t^2]^2  \\
\Gamma(t\to H^\pm b)=\frac{G_F m_t}{8\sqrt 2 \pi}[m^2_t|Y_{33}|^2 + m_b^2|X_{33}|^2][1-m_{H^\pm}^2/m_t^2]^2.
\end{eqnarray}

The multiplicative (vertex) QCD corrections to both $t\to W^\pm b$ and
$t\to H^\pm b$ essentially cancel out in the ratio of partial widths \cite{Li:1990cp}.
In the phase-space function of both decays we
neglect $m_b$, and in the terms $m^2_t|Y_{33}|^2$ and $m_b^2|X_{33}|^2$
we use $m_t=175$ GeV and $m_b$ evaluated at the scale $m_{H^{\pm}}$ (i.e., $m_b\sim 2.95$ GeV).

\begin{figure}[t]
\begin{center}
\includegraphics[origin=c, angle=0, scale=0.5]{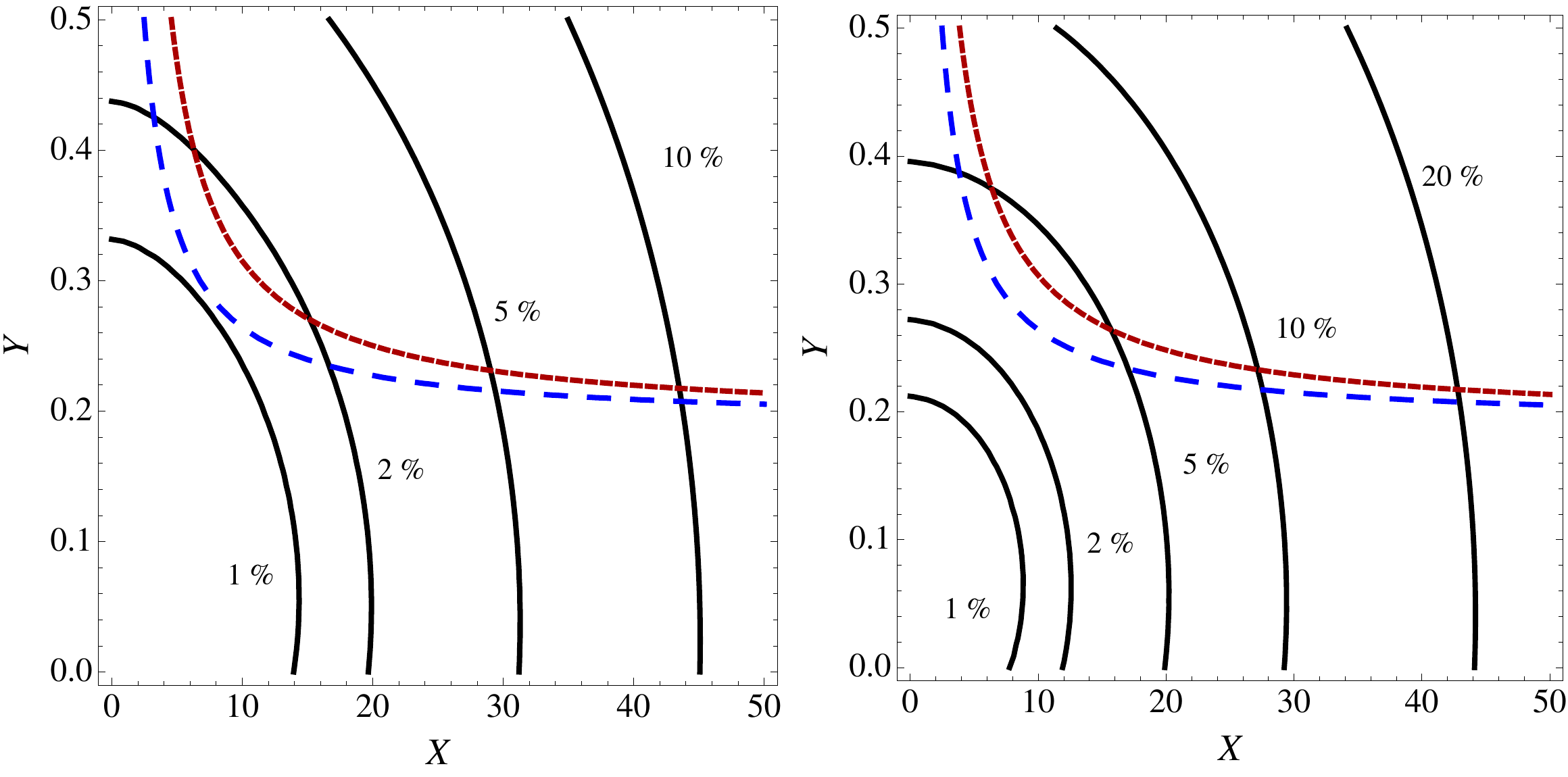}
\vspace*{-5mm}
\caption{Contours of the sum of BR$(t\to H^\pm b)\times {\rm BR}(H^\pm\to cs$)
and BR$(t\to H^\pm b)\times {\rm BR}(H^\pm\to cb$)
in the plane [$X$, $Y$] with $|Z|=0.1$, where $m_{H^\pm}=120$ GeV (left panel) and $m_{H^\pm}=80$ GeV (right panel).}
\label{fig:brcbcs}
\end{center}
\end{figure}

In Fig.~\ref{fig:brcbcs} we show contours of the sum of
\begin{equation}
{\rm BR}(t\to H^\pm b)\times [{\rm BR}(H^\pm\to cs) + {\rm BR}(H^\pm\to cb)]
\label{cbcs}
\end{equation}
in the plane of $[X,Y]$ for $m_{H^\pm}=120$ GeV
and $m_{H^\pm}=80$ GeV, setting $|Z|=0.1$. The cross section in eq.~(\ref{cbcs})
is the signature to which the current search strategy at
the Tevatron and the LHC is
sensitive, i.e., one $b$-tag (LHC \cite{ATLAS:search}) or two $b$-tags (Tevatron \cite{Aaltonen:2009ke})
are applied to the jets originating from the $t\overline t$ decay, but no $b$-tag is applied to the jets originating from $H^\pm$.
For the case of [BR($H^\pm \to cs)$+BR($H^\pm \to cb)$]=100\% the current experimental limits for $m_{H^\pm}=120$ GeV are
${\rm BR}(t\to H^\pm b) <0.14$ from ATLAS
with 0.035 fb$^{-1}$ \cite{ATLAS:search},
${\rm BR}(t\to H^\pm b) <0.12$ from CDF with 2.2 fb$^{-1}$ \cite{Aaltonen:2009ke}, and  ${\rm BR}(t\to H^\pm b) <0.22$ from
D0 with 1 fb$^{-1}$ \cite{:2009zh}. In Fig.~\ref{fig:brcbcs} (left) for $m_{H^\pm}=$120 GeV these upper limits would exclude
the parameter space of $|X|>50$ and small $|Y|$ which is not excluded by the constraint from $b\to s\gamma$.
For the mass region $80\, {\rm GeV} < m_{H^\pm} < 90$ GeV there is only a limit from the D0 search in \cite{:2009zh}, which
gives BR$(t\to H^\pm b)<0.21$. From Fig.~\ref{fig:brcbcs} (right), for $m_{H^\pm}=80$ GeV,
one can see that this limit excludes the parameter space with $|X|>45$ and small $|Y|$.

In Fig.~\ref{fig:brcbcs}  we show contours of $1\%$,
which might be reachable in the 8 TeV run of the LHC.
Simulations by ATLAS (with $\sqrt s=7$ TeV) for $H^\pm\to cs$ \cite{Ferrari:2010zz} have shown that the LHC
should be able to probe values of BR$(t\to H^\pm b)>0.05$ with 1 fb$^{-1}$
for $m_{H^\pm} > 110$ GeV, with the
greatest sensitivity being around $m_{H^\pm}=130$ GeV.
For the operation with $\sqrt s=8$ TeV and an anticipated integrated luminosity of 15 fb$^{-1}$
one expects increased sensitivity (e.g. BR$(t\to H^\pm b)>0.01$ for $m_{H^\pm} > 110$ GeV),
although the region $80 \,{\rm GeV} < m_{H^\pm} < 90$ GeV might remain difficult to probe with the strategy of
reconstructing the jets from $H^\pm$. An alternative way to probe  the
region $80 \,{\rm GeV} < m_{H^\pm} < 90$ GeV is to use
the search strategy by D0 in \cite{:2009zh}, and presumably the LHC could improve on the Tevatron
limit on BR$(t\to H^\pm b) < 0.21$ for this narrow mass region.
From  Fig.~\ref{fig:brcbcs} (left) (for $m_{H^\pm}=120$ GeV) one can see that the region
of $|Y|> 0.32$ and $|X|< 14$, which is not excluded by $b\to s \gamma$, would be probed
if sensitivities to the BR$(t\to H^\pm b)>0.01$ were achieved.
However, a large part of the
region roughly corresponding to $|Y|<0.32$ and $|X|< 20$ (which is also not excluded by $b\to s \gamma$) would require a
sensitivity to BR$(t\to H^\pm b)< 0.01$ in order to be probed
with the current search strategy for $t\to H^\pm b$ and this is probably unlikely in the 8 TeV run of the LHC.

\begin{figure}[t]
\begin{center}
\includegraphics[origin=c, angle=0, scale=0.5]{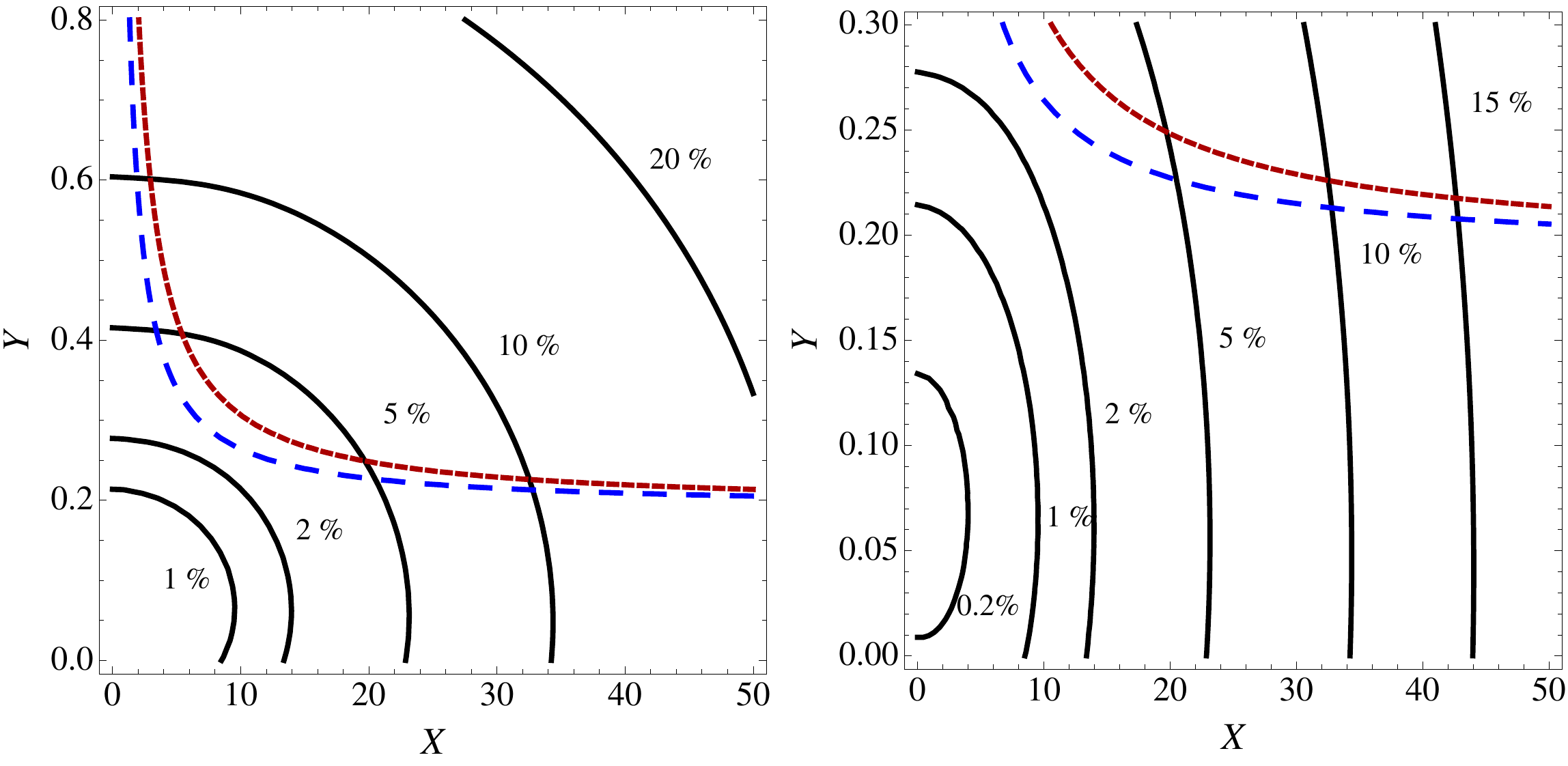}
\vspace*{-5mm}
\caption{Contours of BR$(t\to H^\pm b)\times {\rm BR}(H^\pm\to cb$) in the plane [$X$, $Y$] with $|Z|=0.1$ for
$m_{H^\pm}=80$ GeV.
 The constraint $b\to s \gamma$
is shown as $|fB1(X,Y)| < 1.1 $ for $ Re[fB1(X,Y)]<0$ (red-dashed),
and $|fB1(X,Y)| < 0.7 $ is when $ Re[fB1(X,Y)]<0$ (blue-dashed).
We take $m_s(Q=m_{H^\pm})=0.055$ GeV and show the range $0 < |Y| < 0.8$  (left panel) and $0 < |Y| < 0.3$
(right panel). }
\label{fig:brtbcb80}
\end{center}
\end{figure}

\begin{figure}[t]
\begin{center}
\includegraphics[origin=c, angle=0, scale=0.5]{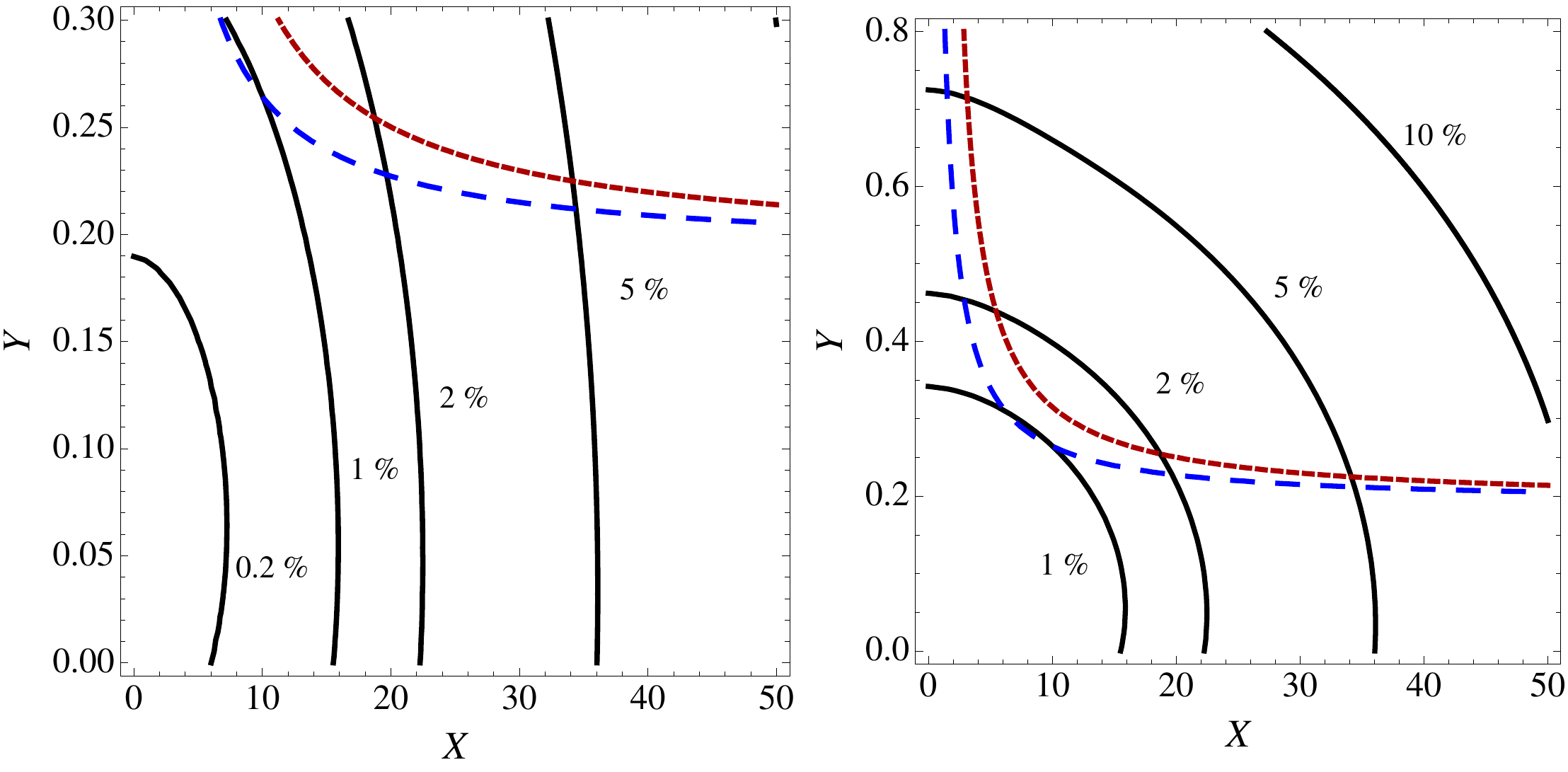}
\vspace*{-5mm}
\caption{Contours of BR$(t\to H^\pm b)\times {\rm BR}(H^\pm\to cb$) in the plane [$X$, $Y$] with $|Z|=0.1$ for
$m_{H^\pm}=120$ GeV.
 The constraint $b\to s \gamma$
is shown as $|fB1(X,Y)| < 1.1 $ for $ Re[fB1(X,Y)]<0$ (red-dashed),
and $|fB1(X,Y)| < 0.7 $ is when $ Re[fB1(X,Y)]<0$ (blue-dashed).
We take $m_s(Q=m_{H^\pm})=0.055$ GeV and show the range $0 < |Y| < 0.8$  (left panel) and $0 < |Y| < 0.3$
(right panel). }
\label{fig:brtbcb120}
\end{center}
\end{figure}

Increased sensitivity to the plane of $[X,Y]$ can be achieved by requiring a $b$-tag on the jets
which originate from the decay of $H^\pm$.
In Figs.~\ref{fig:brtbcb80} and \ref{fig:brtbcb120}, for $m_{H^\pm}=80$ GeV and $m_{H^\pm}=120$ GeV,
respectively, we show contours of
\begin{equation}
{\rm BR}(t\to H^\pm b)\times {\rm BR}(H^\pm\to cb) \,\, .
\label{cbcb}
\end{equation}

With the extra $b$-tag, as advocated previously (see eq.~(12) in \cite{Akeroyd:2012yg}), the sensitivity should reach
${\rm BR}(t\to H^\pm b)\times {\rm BR}(H^\pm\to cb) > 0.5\%$, and perhaps as low
as $0.2\%$. In the latter case, one can see from  Figs.~\ref{fig:brtbcb80} (left)
and \ref{fig:brtbcb120} (right)
that a large part of the regions of $|X|<6$ (for $m_{H^\pm}=120$ GeV) and $|X|<4$ (for $m_{H^\pm}=80$ GeV)
could be probed, even for $|Y|<0.2$. Therefore, there would be sensitivity to a sizeable region of the
parameter space of $[X,Y]$ which is not excluded
by $b\to s\gamma$, a result which is in contrast to the above case where no $b$-tag is applied
to the $b$-jets originating from $H^\pm$. We encourage here
a dedicated search for $t\to H^\pm b$
and $H^\pm \to cb$ by the Tevatron and LHC collaborations. Such a search
would be a well-motivated extension and application of the searches which have already been carried out in
\cite{Aaltonen:2009ke} and
\cite{ATLAS:search} and would offer the possibility of increased sensitivity to the fermionic couplings and
mass of $H^\pm$ in not only the MDHM/A2HDM but also the 2HDM-III.

\subsection{Direct $H^\pm$ production}

Another possibility to produce $H^\pm$ states in our scenario is via $c\bar b$ (and $b\bar c$) fusion in hadron-hadron collisions.
Since neither of the antiquarks in the initial state is a valence state, there is  no intrinsic advantage in exploiting proton-antiproton coliisions at  the Tevatron,
further considering the much reduced energy and luminosity available at the FNAL accelerator with respect to the LHC. Hence, we focus our attention onto the latter. In fact, we can anticipate that, amongst all the energy and luminosity stages occurred or foreseen at the CERN machine, only the combinations $\sqrt s=8,14$ TeV and
standard  instantaneous luminosity of order $10^{33}$ cm$^{-2}$ $s^{-1}$ or higher are of relevance here. Furthermore, in the QCD polluted environment of the LHC, it is clear that one ought to attempt extracting a leptonic decay of our light charged Higgs boson. In the light of the results presented in the previous
(sub)sections, the best option is afforded by $H^\pm$ decays into $\tau\nu$ pairs, eventually yielding an electron/muon (generically denoted by $l$) and missing (transverse) energy.
In this case, the background is essentially due to the charged Drell-Yan (DY) channel giving $\tau\nu$ pairs (i.e., via $W^\pm$ production and decay), which is in fact irreducible.

\begin{figure}[htb]
\centering
\includegraphics[width=0.35\linewidth,angle=90]{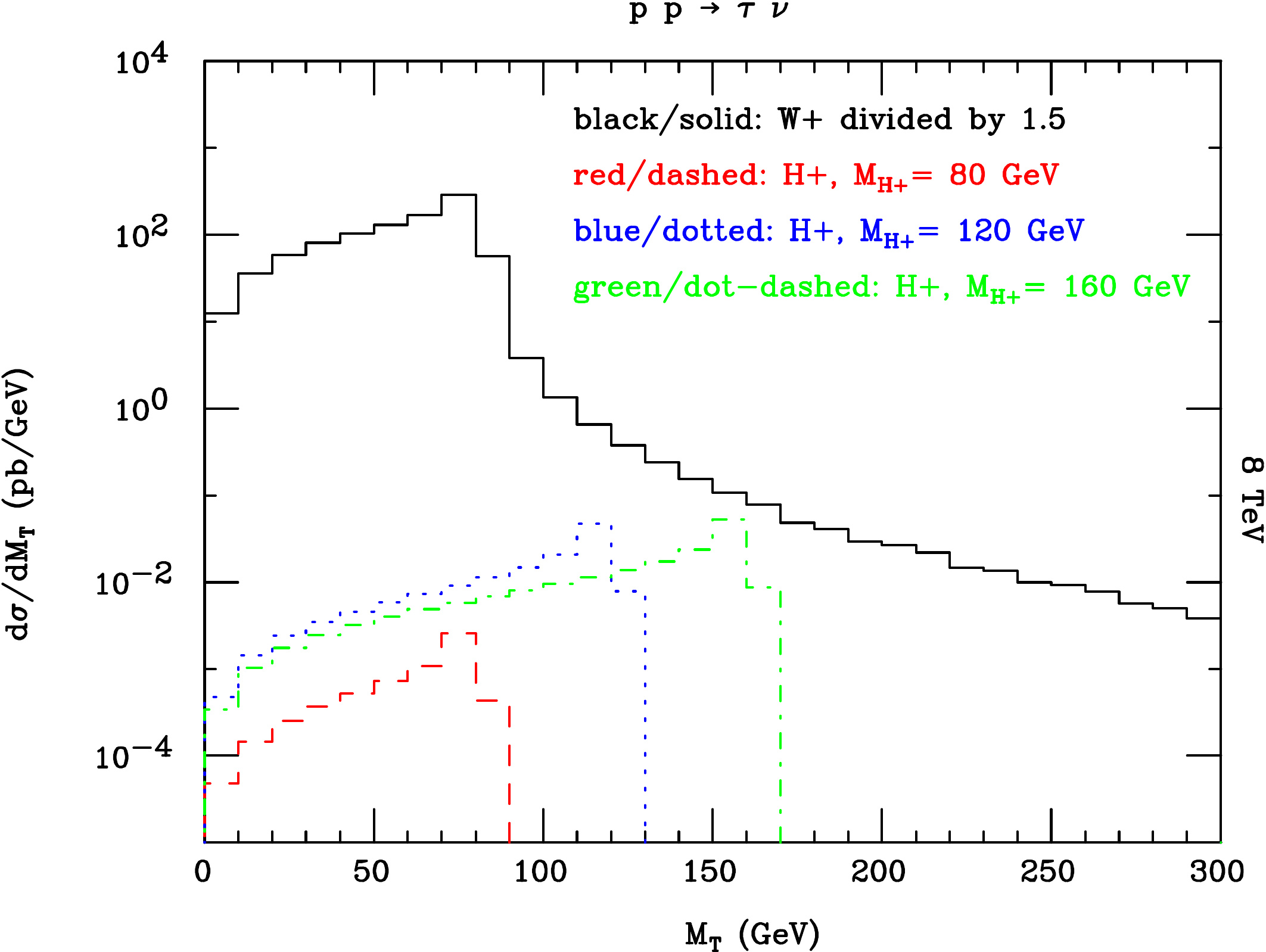}
\includegraphics[width=0.35\linewidth,angle=90]{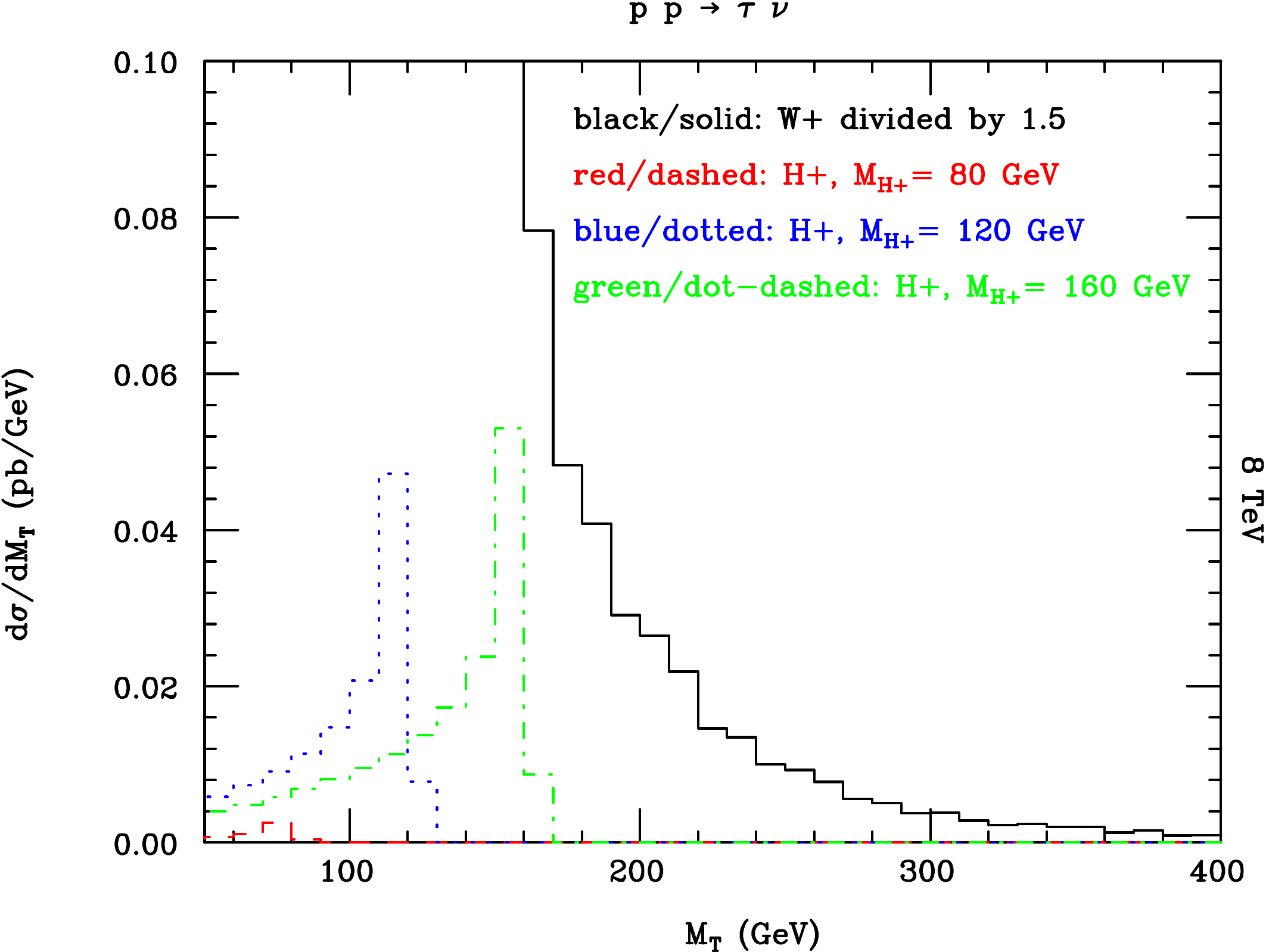}
\caption{Differential distributions in transverse mass $M_T$ for signal and background (the former for three $H^\pm$ mass values) in logaritmic (top) and linear (bottom) scale. Here, $\sqrt s=8$ TeV.}
\protect{\label{fig:cb8}}
\end{figure}

\begin{figure}[htb]
\centering
\includegraphics[width=0.35\linewidth,angle=90]{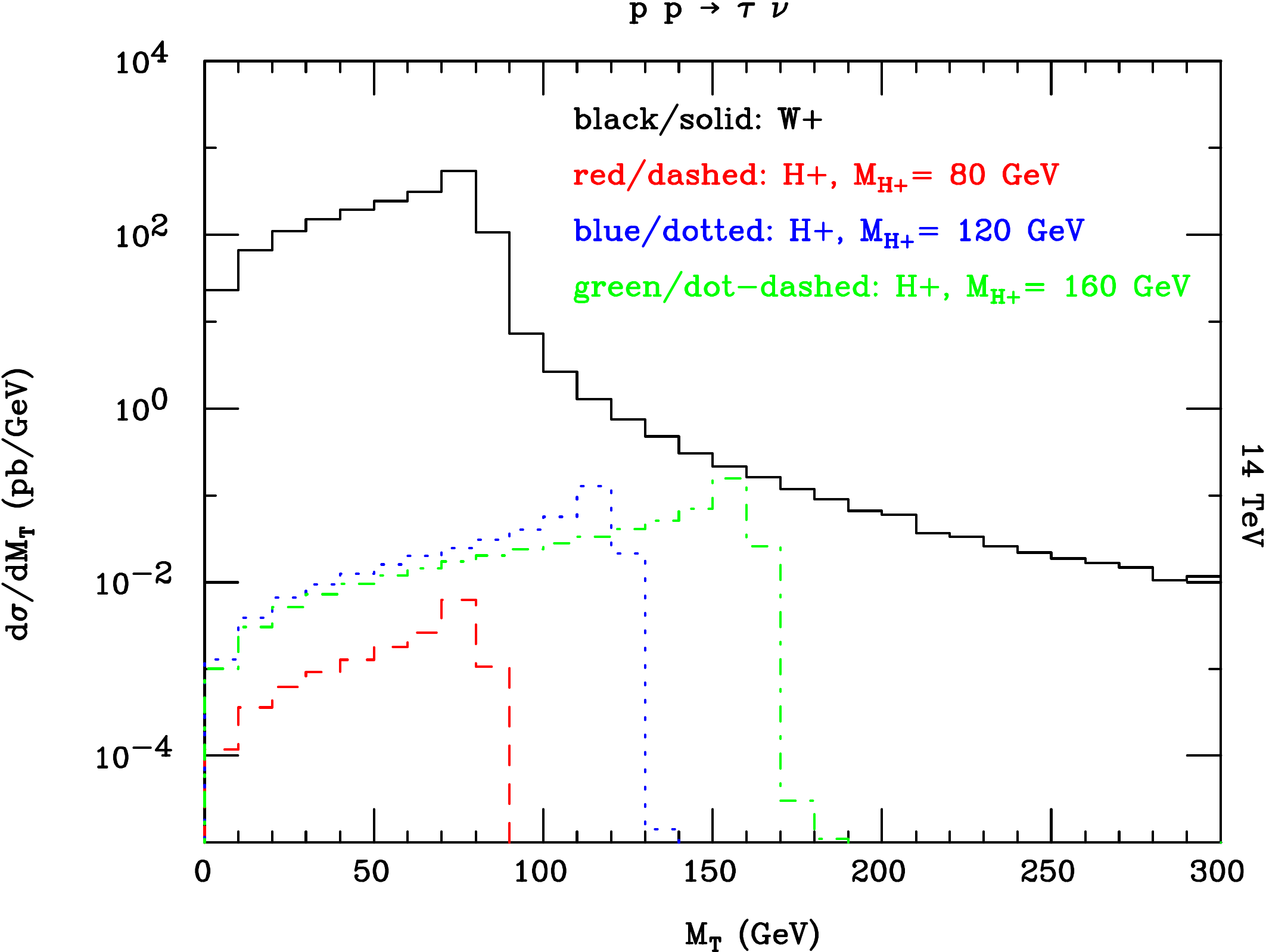}
\includegraphics[width=0.35\linewidth,angle=90]{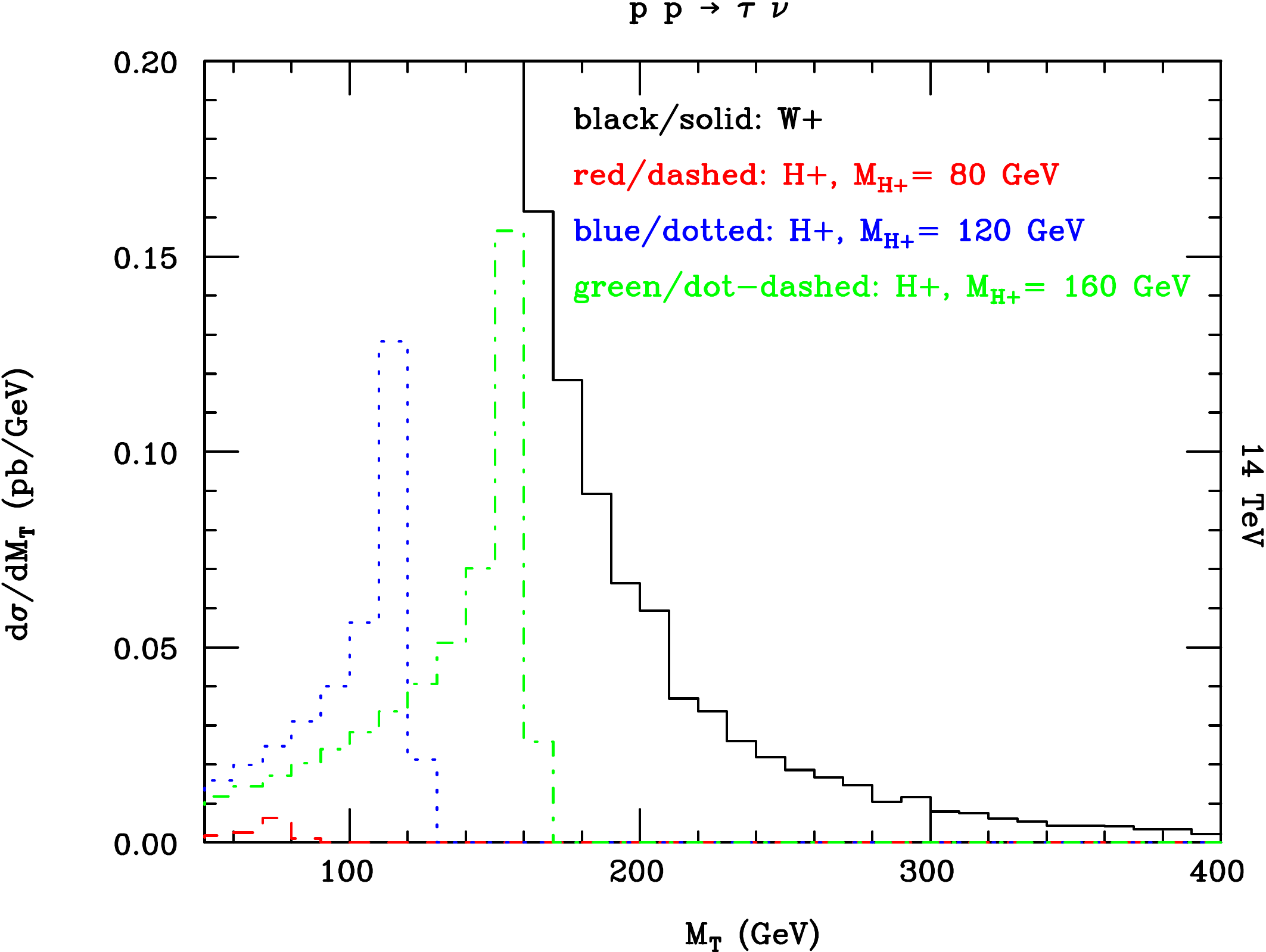}
\caption{The same as Fig.~\ref{fig:cb8}, but for $\sqrt s=14$ TeV.}
\protect{\label{fig:cb14}}
\end{figure}

Figs.~\ref{fig:cb8}--\ref{fig:cb14} display the differential distribution for the signal $S$ (i.e. $c\bar b\to H^+\to \tau^+\nu~+~c.c.$) and the
background $B$ (i.e., $c\bar b\to W^+\tau^+\nu~+~c.c$), the former for three mass choices ($m_{H^\pm}=80, 120$ and
160 GeV)\footnote{Notice that interference effects are negligible between the two, owing to the very
narrow width of the charged Higgs boson.}. As the invariant mass of the final state is not reconstructible, given the missing longitudinal momemtum,
we plot the transverse mass $M_T\equiv\sqrt{(E^T_l+E^T_{\rm miss})^2
             -(p^x_{l}+p^x_{\rm miss})^2
             -(p^y_{l}+p^y_{\rm miss})^2}$,
where $E^T$ represents missing energy/momentum (as we consider the electron and muon massless) in the transverse plane and $p_{x,y}$ are the two components therein
(assuming that the proton beams are directed along the $z$ axis). Clearly, the backgound dominance
over the signal is evident whenever $m_{H^\pm}\approx M_{W^\pm}$.
However, the larger the charged Higgs boson mass with respect to the gauge boson one, the more important the signal becomes relatively to the background.

\begin{figure}[htb]
\centering
\includegraphics[width=0.35\linewidth,angle=90]{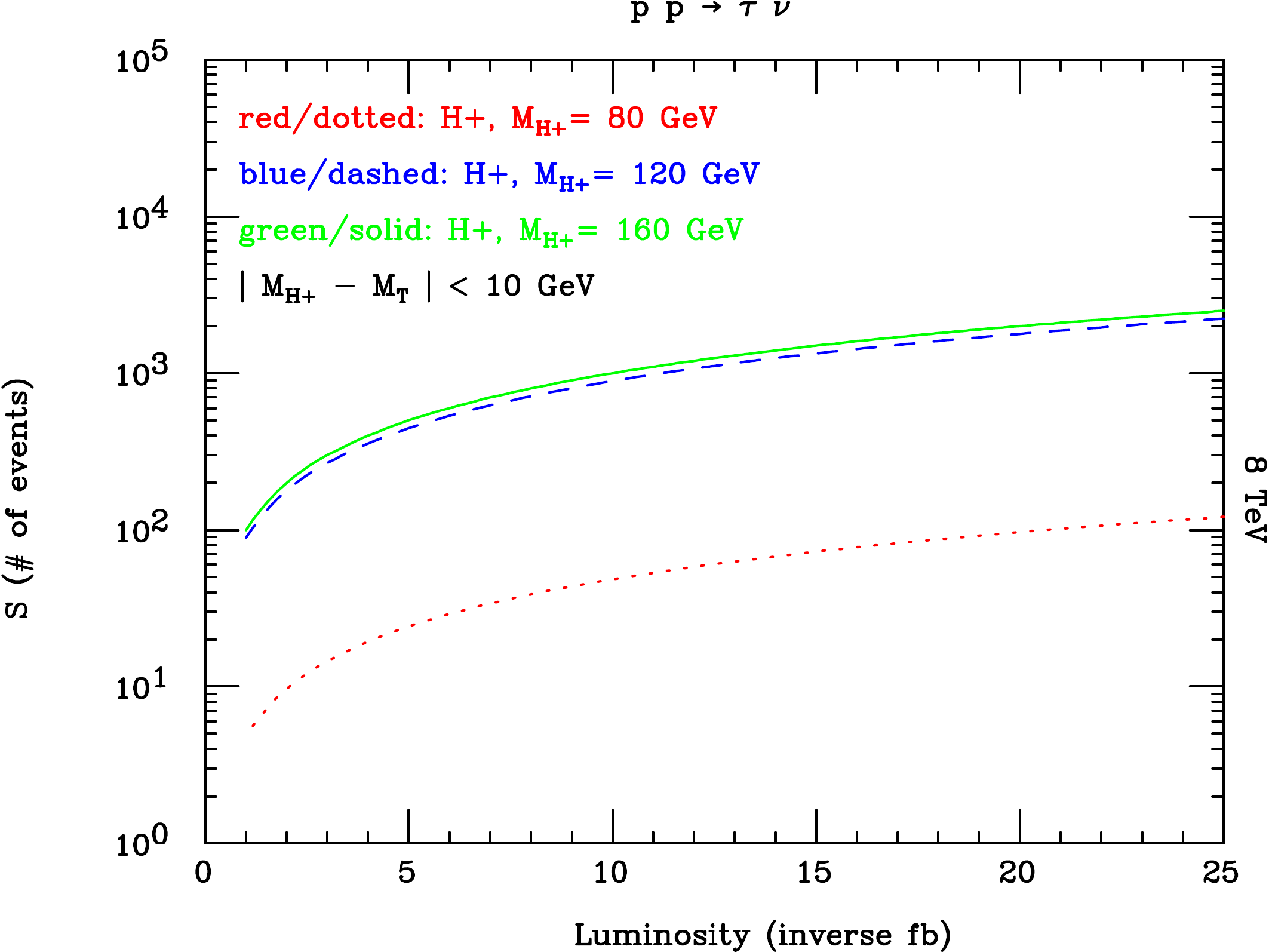}
\includegraphics[width=0.35\linewidth,angle=90]{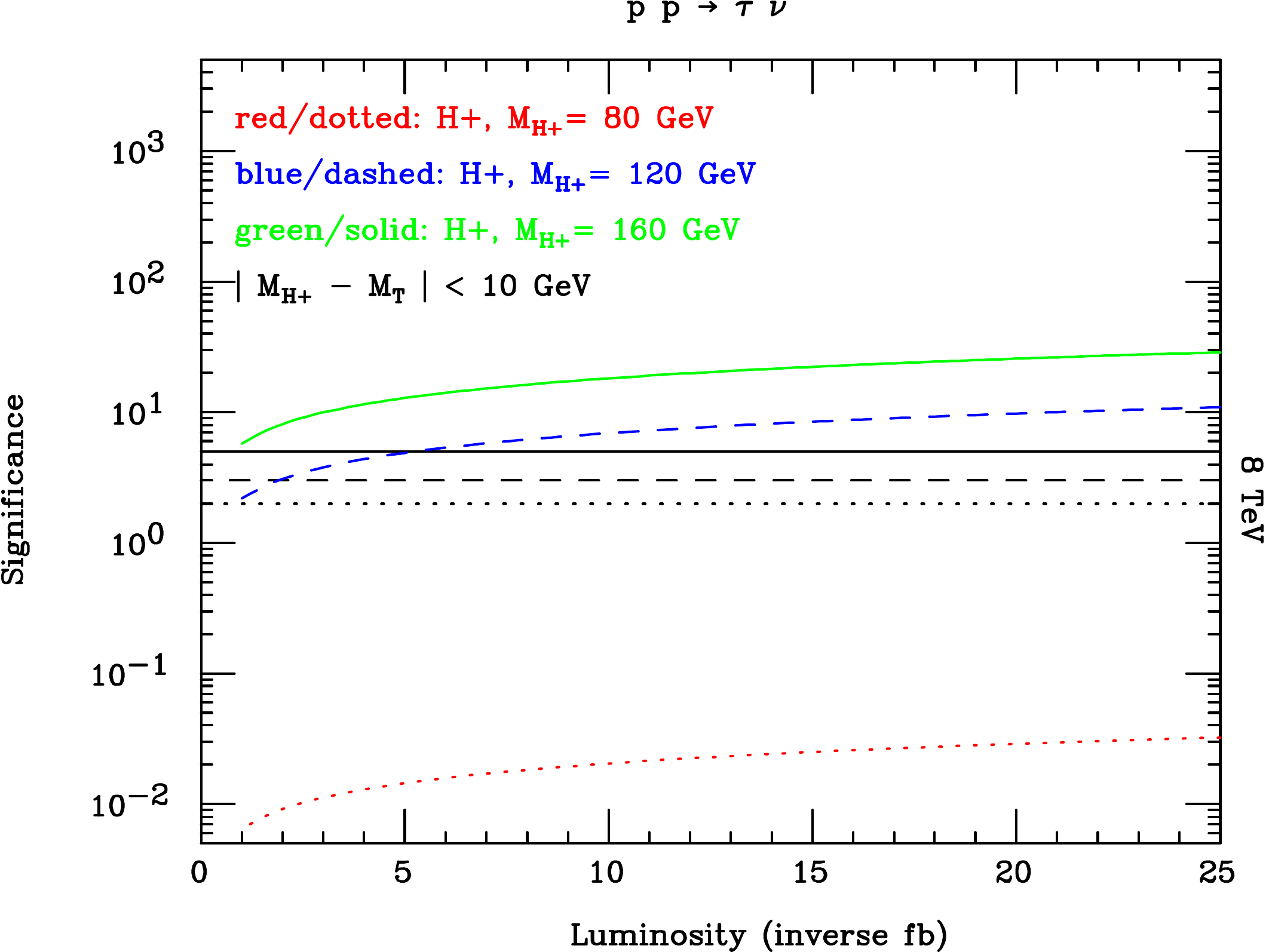}
\caption{Number of events (top) and significance (bottom) of the signal (for three $H^\pm$ mass values) as a function of the
luminosity. Here, $\sqrt s=8$ TeV. The horizontal solid(dashed)[dotted] line corresponds to a significance of 5(3)[2].}
\protect{\label{fig:Lumi8}}
\end{figure}

\begin{figure}[htb]
\centering
\includegraphics[width=0.35\linewidth,angle=90]{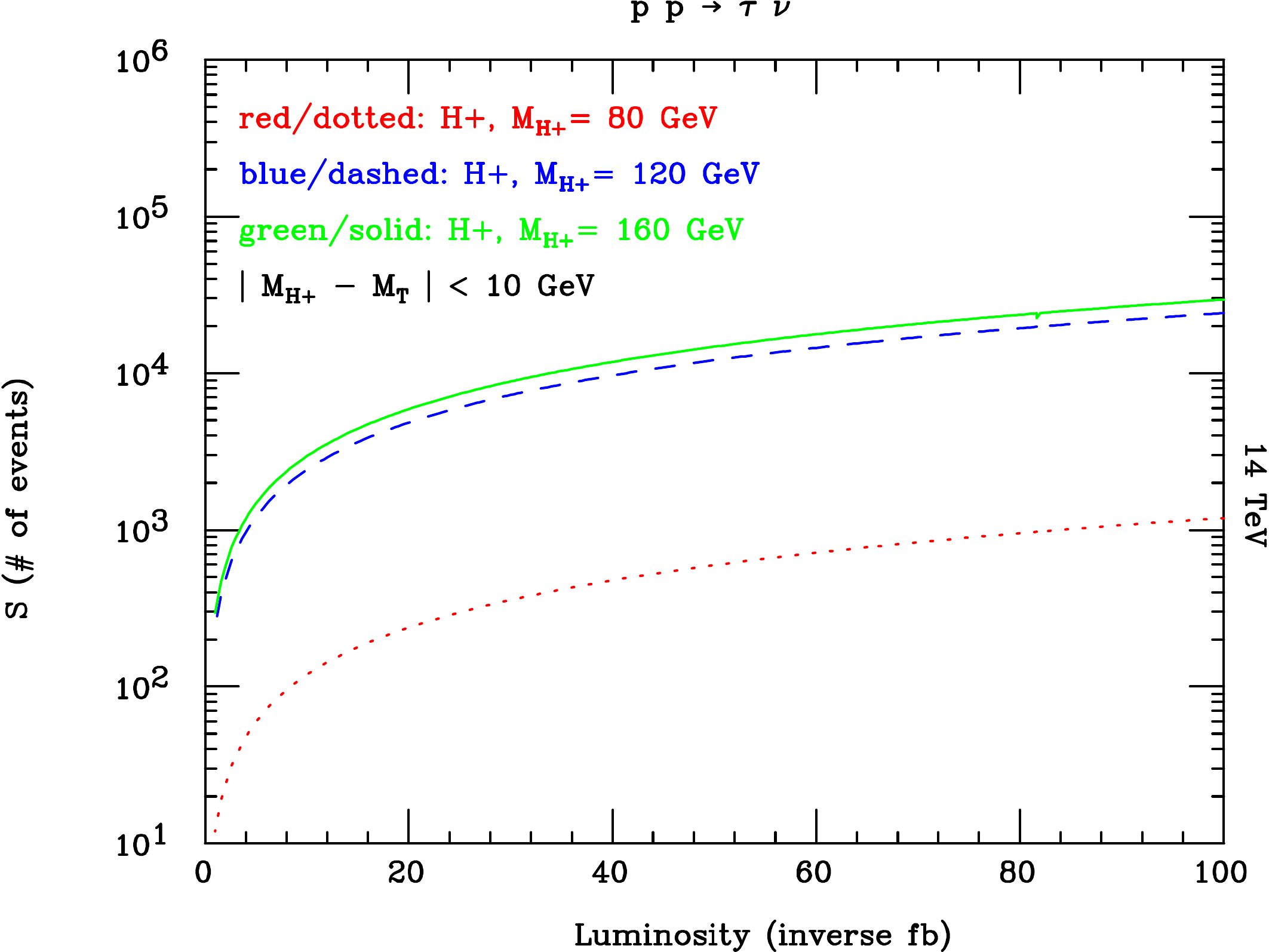}
\includegraphics[width=0.35\linewidth,angle=90]{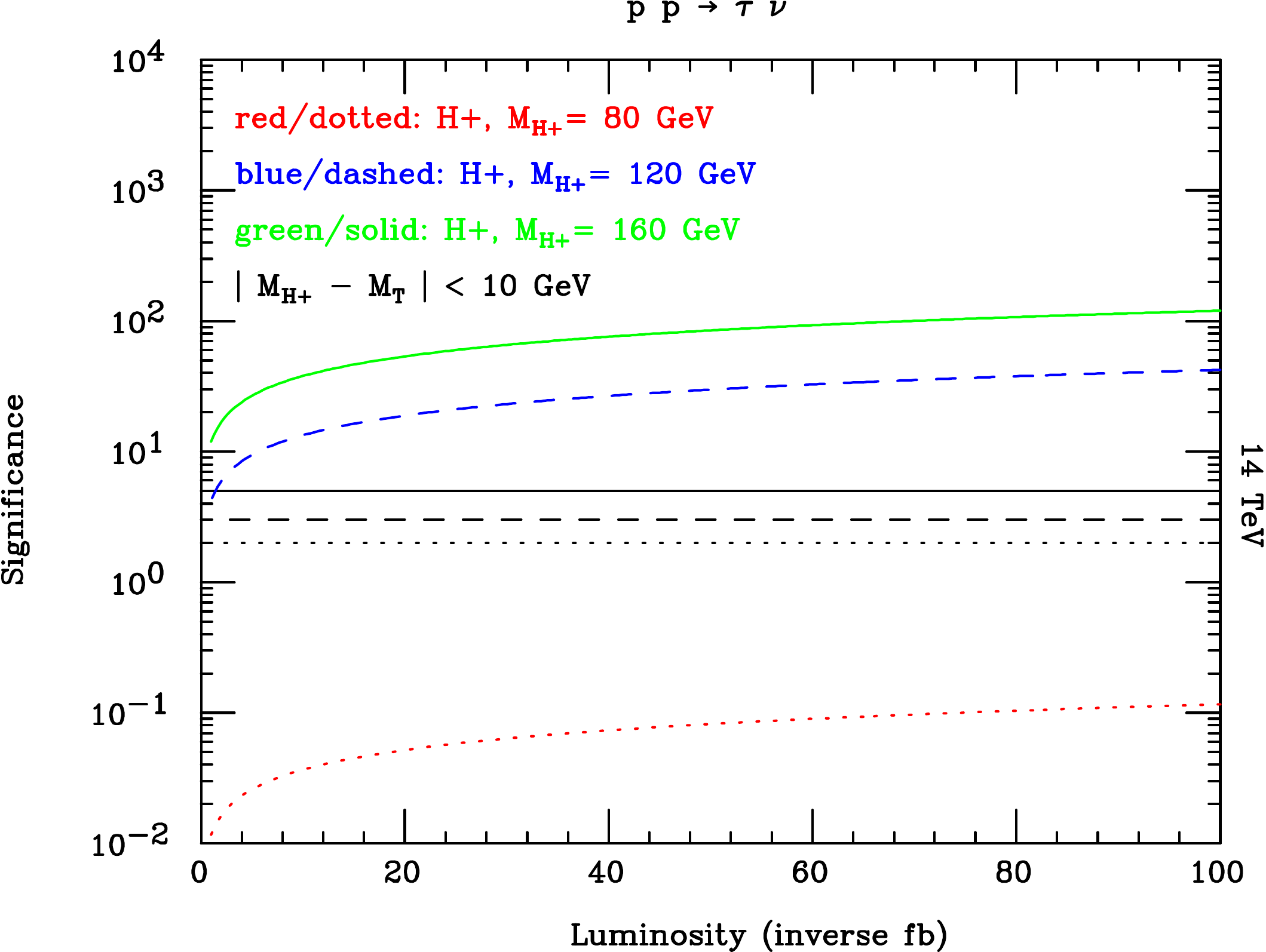}
\caption{The same as Fig.~\ref{fig:Lumi8}, but for $\sqrt s=14$ TeV.}
\protect{\label{fig:Lumi14}}
\end{figure}

In order to establish whether it is possible to extract the direct $H^\pm$ signal, or indeed constrain it, we show in Figs.~\ref{fig:Lumi8}--Figs.~\ref{fig:Lumi14}
the significance of the signal, defined as $S/\sqrt B$, as a function of the collider integrated luminosity $L$, where $S=L~\sigma(c\bar b\to H^+\to\tau^+\nu~+~c.c.)$
and $B=L~\sigma(c\bar b\to W^+\tau^+\nu~+~c.c.)$. Here, we have restricted both the signal and background yield to the mass regions
$|m_{H^\pm}-M_T|<10$ GeV, where, again, we have taken $m_{H^\pm}=80, 120$ and 160 GeV. The $\tau$ decay rates into electron/muons are included and
we assumed 90\% efficiency in $e/\mu$-identification. It is clear that exclusion (significance equal to 2),
evidence (significance equal to 3) and discovery (significance equal to 5) can all be attained at both the 8 and 14 TeV energy stages, for accessible
luminosity samples,
so
long that the $H^\pm$ mass is significantly larger than the $W^\pm$ one. In both machine configurations, corresponding event rates are always substantial.

\section{Conclusions}

Light charged Higgs bosons ($H^\pm$) are being searched for in the decays of top quarks
($t\to H^\pm b$) at the Tevatron and LHC. Separate searches are being
carried out for the decay channels $H^\pm \to cs$ and
$H^\pm \to \tau\nu$, with comparable sensitivity to the mass and fermionic couplings of $H^\pm$.
The searches for $H^\pm \to cs$ in \cite{Aaltonen:2009ke} and
\cite{ATLAS:search} look for a peak at $m_{H^\pm}$ in the dijet invariant mass distribution,
with the assumption that neither of the quarks is a $b$ quark.

In some models with two or more Higgs doublets (the 2HDM and
a MHDM with three or more scalar doublets) the BR for
$H^\pm \to cb$ can be as large as $80\%$. Here, in our model (2HDM-III),
BR($H^\pm \to cb$) could be as large as $90\%$.
Moreover, such a $H^\pm$
could be light enough to be
produced via $t\to H^\pm b$ as well as respect the stringent constraints from $b\to s\gamma$ on both $m_{H^\pm}$ and
the fermionic couplings of $H^\pm$. This is in contrast to the $H^\pm$ state in other 2HDMs for which a large
BR for $H^\pm \to cb$ is possible but for which one expects $m_{H^\pm} > m_t$ in order to comply with
the measured value of $b\to s\gamma$. Along the lines indicated by previous literature,
in the context of the 2HDM-III with a four-zero Yukawa texture,
we suggested that a dedicated search for $t\to H^\pm b$ and $H^\pm \to cb$ would
probe values of the fermionic couplings of $H^\pm$ that are currently not testable.
Such a search would require a $b$-tag of one of the jets originating from $H^\pm$ and would further allow one to reach a
higher sensitivity to a smaller value of the BR($t\to H^\pm b$) than that obtained in the ongoing searches, which
currently do not make use of this additional $b$-tag.

Finally, we have  shown that a $H^\pm$ state of the 2HDM-III can also be produced
directy from $cb$ fusion, followed by a decay into $\tau\nu$ pairs, assuming semi-leptonic
decays (to contrast the otherwise overwhelming QCD background), albeit only at the LHC,
which should have sensitivity to it already at 8 TeV and, indeed, full coverage at 14 TeV, so
long that $m_{H^\pm}$ is 120 GeV or above.

We have reached these conclusions over the 2HDM-III parameter space that we have shown to
survive a long list of constraints emerging from $B$-physics, namely:
$\mu -e$ universality in $\tau$ decays, several leptonic $B$-decays ($B \to \tau \nu$, $D \to \mu \nu$ and $D_s \to {l} \nu$), the semi-leptonic transition $B\to D \tau \nu$, plus
$B \to X_s \gamma$, including $B^0-\bar B^0$ mixing, $B_s \to \mu^+ \mu^-$ and the radiative decay $Z\to b \bar{b}$.

The outlook is therefore clear. Depending on the search channel, both the Tevatron and the LHC
have the potential to constrain or else discover the 2HDHM-III supplemented by a four-zero
Yukawa texture including non-vanishing off-diagonal terms in the Yukawa matrices. We are now
calling on our experimental colleagues to achieve this, as the search strategies recommended here
are easily implementable and pursuable.

 \section*{Acknowledgements}
  S.M. is supported in part through the NExT Institute.
J. H.-S. thanks the University of Southampton and the Rutherford Appleton Laboratory for hospitality
during his visit to the NExT Institute where  part of this work was carried out.
J. H.-S, A. R.-S and R. N.-P acknowledge the financial
support of SNI, CONACYT, SEP-PROMEP and VIEP-BUAP.
We all thank A.G. Akeroyd for innumerable constructive and useful discussions.



\end{document}